\documentclass{emulateapj}
\usepackage{amssymb}
\usepackage{epsfig}

\begin{document}

\title{{\em Spitzer} Quasar and ULIRG Evolution Study (QUEST). IV. \\
  Comparison of 1-Jy Ultraluminous Infrared Galaxies with
  Palomar-Green Quasars}

\author{S. Veilleux\footnote{Also: Max-Planck-Institut f\"ur
    extraterrestrische Physik, Postfach 1312, D-85741 Garching,
    Germany}, D. S. N. Rupke\footnote{New address: Institute for
    Astronomy, University of Hawaii, 2680 Woodlawn Drive, Honolulu, HI
    96822.}, and D.-C. Kim\footnote{New address: Department of
    Astronomy, University of Virginia, Charlottesville, VA
    22904-4325.}}  \affil{Department of Astronomy, University of
  Maryland, College Park, MD 20742-2421; \\ veilleux@astro.umd.edu,
  drupke@ifa.hawaii.edu, ddk3wc@mail.astro.virginia.edu}

\author{R. Genzel, E. Sturm, D. Lutz, A. Contursi, M. Schweitzer, and
  L. J. Tacconi} \affil{Max-Planck-Institut f\"ur extraterrestrische
  Physik, Postfach 1312, D-85741 Garching, Germany; genzel@mpe.mpg.de,
  sturm@mpe.mpg.de, lutz@mpe.mpg.de, contursi@mpe.mpg.de,
  schweitzer@mpe.mpg.de, linda@mpe.mpg.de}

\author{H. Netzer and A. Sternberg}
\affil{School of Physics and Astronomy and The Wise Observatory, The
  Raymond and Beverly Sackler Faculty of Exact Sciences, Tel-Aviv
  University, Tel-Aviv 69978, Israel; netzer@wise1.tau.ac.il,
  amiel@wise.tau.ac.il}

\author{J. C. Mihos}
\affil{Department of Astronomy, Case Western Reserve University, 10900
  Euclid Avenue, Cleveland, OH 44106; mihos@case.edu}

\author{A. J. Baker}
\affil{Department of Physics and Astronomy, Rutgers, the State
  University of New Jersey, Piscataway, NJ 08854-8019;
  ajbaker@physics.rutgers.edu}

\author{J. M. Mazzarella and S. Lord}
\affil{IPAC, California Institute of Technology, MS 100-22, Pasadena,
  CA 91125; mazz@ipac.caltech.edu, lord@ipac.caltech.edu}

\author{D. B. Sanders, A. Stockton, R. D. Joseph, and J. E. Barnes}
\affil{Institute for Astronomy, University of Hawaii, 2680 Woodlawn
  Drive, Honolulu, HI 96822; sanders@ifa.hawaii.edu,
  stockton@ifa.hawaii.edu, joseph@ifa.hawaii.edu,
  barnes@ifa.hawaii.edu}

\author{}
\affil{}

\begin{abstract}

  We report the results from a comprehensive study of 74 ultraluminous
  infrared galaxies (ULIRGs) and 34 Palomar-Green (PG) quasars within
  $z\sim0.3$ observed with the {\em Spitzer} Infrared Spectrograph
  (IRS). The contribution of nuclear activity to the bolometric
  luminosity in these systems is quantified using six independent
  methods that span a range in wavelength and give consistent results
  within $\sim$ $\pm$10$-$15\% on average.  This agreement suggests
  that deeply buried AGN invisible to {\em Spitzer} IRS but bright in
  the far-infrared are not common in this sample. The average derived
  AGN contribution in ULIRGs is $\sim$35$-$40\%, ranging from
  $\sim15-35$\% among ``cool'' ($f_{25}/f_{60}\le0.2$) optically
  classified HII-like and LINER ULIRGs to $\sim$50 and $\sim$75\%
  among warm Seyfert 2 and Seyfert 1 ULIRGs, respectively. This number
  exceeds $\sim$80\% in PG~QSOs.  ULIRGs fall in one of three distinct
  AGN classes: (1) objects with small extinctions and large PAH
  equivalent widths are highly starburst-dominated; (2) systems with
  large extinctions and modest PAH equivalent widths have larger AGN
  contributions, but still tend to be starburst-dominated; and (3)
  ULIRGs with both small extinctions and small PAH equivalent widths
  host AGN that are at least as powerful as the starbursts. The AGN
  contributions in class 2 ULIRGs are more uncertain than in the other
  objects, and we cannot formally rule out the possibility that these
  objects represent a physically distinct type of ULIRGs. A
  morphological trend is seen along the sequence $(1)-(2)-(3)$, in
  general agreement with the standard ULIRG $-$ QSO evolution scenario
  and suggestive of a broad peak in extinction during the intermediate
  stages of merger evolution.  However, the scatter in this sequence,
  including the presence of a significant number of AGN-dominated
  systems prior to coalesence and starburst-dominated but fully merged
  systems, implies that black hole accretion, in addition to depending
  on the merger phase, also has a strong chaotic/random component, as
  in local AGN.

\end{abstract}

\keywords{galaxies: active -- galaxies: interactions -- galaxies:
  quasar -- galaxies: starburst -- infrared: galaxies}

\clearpage

\section{Introduction}

More than twenty years ago, Sanders et al.\ (1988a, 1988b) proposed
the existence of an evolutionary connection between ultraluminous
infrared galaxies (ULIRGs; log[$L$(IR)$/L_\odot$] $\ge$
12)\footnote{Throughout the paper, $L$(IR) and $L$(FIR) refer to the
$8-1000$ \micron\ and $40-120$ \micron\ luminosities as defined in
Sanders \& Mirabel (1996).} and quasars. Their imaging and
spectrophotometric data on ten local ULIRGs were interpreted to imply
that ULIRGs are dust-enshrouded quasars formed through the strong
interaction or merger of two gas-rich spirals. This merging process
had long been suspected to lead to the formation of an elliptical
galaxy (e.g., Toomre \& Toomre 1972) and subsequent numerical
simulations have lent support to this idea (e.g., Barnes 1989;
Kormendy \& Sanders 1992; Springel et al.\ 2005; Bournaud et al.\
2005; Naab et al.\ 2006). ULIRGs have been found since then to be an
important population in the distant universe, a major contributor to
the cosmic star formation at $z \ga 1-2$ (see, e.g., reviews by
Sanders \& Mirabel 1996; Blain et al.\ 2002; Lonsdale et al.\ 2006).
The evidence is growing that the majority of the high-z ULIRGs are fed
by continuous gas accretion, and only about one third are gas-rich
(``wet'') mergers (Daddi et al.\ 2007; Shapiro et al.\ 2008; Genzel et
al.\ 2006, 2008; F\"{o}rster Schreiber et al.\ 2006, 2009; Genel et
al.\ 2008; Dekel \& Birnboim 2008; Dekel et al.\ 2009). However, the
fraction of ULIRGs involved in mergers appears to increase steeply at
higher infrared luminosity (e.g., submillimeter-selected galaxies;
Tacconi et al.\ 2006, 2008).

Considerable effort has been devoted in the past decade to understand
ULIRGs and quasars in the local universe, where galaxy merging and its
relation to starbursts and AGN can be studied in greater detail than
in the distant universe.  Our group is conducting a comprehensive,
multiwavelength imaging and spectroscopic survey of local ULIRG and
QSO mergers called QUEST -- {\em Q}uasar and {\em U}LIRG {\em
E}volution {\em ST}udy. QUEST has already provided crucial new
insights into merger morphology, kinematics, and evolution: we now
know that ULIRGs are advanced mergers of gas-rich, disk galaxies
sampling the Toomre merger sequence beyond the first peri-passage
(Veilleux et al.\ 2002).  The near-infrared (NIR) light distributions
in many ULIRGs, particularly those with AGN-like optical and infrared
characteristics, show prominent early-type morphology ($R^{1/4}$ law;
Wright et al. 1990; Scoville et al.\ 2000; Veilleux et al.\ 2002,
2006).  The hosts of ULIRGs lie close to the locations of
intermediate-size ($\sim$ 1 -- 2 $L^*$) spheroids in the photometric
projection of the fundamental plane of ellipticals, although there is
a tendency for the ULIRGs with small hosts to be brighter than normal
spheroids.  Excess emission from a merger-triggered burst of star
formation in the ULIRG hosts may be at the origin of this difference.

NIR stellar absorption spectroscopy with the VLT and Keck
has also been carried out by our group to constrain the host dynamical
mass for many of these ULIRGs.  The analysis of these data (Dasyra et
al.\ 2006a, 2006b) built on the analyses of Genzel et al.\ (2001) and
Tacconi et al.\ (2002) and revealed that the majority of ULIRGs are
triggered by almost equal-mass major mergers of 1.5:1 average ratio,
in general agreement with Veilleux et al.\ (2002). In Dasyra et al.\,
we also found that coalesced ULIRGs resemble intermediate mass
ellipticals/lenticulars with moderate rotation, in their velocity
dispersion distribution, their location in the fundamental plane and
their distribution of the ratio of rotation/velocity dispersion
[v$_{\rm rot}$ sin(i)/$\sigma$].  These results therefore suggest that
ULIRGs form moderate mass ($m^\ast \sim 10^{11}$ M$_\odot$), but not
giant (5 -- 10 $\times$ 10$^{11}$ M$_\odot$) ellipticals.  Converting
the host dispersion into black hole mass with the aid of the M$_{\rm
  BH}$ -- $\sigma_*$ relation (e.g., Gebhardt et al.\ 2000, Ferrarese
\& Merritt 2000) yields black hole mass estimates ranging from
10$^{7.0}$ $M_\odot$ to 10$^{8.7}$ $M_\odot$, with slightly larger
values in coalesced ULIRGs than in binaries. BH masses derived from
similar data on a dozen PG QSOs agree with those of coalesced ULIRGs
(Dasyra et al.\ 2007). A recent analysis of HST/NICMOS data on several
of these PG~QSOs appears to support this conclusion (Veilleux et al.\
2009; see also Surace et al.\ 2001; Guyon et al.\ 2006).

QUEST has also provided new quantitative information on the importance
of gas flows in and out of ULIRGs.  Direct evidence for powerful
galaxy-scale winds has been found in most ULIRGs (e.g., Rupke et al.\
2002, 2005a, 2005b, 2005c; see also Martin 2005 and Veilleux et al.\
2005), and the metal underabundance and smaller yield measured in the
cores of these objects (Rupke et al.\ 2008) point to strong
merger-induced gas inflows in the recent past as predicted by
numerical simulations (e.g., Barnes \& Hernquist 1996; Mihos \&
Hernquist 1996; Iono et al.\ 2004; Naab et al.\ 2006).

The last two crucial issues addressed by QUEST are the nature of the
energy production mechanism -- starburst or AGN -- in ULIRGs and QSOs
and the importance of dust extinction along the merger sequence. Early
optical and NIR spectroscopy has revealed trends of increasing AGN
dominance among ULIRGs with the largest infrared luminosity, warmest
25-to-60 $\mu$m color, and latest merger phase (e.g., Veilleux et al.\
2006 and references therein), but these results are potentially biased
by dust obscuration.  Mid-infrared (MIR) spectroscopy with the {\em
  Infrared Space Observatory} ({\em ISO}) (e.g., Genzel et al.\ 1998;
Lutz et al.\ 1998b; Lutz et al.\ 1999; Rigopoulou et al. 1999; Tran et
al.\ 2001) has provided crucial new information on the energy source
in ULIRGs, less affected by the effects of dust, although the number
of objects in the sample was limited by the relatively modest
sensitivity of {\em ISO}.

The most recent progress in this area of research occurred with the
advent of the {\em Spitzer Space Telescope}. The Infrared Spectrograph
(IRS) Guaranteed-Time and General Observation (GTO and GO) programs
have provided a wealth of new information on the physical properties
of local ULIRGs (e.g., Armus et al.\ 2004, 2007; Farrah et al.\ 2007;
Desai et al.\ 2007; Hao et al. 2007; Higdon et al.\ 2006; Imanishi et
al.\ 2007; Lahuis et al.\ 2007; Spoon et al.\ 2007; CAo et al.\
2008). In the present paper, we revisit a carefully selected subset of
these data and combine them with GO-1 IRS data acquired by our QUEST
program, to calculate AGN contributions to the total {\em bolometric}
luminosities, quantify the issue of ULIRG evolution along the merger
sequence, and to determine where, if at all, quasars fit in this
picture.

Our earlier papers in this series have focussed exclusively on the
QSOs. In Schweitzer et al.\ (2006; Paper I), we showed that starbursts
are responsible for at least $\sim$30\%, but likely most, of the FIR
luminosity of PG~QSOs.  We argued in Netzer et al.\ (2007; Paper II)
that both strong- and weak-FIR emitting sources have the same, or very
similar, intrinsic AGN spectral energy distributions (SEDs).  In
Schweitzer et al.\ (2008; Paper III), we found that emission from dust
in the innermost part of the narrow-line region is needed in addition
to the traditional obscuring torus in order to explain the silicate
emission in these QSOs. The present paper reports the results from our
analysis of the continuum, emission line, and absorption line
properties of 74 ULIRGs and 34 QSOs.  In Section 2, we describe the
sample. Next, we discuss the observational strategy of our program and
the methods we used to obtain, reduce, and analyze the IRS spectra,
including the archived data (Section 3, Section 4, and Section 5,
respectively). The results are presented in Section 6 and tested
against the evolution scenario of Sanders et al.\ (1988a, 1988b) in
Section 7. The main results and conclusions are summarized in Section
8.  Throughout this paper, we adopt $H_0$ = 70 km s$^{-1}$ Mpc$^{-1}$,
$\Omega_M$ = 0.3, and $\Omega_\Lambda$ = 0.7.

\section{Sample}

\begin{figure*}[ht]
\epsscale{0.9}
\plotone{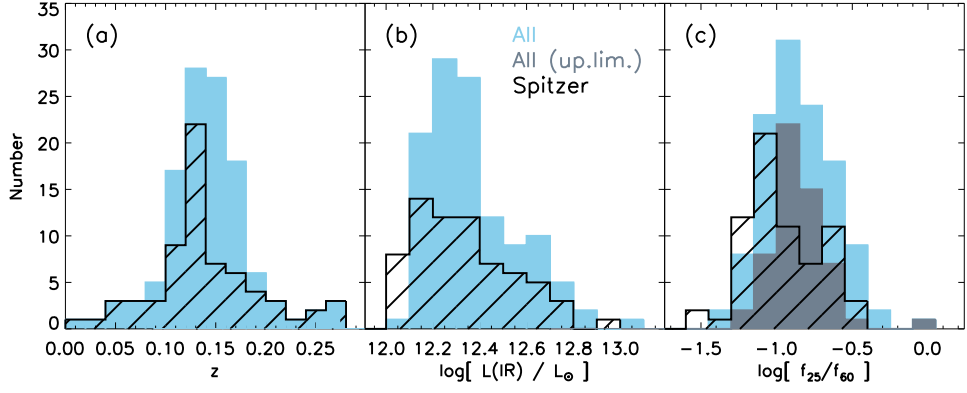}
\caption{ Distributions of ($a$) redshifts, ($b$) infrared
  luminosities, and ($c$) 25-to-60\micron\ flux ratios of the {\em
    Spitzer} 1 Jy ULIRGs (black hatched histogram) {\em vs.\/} entire 1 Jy
  sample (blue histogram).  In panels ($b$) and ($c$), {\it IRAS} 12
  and 25\micron\ fluxes are used for the entire 1 Jy sample.  Many of
  these fluxes are upper limits, which we label explicitly in panel
  ($c$).  {\it IRAS}-type {\it Spitzer} 12 and 25\micron\ fluxes are
  used for the {\em Spitzer} subsample.  The new flux measurements account
  for the imperfect overlap in panels ($b$) and ($c$).  The {\em Spitzer}
  ULIRGs are representative of the 1-Jy sample in both range and
  distribution of properties.}
\label{fig:basic-ulirgs}
\end{figure*}

The basic properties of the ULIRGs and quasars in our sample are
listed individually in Table \ref{tab:sample}. For a summary of the
properties of the ULIRGs by spectral types, infrared colors and
luminosities, and morphology, see Table $\ref{tab:agnfrac_avg}$. The
ULIRG component of our program focuses on the 1-Jy sample, a complete
flux-limited sample of 118 ULIRGs selected at 60 $\mu$m from a
redshift survey of the {\em IRAS} faint source catalog (Kim \& Sanders
1998). All 1-Jy ULIRGs have $z < 0.3$.  Twenty-nine objects were
observed under our own Cycle 1 medium-size program (\#3187; PI
Veilleux; Note that the 1-Jy ULIRG Mrk~1014 is also
PG~0157+001). These objects were selected to be representative of the
1-Jy sample as a whole in terms of redshift, luminosity, and {\em
IRAS} 25-to-60 $\mu$m colors.  These data were supplemented by
archival IRS spectra of 39 other galaxies from the 1-Jy sample, and 5
archival IRS spectra of infrared-luminous galaxies from the Revised
Bright Galaxy Sample (RBGS, Sanders et al. 2003; these objects are
UGC~05101, F10565+2448, F15250+3609, NGC~6240, and F17208$-$0014).
Most of the archival spectra are from GTO program \#105 (PI Houck),
and three are from GO program \#20375 (PI Armus).  These spectra cover
bright sources in the 1~Jy sample, while ours are deeper exposures of
fainter ones.  Together, they represent almost 2/3 of the 1~Jy sample.
The 5 RBGS spectra represent well-studied benchmarks from the local
universe.  Figure \ref{fig:basic-ulirgs} shows the distributions of
redshifts, infrared luminosities, and 25-to-60 $\mu$m {\em IRAS}
colors for the combined set of ULIRGs compared with that of the entire
1-Jy sample.  We confirm that the ULIRGs in our study are
representative of the range of properties of the 1-Jy sample.  Optical
spectral types, which are referred to extensively in this paper, are
taken from Veilleux et al. 1999a and Rupke et al. 2005a for the 1~Jy
sample, and Veilleux et al. 1995 for the 5 RBGS objects.

\begin{figure*}[ht]
\epsscale{0.9}
\plotone{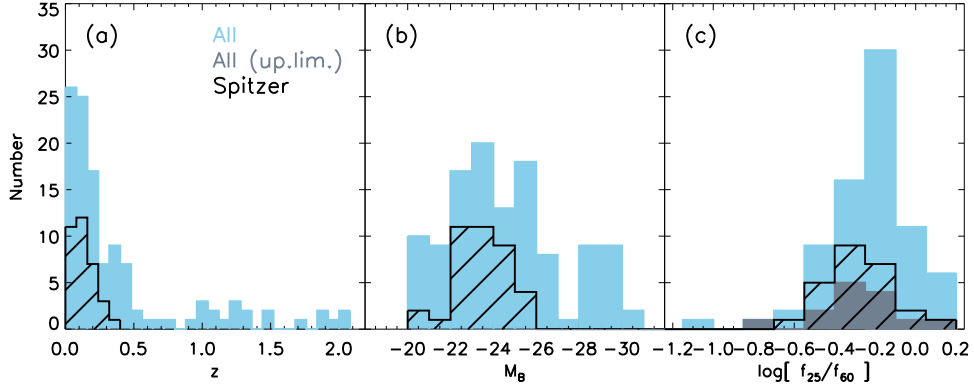}
\caption{ Distributions of ($a$) redshifts, ($b$) B-band absolute
  magnitudes, and ($c$) 25-to-60 $\mu$m flux ratios of the {\em Spitzer}
  PG~QSOs (black hatched histogram) {\em vs.\/} entire PG~QSO sample
  (blue histogram).  As in Figure \ref{fig:basic-ulirgs}, {\it IRAS}
  25\micron\ fluxes are used for the entire sample, and {\it
    IRAS}-type {\it Spitzer} fluxes for the current subsample.  The
  {\em Spitzer} QSOs sample the low redshift and low luminosity ends of
  the PG~QSO sample.}
\label{fig:basic-qsos}
\end{figure*}

\begin{figure*}[ht]
\epsscale{0.5}
\plotone{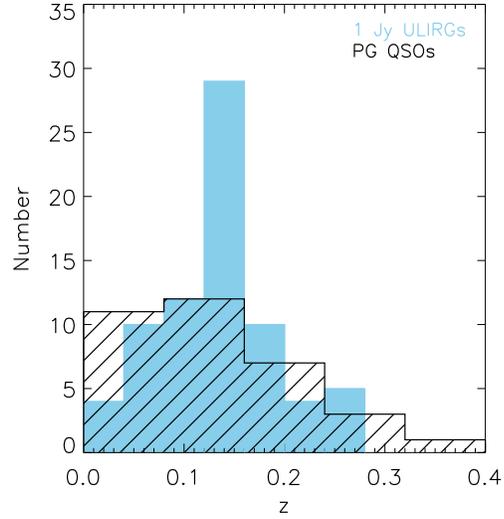}
\caption{ Redshift distributions of the {\em Spitzer} ULIRGs (blue
  histogram) and PG~QSOs (black hatched histogram). The two samples
  are well matched in redshift. }
\label{fig:z}
\end{figure*}

The original QUEST sample of quasars has already been discussed in
detail in Papers I and II and this discussion will not be repeated
here.  Suffice it to say that the original QUEST sample contains 25 $z
\la 0.3$ quasars, including 24 Palomar-Green (PG) quasars from the
Bright Quasar Sample (Schmidt \& Green 1983) and another one
(B2~2201+31A = 4C 31.63) with a $B$ magnitude that actually satisfies
the PG QSO completeness criterion of Schmidt \& Green (1983).  Nine
other PG~QSOs (PG~0050+124 = I~Zw~1, PG~0804+761, PG~1119+120 =
Mrk~734, PG~1211+143 = Mrk~841, PG~1244+026 [NLS1], PG~1351+640,
PG~1448+273 [NLS1], and PG~1501+106) observed under different Spitzer
programs were later added to the quasar sample.  Figure
\ref{fig:basic-qsos} emphasizes the fact that the quasars in our study
cover the low redshift and low B-band luminosity ends of the PG~QSO
sample, while Figure \ref{fig:z} shows that the ULIRGs and quasars in
our study are well matched in redshift.  Finally, note that two
ULIRGs, Mrk~1014 and 3C~273, are also PG QSOs; we treat them as ULIRGs
for the purposes of this study.

High-quality optical and NIR images obtained from the ground
and with {\em HST} are available for all ULIRGs and quasars in the
present sample (e.g, Surace \& Sanders 1999; Scoville et al.\ 2000;
Surace et al.\ 1998, 2001; Guyon et al.\ 2006; Veilleux et al.\ 2002,
2006, 2009).  In addition, high-quality optical spectra exist for all
1-Jy ULIRGs (e.g., Veilleux et al.\ 1999a; Farrah et al.\ 2005; Rupke
et al.\ 2005a, 2005b, 2005c) and PG QSOs (Boroson \& Green 1992), and
a large subset of these objects also have been the targets of
NIR JHK-band spectroscopy by our group over the years (e.g.,
Veilleux et al.\ 1997, 1999b; Dasyra et al.\ 2006a, 2006b, 2007) as
well as some L-band spectroscopy (e.g., Imanishi et al.\ 2006a;
Risaliti et al.\ 2006; Imanishi et al.\ 2008; Sani et al. 2008).
These ancillary data will be used for our interpretation of the {\em
  Spitzer} data in Section 6 and Section 7.

\section{Observations}

Galaxies from our own program (\#3187; PI Veilleux) were observed in
the IRS modules SL, SH, and LH, using staring mode (Houck et
al. 2004).  Together, these modules cover observed wavelengths of
$5-35$~\micron.  The high resolution data at observed wavelengths of
$10-35$~\micron\ (SH and LH modules, with resolution $R\sim600$) allow
sensitive measurements of important atomic and molecular emission
lines. 

For targeting, moderate-accuracy IRS blue peak-ups were performed on
the targets themselves rather than offsetting from 2MASS stars.  This
peak-up method is justified given the compact MIR continua in
these systems (e.g., Soifer et al. 2000; Surace et al. 2006).

For the four binary ULIRGs with nuclear separations exceeding
3\arcsec\ (F01166$-$0844, F10190$+$1322, F13454$-$2956, and
F21208$-$0519), a unique observation was made of each nucleus.
However, in 3 of these cases (F10190$+$1322 being the exception),
aperture effects due to the larger slit sizes of the long-wavelength
modules allowed accurate measurements of only one of the two nuclei.

The observational setup used for the archival IRS spectra is described
in detail in Armus et al. (2007) and references therein.  It is
essentially the same as the one we used for our own program so direct
comparison between the two data sets is permissible.

Some objects in our sample have full low-resolution spectra ({\em
i.e.}, including both the SL and LL modules, covering $5-35$ \micron).
We have used only the high-resolution data for spectral line
measurements (except for the [Ne~VI] line, which falls in the SL
module).  The LL data was used primarily for checking flux
calibration.  However, when only SL$+$LL data was available, or when
the high-resolution data was of low S/N, the full low-resolution spectrum
was used in the continuum fitting.

{\it Spitzer} proposal ID numbers and exposure times for each IRS
module are listed for all galaxies in Table \ref{tab:obs}.

\section{Data  Reduction} 

For the majority of QUEST sources, we started with BCD data processed
by version S12.0 of the IRS pipeline.  For the non-QUEST 1~Jy ULIRGs
that were reduced at a later date, data from pipelines S12, S13, or
S15 were used.  Comparisons among these pipelines show only minor
differences that do not impact our measurements.

The data were first corrected for rogue pixels using an automatic
search-and-interpolate algorithm (which was supplemented by visual
examination).  For the SL module, background light was then subtracted
by differencing the two nod positions.  For the SL and LL modules, the
data was extracted prior to coadding.  For SH and LH, we coadded
exposures for a given nod position prior to extraction.

The one-dimensional spectra were extracted using the Spectroscopic Modeling
Analysis and Reduction Tool (SMART; Higdon et al. 2004).  The
extraction apertures were tapered with wavelength to match the
point-spread function for SL and LL data and encompassed the entire
slit for SH and LH data.  The correction/extraction process was
iterated until we were assured that the majority of hot pixels had
been removed.

For SL, we combined the two one-dimensional nod spectra for the orders
SL1 and SL2 separately and then stitched the orders by trimming a few
pixels from one or the other order.  (We discarded SL3 because of flux
discrepancies.)  For SH and LH, we combined the two nods and then
stitched the orders together by applying multiplicative offsets for
each order that were linear in flux density {\em vs.\/} wavelength
(effectively removing a tilt artifact from certain orders where
necessary).

We subtracted zodiacal light from the high-resolution data using a
blackbody fit to the Spitzer Planning Observations Tool (SPOT)
zodiacal estimates at 10, 20, and 35$\micron$.

Finally, we matched modules in flux to form complete $5-35$~\micron\
spectra.  Because ULIRGs are compact mid-IR sources, different modules
in general agree well in flux at wavelengths where they overlap.
Where there was disagreement, we used available low-resolution spectra
(SL $+$ LL) to improve the zodiacal subtraction, since these spectra
were sky-subtracted using simultaneous sky observations.  Where this
was not possible, we used small additive offsets.

\begin{figure*}[ht]
\epsscale{1.1}
\plotone{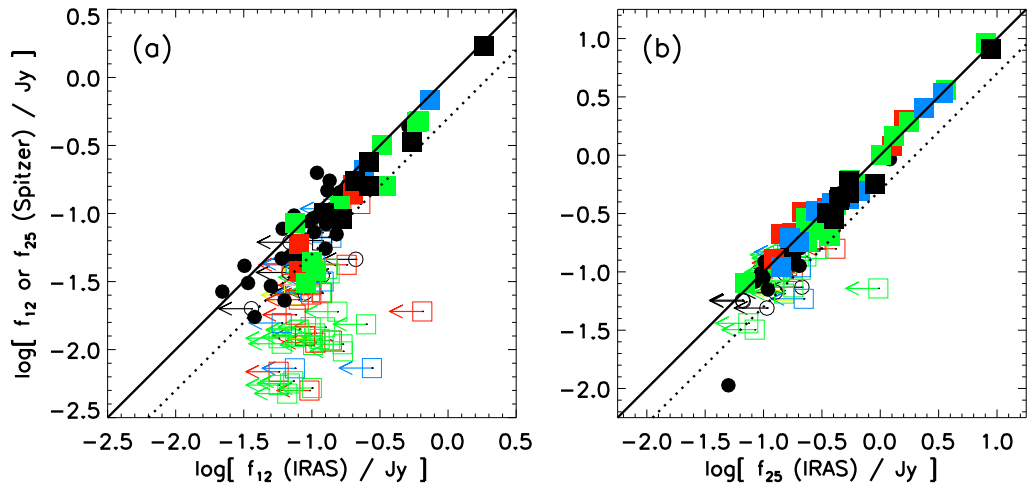}
\caption{ Comparisons of {\em IRAS} ($a$) 12 and ($b$) 25 $\mu$m flux
  densities with {\em Spitzer}-derived quantities. Squares represent
  ULIRGs and circles are QSOs. Open symbols are upper limits.  The
  colors of the symbols reflect the optical spectral types: red,
  green, blue, and black squares are HII-like, LINER, Seyfert 2, and
  Seyfert 1 ULIRGs, respectively. The solid line is perfect agreement,
  and the dotted line is for $f(IRAS)/f(Spitzer)=2$.  Excellent
  agreement is seen at the 5\% level at 25 $\mu$m and at the 25\%
  level at 12 $\mu$m.}
\label{fig:fluxcomp}
\end{figure*}

To check the flux calibration, we computed synthetic {\it IRAS} flux
densities at 12 and 25~\micron\ using the {\it Spitzer} data by
averaging the flux densities over the {\it IRAS} bandpasses.  At
25~\micron, the agreement with {\it IRAS} is excellent
(Figure \ref{fig:fluxcomp}).  The median {\it IRAS}-to-{\it Spitzer}
flux density ratio is 1.04, with a standard deviation of 0.3.  At
12~\micron, most of the {\it IRAS} fluxes are upper limits, but the
agreement is still decent ($f_{12}^{IRAS}/f_{12}^{Spitzer}\sim1.25$ on
average for sources with $f_{12} > 0.1$ Jy).  The cause of the small
discrepancy at 12\micron\ is unclear, but may result from
Eddington-Malmquist bias.

{\em In this paper, we adopt the Spitzer-derived 12 and 25 $\mu$m
  fluxes to avoid the use of IRAS upper limits, a particularly severe
  problem for the LINER and HII-like ULIRGs of our sample.}

\section{Data Analysis}

\subsection{Emission Lines}

In Tables \ref{tab:eml1} and \ref{tab:eml2}, we list atomic and
molecular emission-line fluxes measured from our spectra.
Measurements were made with the IDEA tool in SMART; we fitted Gaussian
profiles atop a linear continuum.  Upper limits were determined by
assuming an unresolved line.  The available resolution allowed us to
decompose close line blends, including the important [\ion{Ne}{5}]
$14.32$~\micron\ / [\ion{Cl}{2}] $14.36$~\micron\ and [\ion{O}{4}]
$25.89$~\micron\ / [\ion{Fe}{2}] $25.99$~\micron\ blends.

\subsection{Continuum and Dust Features}

A vitally important task was to properly model the sum of the
blackbody continuum emission, which is punctuated by deep extinction
and absorption features, and the full-featured small dust grain
continuum.  The primary goal of this modelling was to accurately
extract the fluxes of absorbed PAH features, but we also gained useful
information about the continuum.  Because we are not concerned with
detailed physics, we have chosen a simple, but robust and
empirically-motivated, method.

Before fitting, we measured narrow-band flux densities (of width
3.3\%\ of the central wavelength) at regularly-spaced intervals across
the continuum, avoiding deep absorption features.  These are listed in
Table \ref{tab:cont}.

\begin{figure*}[ht]
\epsscale{1.0}
\plotone{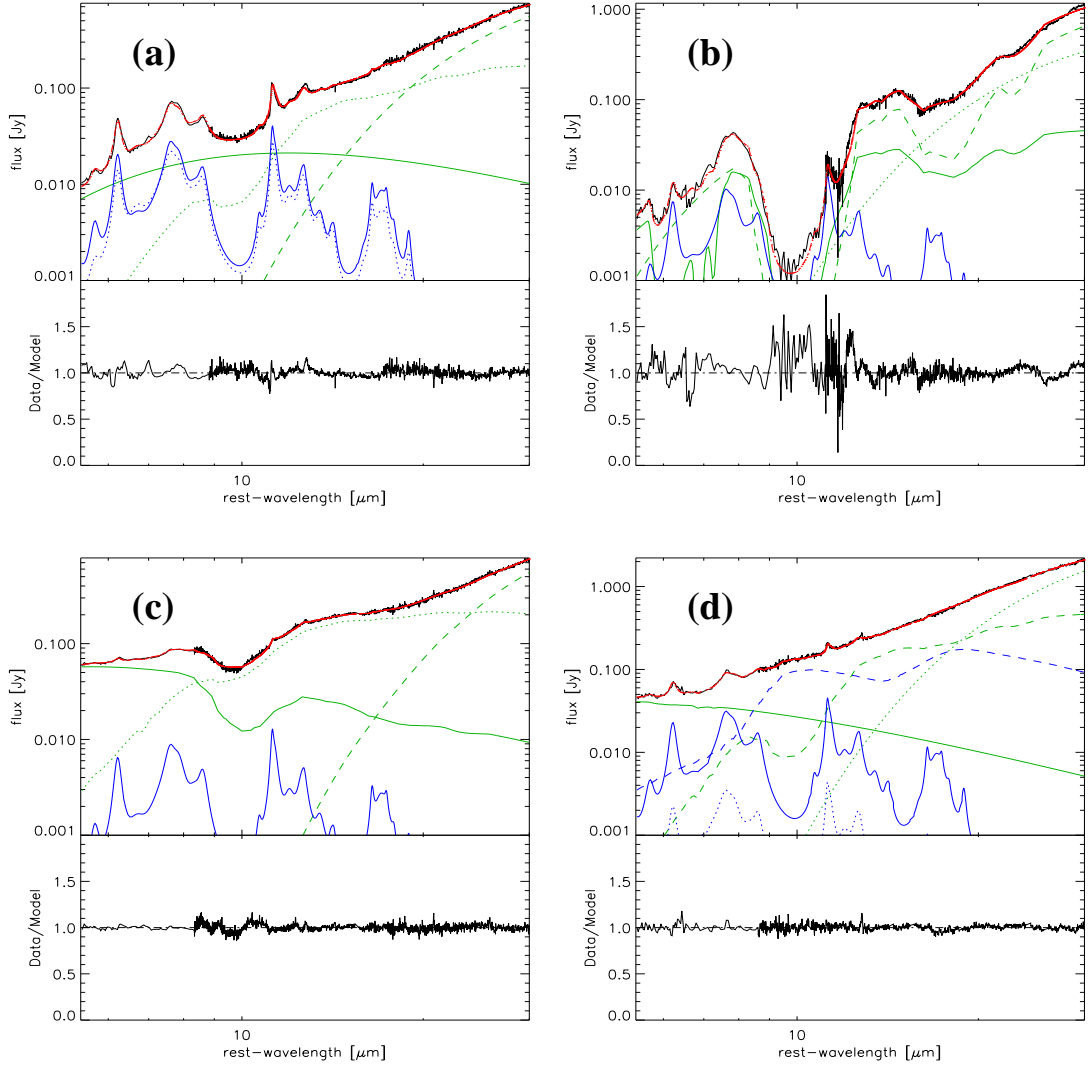}
\caption{ Four examples of spectral decompositions: ($a$)
  PAH-dominated ULIRG, F15206+3342, ($b$) Highly obscured ULIRG,
  F00091$-$0738, ($c$) AGN-dominated ULIRG without silicate emission,
  F11119+3257, and ($d$) AGN-dominated ULIRG/PG~QSO with silicate
  emission, Mrk~1014 = PG~0157+001. In each case, the top panel shows
  the fit to the IRS spectrum, while the bottom panel shows the
  data-to-model flux ratio. The data are in black, the overall fit is
  in red, the three blackbody components are in green, the two PAH
  templates are in blue, and the silicate emission component is the
  long-dash blue line. See Section 5.2 for more detail on the fits. }
\label{fig:fit}
\end{figure*}

To fit the MIR spectra, we used the IDL package developed to model the
blackbody and silicate emission of QUEST quasars (Paper III). We refer
to this paper for the basic fitting details.  Here we describe some
unique features necessary for fitting the spectra of PAH-strong and
sometimes deeply-absorbed ULIRGs. Typical fits are shown in Figure
\ref{fig:fit}.

\begin{enumerate}

\item A modified version of the Chiar \& Tielens (2006) Galactic
  Center extinction curve was used.  To the basic extinction profile,
  we added extinction from the water ice plus hydrocarbon feature at
  $5.7-7.8$ \micron.  The profile is taken from observations of
  F00183$-$7111 (Spoon et al. 2004).  In place of the broad silicate
  features at 9.7 and 18 \micron\, we substituted features that were
  empirically derived from the deeply absorbed and almost completely
  PAH-free ULIRG, F08572$+$3915.  The silicate profile of this galaxy
  provides a universally good fit to the ULIRGs in our sample.
  However, using the original Chiar \& Tielens silicate profile, or
  one from a less deeply-absorbed system, yields poor fits at high
  optical depths in a number of deeply absorbed systems.  The
  strengths of the water ice $+$ hydrocarbon feature and silicate
  absorption $+$ overall extinction curve were allowed to vary
  independently.  Foreground and mixed dust screens were both tried,
  and we found that foreground screens fit better.

\item We fit three blackbodies to almost all spectra.  These
  blackbodies represent a convenient parameterization of the
  MIR continuum.  Due to the absence of wavelengths probing
  the hottest and coldest dust, the temperatures of these blackbodies
  do not represent actual dust components.  Nonetheless, their values
  provide useful guidance in understanding the shape of the continuum.
  For most sources, three is the minimum number of blackbodies that
  produce an acceptable fit.  However, the use of two or four
  blackbodies significantly improved the fit in six cases.

\item The temperatures, water ice $+$ hydrocarbon absorption, and
  extinctions of the blackbodies were in general allowed to freely
  vary (see results in Table \ref{tab:abs}).  However, in some cases
  the water ice plus hydrocarbon absorption and/or the extinction had
  to be fixed to zero in a particular component, when it was apparent
  that the fitted value was unphysical.  In a handful of cases, the
  temperature of the hottest component was also poorly constrained by
  the fit.  Experimentation and by-eye examination suggests that these
  unconstrained temperatures are not much larger than 1000~K, to which
  we fixed them.  The actual temperatures are probably in the range
  $\sim 700-2000$ K.

\item PAH emission was modeled using the average MIR spectra derived
  from the SINGS program galaxies (Smith et al. 2007).  Pure PAH
  templates were created by running the PAHFIT program (Smith et
  al. 2007) to extract the PAH emission features from the four average
  SINGS spectra (including the 17 $\mu$m emission band; van Kerckhoven
  et al.\ 2000; Peeters et al.\ 2004).  The PAH model underlying
  PAHFIT is a series of Drude profiles; the 6.2 and 7.7 \micron\
  features, which we discuss in \S 6.3, consist of 1 and 3 Drude
  components, respectively.  Note that the templates output by PAHFIT
  consist only of PAH emission, without an associated continuum from
  star formation regions; this continuum is fit by the blackbodies
  mentioned above.  We chose two of these templates (from Smith's 3rd
  and 4th average SINGS spectra) that provided the largest range in
  the ratio of the 7.7~\micron\ and 11.2~\micron\ PAH features.  The
  overall strengths of these two pure PAH templates were allowed to
  vary in the fit.

  The fit results, listed in Table \ref{tab:pahs}, were not sensitive
  to the choice of pure PAH template.  Only a small range ($\sim$ 0.13
  dex) in PAH 6.2/7.7 \micron\ and 7.7/11.2 \micron\ ratios was
  allowed by the two PAH templates and most galaxies are dominated by
  one or the other template.  We found that using templates produced
  from SINGS spectra of galaxies with very low values of the
  7.7/11.2~\micron\ ratio (J.~D. Smith 2007, pvt. comm.) yielded poor
  fits.  Thus, the strong suppression of the 6.2 or 7.7~\micron\
  complex observed in galaxies that host low-luminosity AGN (Sturm et
  al. 2006; Smith et al. 2007) and in some star-forming regions (Hony
  et al. 2001) does not occur in ULIRGs.

  We found that allowing PAH extinction below the level of $A_V \sim
  10$ did not significantly affect the fits (i.e., the results were
  basically indistinguishable from the $A_V = 0$ case).  We thus left
  the PAHs unextincted in most cases.  Adding larger amounts of
  extinction had the effect of raising the 7.7/11.2~\micron\ ratio and
  almost always made the fits worse.  However, in four cases we
  allowed the PAH features to be extincted at or above the $A_V \sim
  10$ level because the fit was significantly improved.  These four
  cases are as follows, with PAH extinction in parentheses (two values
  indicate two fitted components): F00397$-$1312 ($A_V = 44$),
  F01494$-$1845 ($A_{V} =$ 10/14), F20414$-$1651 ($A_{V} =$ 0/22),
  F21208$-$0519:N ($A_V = 11$).  These values correspond well with the
  effective continuum extinction (Section 6.2) in these sources: $A_V\sim$
  50, 18, 18, and 15, respectively.  We discuss PAH extinction further
  in Section  6.3.

\item For four galaxies (Mrk~1014, F07598$+$6508, 3C~273, and
  F21219$-$1757) we included silicate emission components, as described
  in Paper III.

\item Due to the absence of pipeline error spectra, the mixing of data
  over a significant flux range, and the different dispersion among
  different IRS modules, careful weighting had to be performed during
  the fits (see Paper III for details).  We tried three methods: (1)
  weighting based on the actual fluxes; (2) weighting based on a
  power-law fit to the fluxes; and (3) an average of the two.  In
  almost all cases, the third method produced the best fits.  However,
  in isolated cases we used one of the other weighting schemes if it
  was clearly superior.

\end{enumerate}

Along with the ULIRGs in our sample, we also did three-blackbody fits,
with both silicate and PAH emission included, to the three average PG
QSO spectra from Paper II.  These spectra were divided by FIR strength
into FIR-strong QSOs, FIR-weak, and FIR-undetected.  These new fits
differ from the fits in Papers I and III in that Paper I fit only the
PAHs using simple Lorentzian fits to a few individual features, while
Paper III presented more formal fits but did not include PAH emission.
Here we include all components, and make sure that the PAH fits of the
QSOs are done using the same procedures applied here to the ULIRGs.

\section{Results}

\begin{figure*}[ht]
\epsscale{0.9}
\plotone{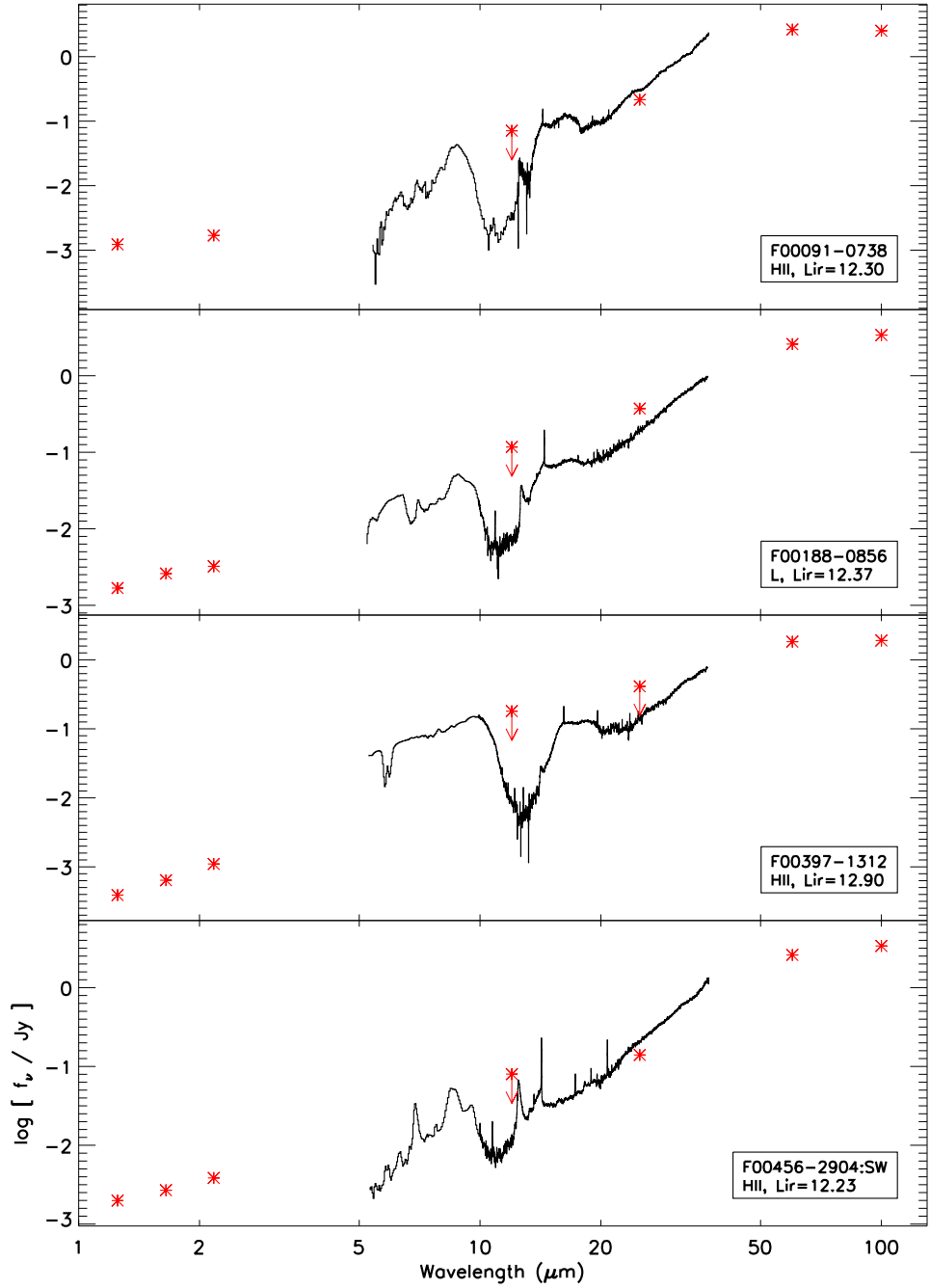}
\caption{ IRS spectra of all ULIRGs considered in our study, in order
  of increasing right ascension. The sources of the overlaid
  photometry, shown as red stars, are discussed in Section
  6. F11223$-$1244:W is shown in this figure but was mistakenly
  omitted from the analysis.}
\label{fig:spec}
\end{figure*}

The IRS spectra of all ULIRGs in the current study are presented in
Figure \ref{fig:spec}, with archival photometry overplotted: $J$, $H$,
and $K$, and flux densities at 12, 25, 60, and 100 \micron.  The
NIR photometry is from 2MASS and Sanders et al. (1988a).  The
far-infrared photometry is mostly from the {\em IRAS} Faint Source
Catalog, but also includes some data from Sanders et al.\ (2003) and a
few $ISO$ 12 \micron\ points from Klaas et al.\ (2001). The IRS
spectra of the QSOs were presented in Paper II, so they are not shown
here again.  The basic results from our analysis are listed in Tables
$\ref{tab:eml1}-\ref{tab:pahs}$.

In this section, we first describe the results from our analysis of
the broadband continuum emission in Section 6.1 before discussing the
absorption and emission-line features in Section 6.2 and Sections $6.3-6.6$,
respectively.

\subsection{Broadband Continuum Emission}

\subsubsection{Average Spectra}

We divided the sample into various categories, and produced average
spectra by normalizing individual spectra to the same rest-frame, 15
\micron\ flux density.

\begin{figure*}[ht]
\epsscale{1.0}
\plotone{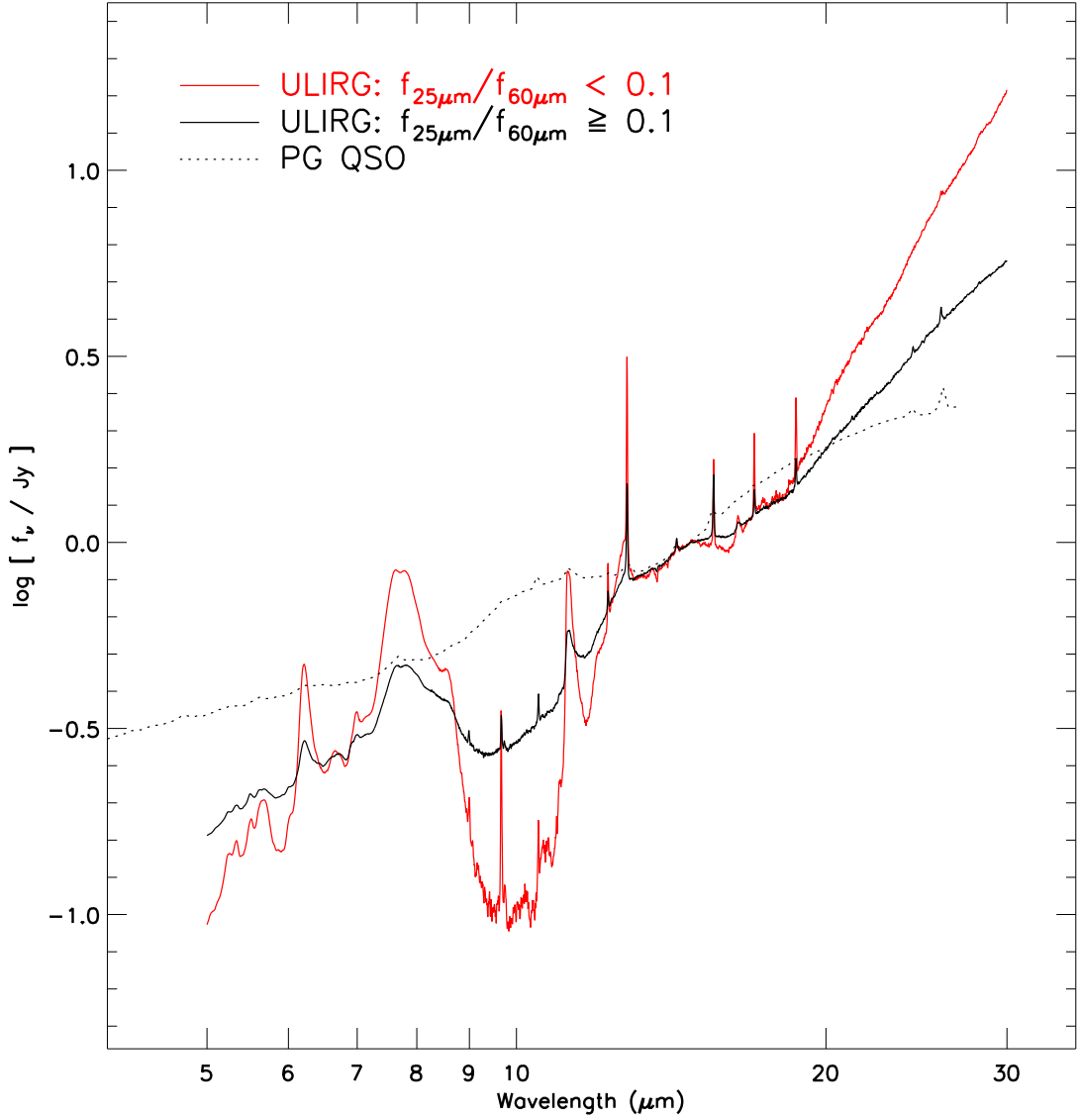}
\caption{ Average IRS spectra for ULIRGs with 25-to-60 $\mu$m flux
  ratios, $f_{25}/f_{60}$, above and below 0.1, compared with the QSOs
  in our sample (Paper II).  The individual spectra in each category
  were normalized to have the same rest-frame 15 $\mu$m flux density.
  Note the progression from cool ULIRGs to warm ULIRGs, and then to
  QSOs.}
\label{fig:avgspec-f25f60}
\end{figure*}

First, we show average spectra of ``cool'' ($f_{25}/f_{60} < 0.1$)
ULIRGs, ``warm'' ($f_{25}/f_{60} \ge 0.1$) ULIRGs, and all PG~QSOs
(Figure \ref{fig:avgspec-f25f60}). The average spectrum of cool ULIRGs
shows a steep 5-30 $\mu$m SED with strong PAH features, H$_2$ lines,
and low-ionization fine structure lines typical of starburst-dominated
systems, while PG~QSOs have a shallow 5-30 $\mu$m SED with silicate
features in emission and relatively strong high-ionization lines and
weak PAH features typical of AGN-dominated systems. The properties of
the average spectrum of warm ULIRGs are intermediate between those of
cool ULIRGs and PG~QSOs.

\begin{figure*}[ht]
\epsscale{1.0}
\plotone{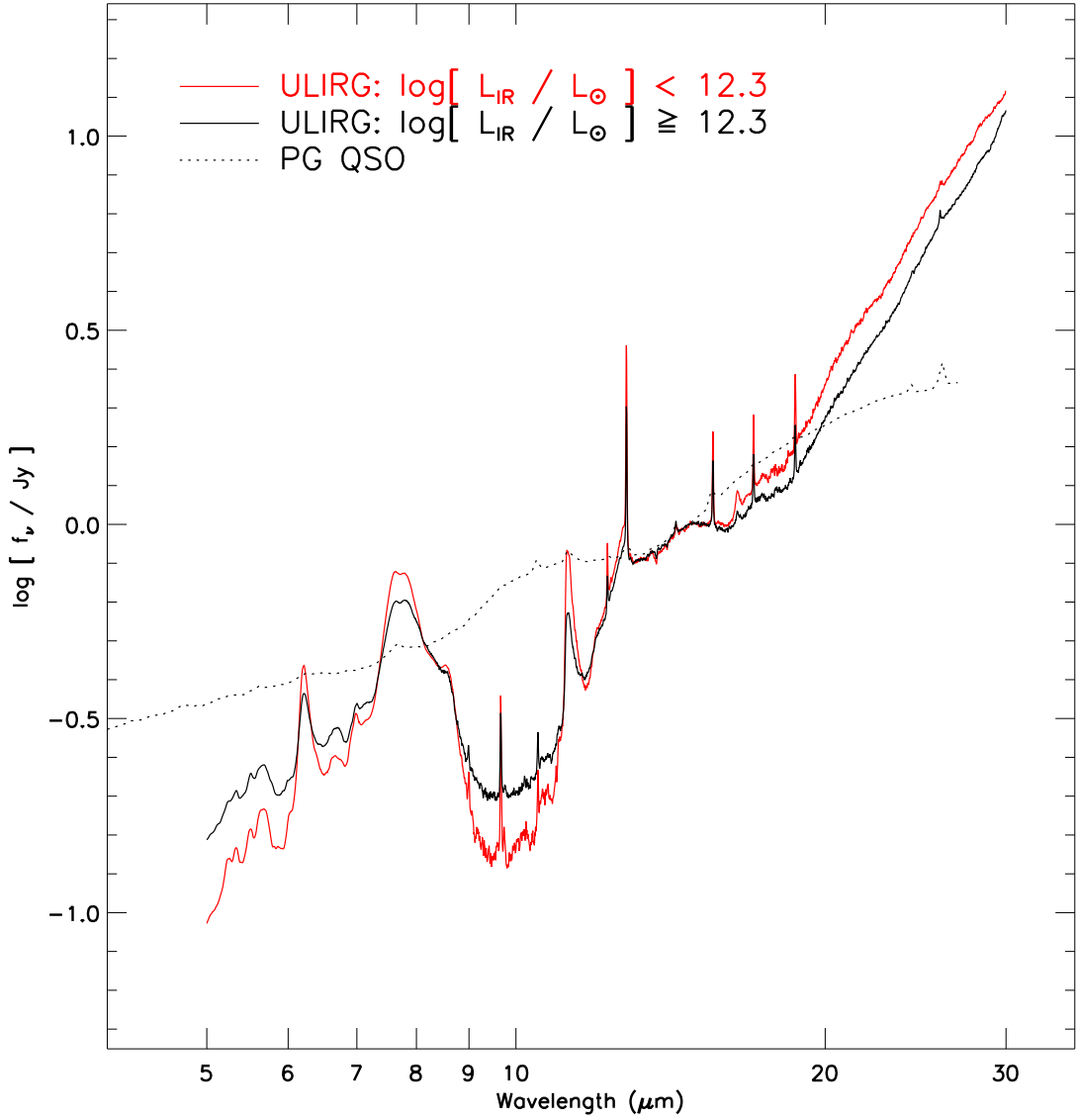}
\caption{ Average IRS spectra for ULIRGs with infrared luminosities
  larger or smaller than 10$^{12.3}$ $L_\odot$, compared with the QSOs
  in our sample.  The individual spectra in each category were
  normalized to have the same rest-frame 15 $\mu$m flux density.}
\label{fig:avgspec-lir}
\end{figure*}

A similar exercise is carried out using the infrared luminosity:
log[$L$(IR)$/L_\odot$] $<$ 12.3 and $\ge$ 12.3 (this threshold
was selected to get roughly equal number of objects in each luminosity
bin).  Figure \ref{fig:avgspec-lir} shows that the contrast between
low- and high-luminosity ULIRGs is nowhere near as large as between
cool and warm ULIRGs. High-luminosity ULIRGs tend to be slightly
warmer than low-luminosity ULIRGs so the differences we see in Figure
8 are readily explained by the correlation between {\em Spitzer} spectral
characteristics and $f_{25}/f_{60}$ discussed in the previous
paragraph.

\begin{figure*}[ht]
\epsscale{1.0}
\plotone{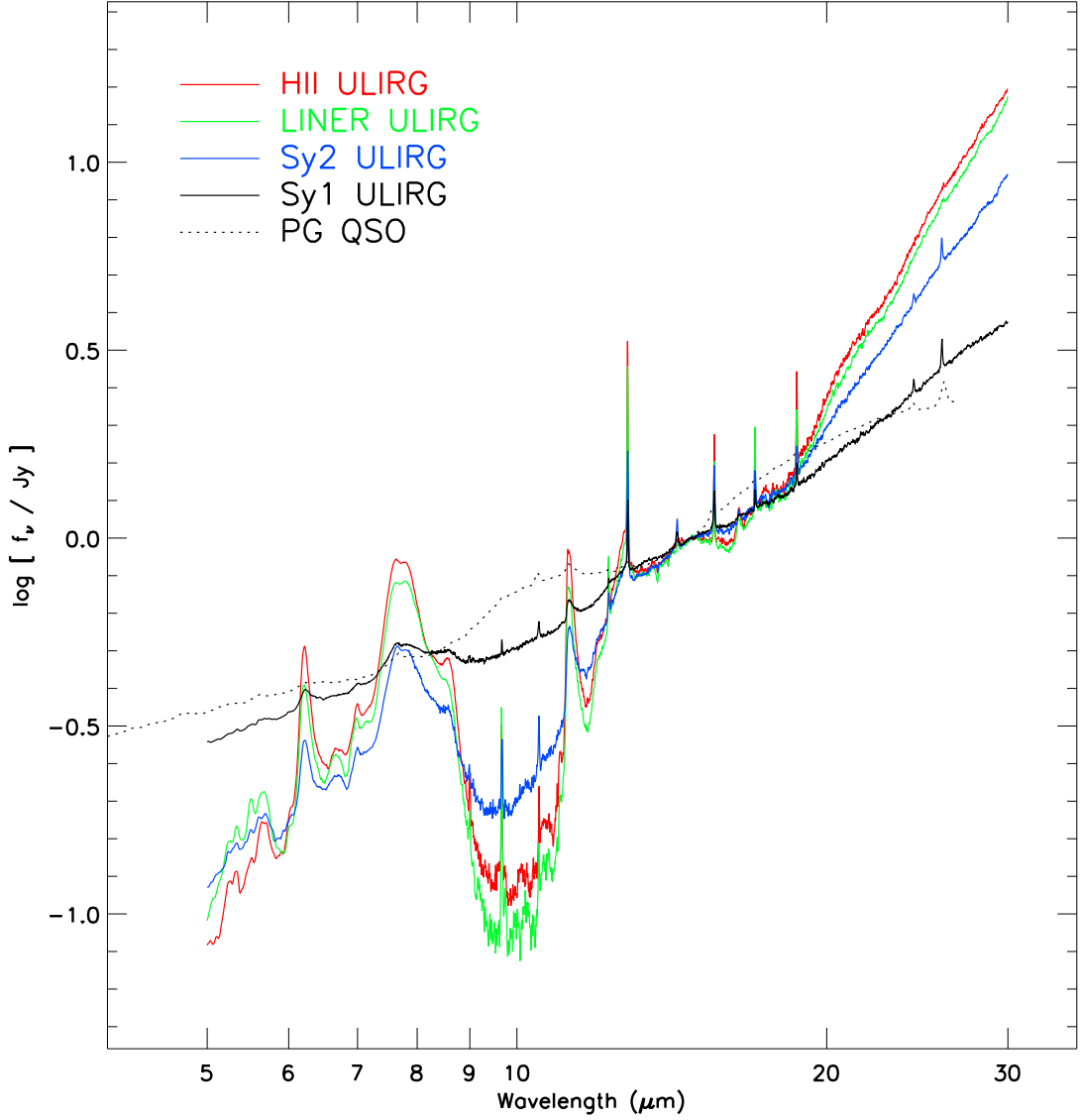}
\caption{ Average IRS spectra for ULIRGs of various optical spectral
  types, compared with the QSOs in our sample.  The individual spectra
  in each category were normalized to have the same rest-frame 15
  $\mu$m flux density.  Note the similarity between the average
  spectrum of Seyfert 1 ULIRGs and that of QSOs.}
\label{fig:avgspec-type}
\end{figure*}

In Figure \ref{fig:avgspec-type}, we divide the 1-Jy ULIRGs according
to their optical spectral type.  The overall 5-30 $\mu$m SED clearly
steepens and the silicate absorption feature and H$_2$ and
low-ionization fine structure emission lines clearly become stronger
as one goes from the QSOs, to the Seyfert 1s, the Seyfert 2s, and
finally to the LINER and H~II-like ULIRGs. The averaged spectra of
these last two classes of ULIRGs are hardly distinguishable from each
other with the possible exception of the silicate absorption through,
where we are S/N-limited (this is consistent with the {\em ISO}-based
results of Lutz et al.\ 1999).

\begin{figure*}[ht]
\epsscale{1.0}
\plotone{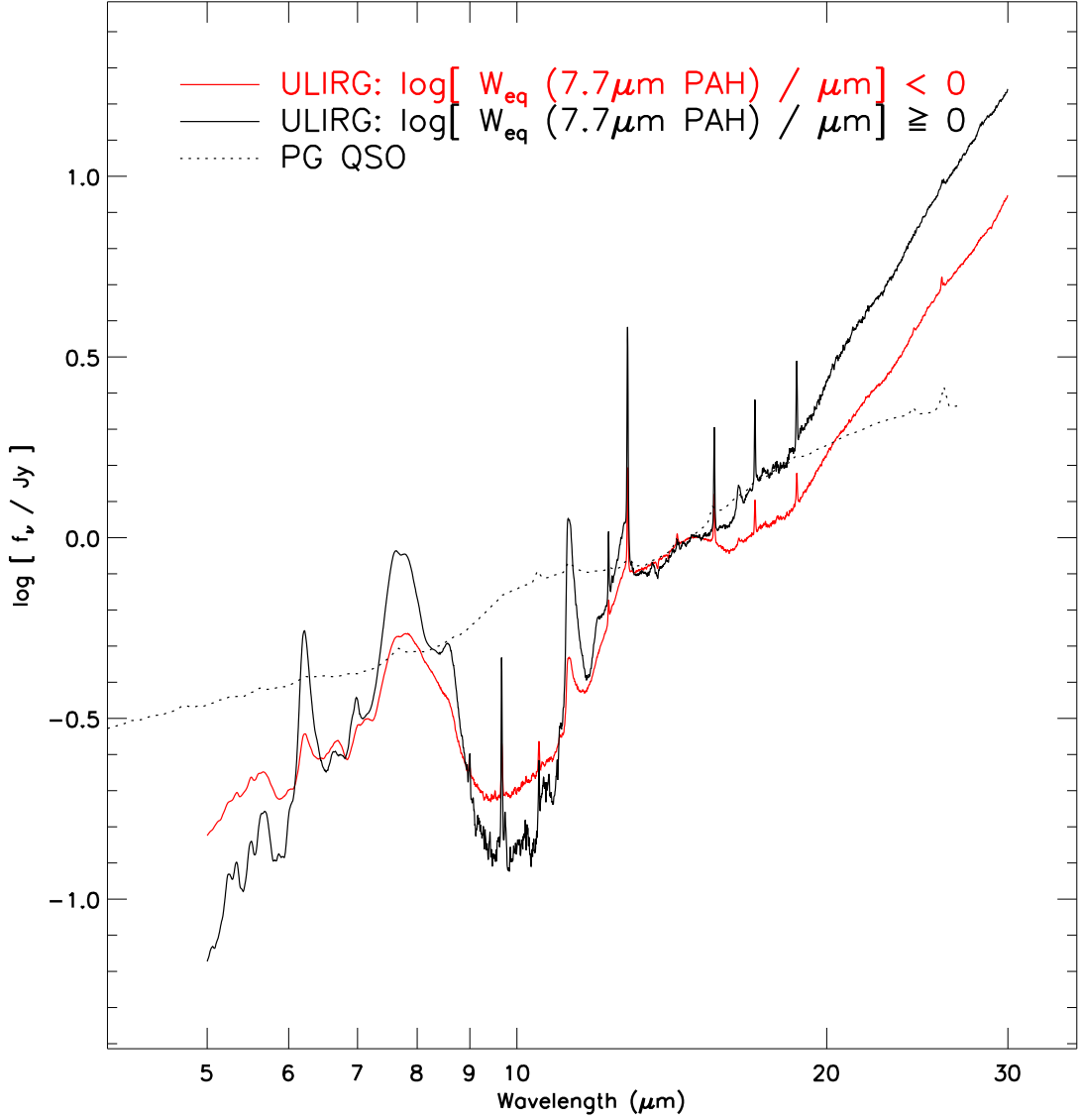}
\caption{ Average IRS spectra for ULIRGs with PAH 7.7 $\mu$m
  equivalent widths larger or smaller than 1 \micron, compared with
  the QSOs in our sample.  The individual spectra in each category
  were normalized to have the same rest-frame 15 $\mu$m flux density.}
\label{fig:avgspec-pah}
\end{figure*}

\begin{figure*}[ht]
\epsscale{1.0}
\plotone{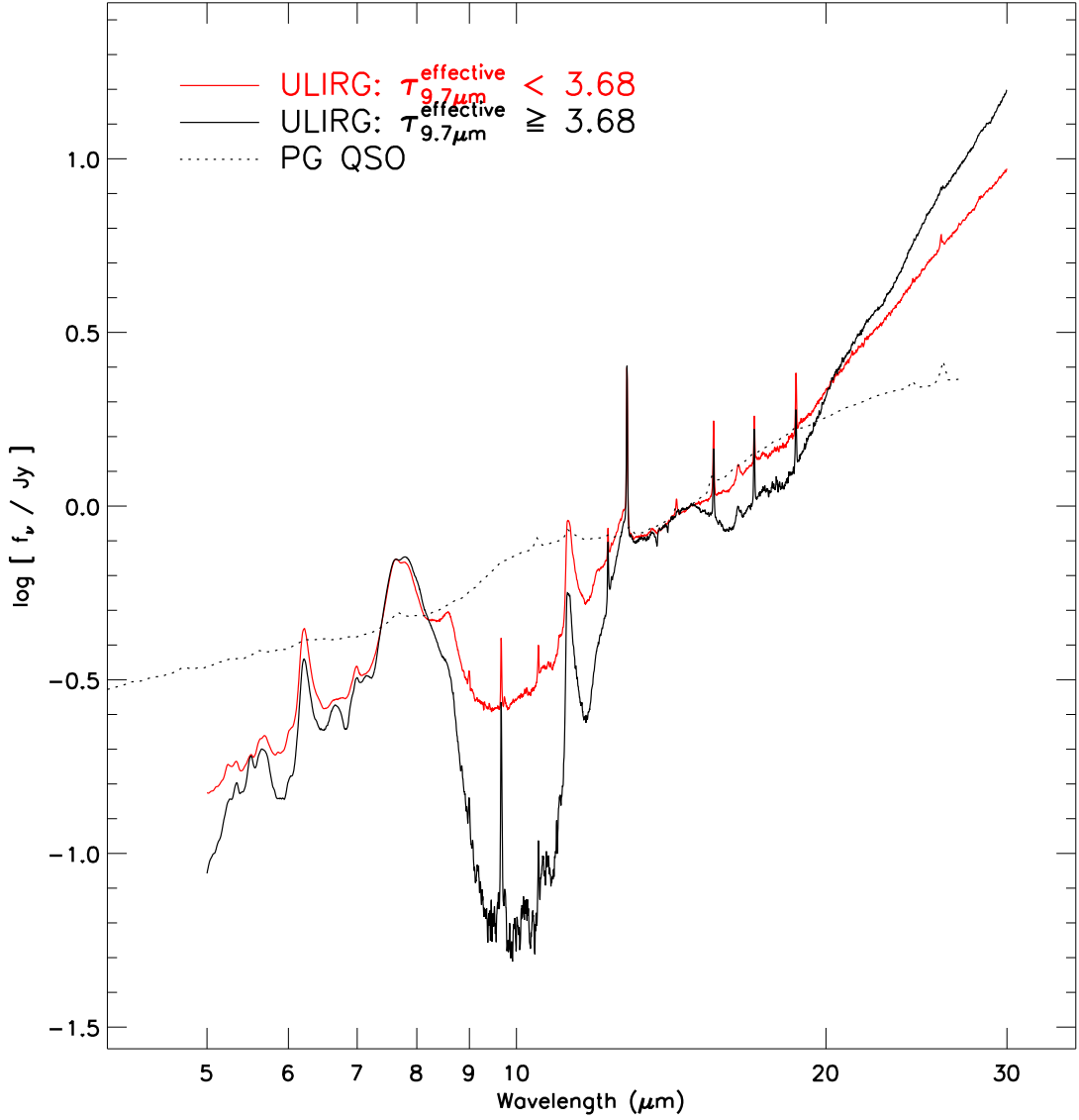}
\caption{ Average IRS spectra for ULIRGs with effective optical depth
  of the 9.7 $\mu$m absorption feature larger or smaller than 3.68
  (the sample median), compared with the QSOs in our sample.  The
  individual spectra in each category were normalized to have the same
  rest-frame 15 $\mu$m flux density.  Significant PAH emission is
  detected in both average ULIRG spectra.}
\label{fig:avgspec-tau}
\end{figure*}

Finally, in Figures \ref{fig:avgspec-pah} and \ref{fig:avgspec-tau},
we divide the ULIRG sample based on the equivalent width of the PAH
7.7 $\mu$m feature and the effective optical depth of the silicate
absorption trough, and compare the results once again to the average
spectrum of PG quasars.  The strength of the PAH feature and effective
optical depth of the silicate feature are derived from the SED
decomposition described in Section 6.2 and Section 6.3.  Strong
absorption features of water ice + hydrocarbons (5.7-7.8 $\mu$m),
silicate (8.5 -- 12 $\mu$m), C$_2$H$_2$ 13.7 $\mu$m, and HCN 14 $\mu$m
are detected in the absorption-dominated ULIRGs, similar in depth to
the features seen in the heavily absorbed spectra of NGC~4418 and
other galaxies including some ULIRGs (e.g., Spoon et al.\ 2001, 2002,
2004, 2006).  Silicate absorption is visible in both PAH-dominated and
PAH-weak systems. Similarly, PAH emission is detected regardless of
the depth of the silicate absorption feature.  As we discuss
quantitatively in Section 6.3, Section 7.1, and Section 7.3, this lack
of a clear trend between PAH strength and silicate absorption is
largely due to the strong-AGN ULIRGs, which have weak PAHs {\em and}
weak silicate absorption (see also Desai et al. 2007 and Spoon et
al. 2007).

\subsubsection{Continuum Diagnostics}

\begin{figure*}[ht]
\epsscale{1.1}
\plotone{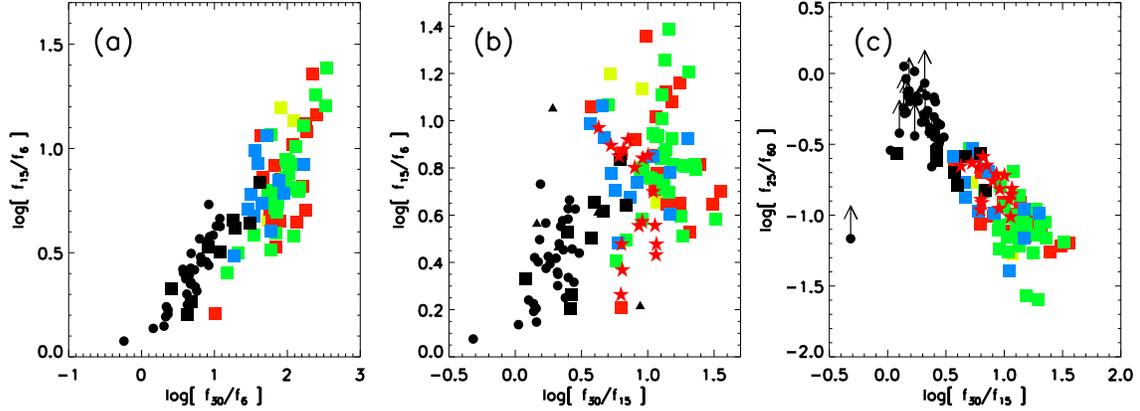}
\caption{ MIR color-color diagrams: ($a$) 15-to-6 {\em vs.\/}
  30-to-6 $\mu$m flux ratios, ($b$) 15-to-6 {\em vs.\/} 30-to-15
  $\mu$m flux ratios, ($c$) 25-to-60 {\em vs.\/} 30-to-15 $\mu$m flux
  ratios.  The meaning of the ULIRG and QSO symbols is the same as in
  Figure \ref{fig:fluxcomp}.  In addition, the red stars and black
  triangles are starburst and Seyfert galaxies observed with $ISO$
  (Verma et al. 2003; Sturm et al. 2002; Brandl et al. 2006). The
  tightest correlation is seen in ($c$).}
\label{fig:mircolor}
\end{figure*}

In Figure \ref{fig:mircolor}, we compare the continuum flux ratios
$f_{15}/f_{6}$, $f_{30}/f_{6}$, $f_{30}/f_{15}$, and $f_{25}/f_{60}$
of all ULIRGs and quasars in the sample. The ``reddening'' of the SED
as one goes from the QSOs to the ULIRGs is evident in all panels of
this figure. The best segregation by optical spectral type is seen
when using $f_{25}/f_{60}$ and $f_{30}/f_{15}$. QSOs, Seyfert 1
ULIRGs, Seyfert 2 ULIRGs, and H~II-like + LINER ULIRGs form a sequence
of increasing 60-to-25 and 30-to-15 $\mu$m flux ratios, the H~II-like
ULIRGs being indistinguishable from the LINER ULIRGs. Interestingly,
the optically-selected starbursts observed with {\em ISO} (Verma et
al. 2003) have 25-to-60 $\mu$m and 30-to-15 $\mu$m flux ratios (Brandl
et al. 2006) that are intermediate between those of H~II-like/LINER
ULIRGs and Seyfert 2 ULIRGs. We return to the $f_{30}/f_{15}$ ratio in
Section 6.1.3 and in Section 7.1, where we discuss MIR spectral
classification.

\begin{figure*}[ht]
\epsscale{0.5}
\plotone{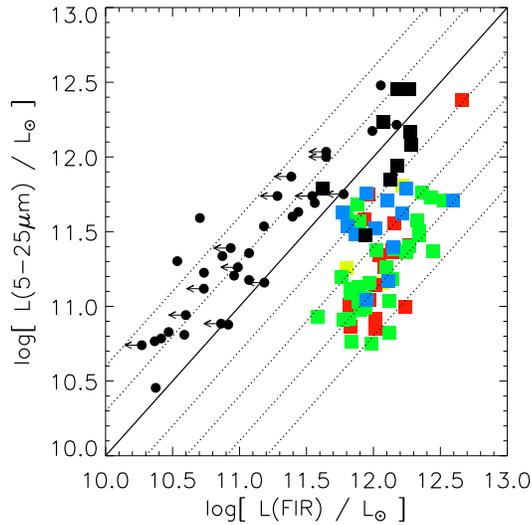}
\caption{ {\em Spitzer}-derived MIR (5 -- 25 $\mu$m) luminosities
  of 1-Jy ULIRGs and PG~QSOs {\em vs.\/} FIR
  luminosities. The meaning of the symbols is the same as in
  Figure \ref{fig:fluxcomp}. The solid line is the line of equality,
  while the dashed lines show the locations of objects with
  FIR luminosities equal to (1/4, 1/2, 2, 4, 8, 16) $\times$
  the MIR luminosities. All ULIRGs, except most of those that
  are optically classified as Seyfert 1s, are MIR
  underluminous relative to QSOs. Seyfert 2 ULIRGs are intermediate
  between QSOs/Seyfert 1 ULIRGs and HII-like/LINER ULIRGs.  The
  HII-like ULIRG at very high infrared luminosity is F00397$-$1312,
  and has heavily extincted PAH emission.}
\label{fig:mirfir}
\end{figure*}

Figure \ref{fig:mirfir} shows that PG~QSOs and optically classified
Seyfert 1, Seyfert 2, and LINER + H~II-like ULIRGs progressively have
weaker MIR emission relative to their FIR emission (see also Figure\ 7
in Paper I). The solid line represents $L$(MIR) $= L$(FIR). QSOs are
well fit, on average, by the dotted line above it: $L$(MIR) $\approx 2
L$(FIR). This line traces AGN-dominated systems and may be used in
principle to estimate the AGN contribution to the ULIRG power. We
return to this point in Section 6.1.3 and Section 7.1 of this paper.
Here we simply note that the extrapolation of this line to higher MIR
luminosities is a good fit to the measurements of some, but not all,
Seyfert 1 ULIRGs. These latter objects are more MIR-luminous than QSOs
but they have only slightly cooler SEDs than QSOs (e.g.,
Figures \ref{fig:avgspec-type} and \ref{fig:mircolor}).

\subsubsection{Results from SED Decomposition}

\begin{figure*}[ht]
\epsscale{1.0}
\plotone{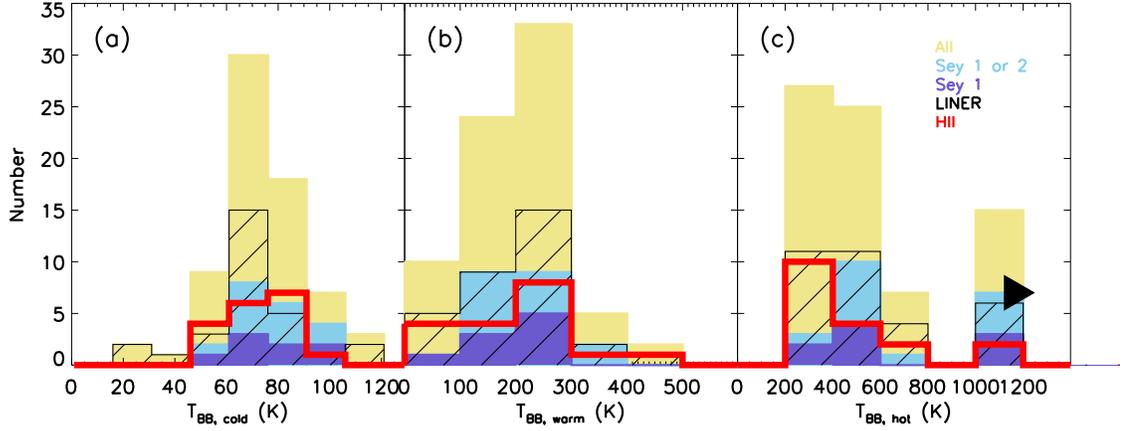}
\caption{ Distributions of the blackbody temperatures for the ($a$)
  cold, ($b$) warm, and ($c$) hot components of the template fits to
  the IRS spectra of the 1-Jy ULIRGs.  The arrowhead in panel ($c$)
  indicates that the somewhat uncertain $\sim$1000 K temperature are
  possibly lower limits (with an estimated range $\sim 700-2000$ K).
  The temperatures of the hot components in HII-like / LINER ULIRGs
  are distinctly lower than in Seyfert ULIRGs.}
\label{fig:bbtemp}
\end{figure*}

\begin{figure*}[ht]
\epsscale{1.0}
\plotone{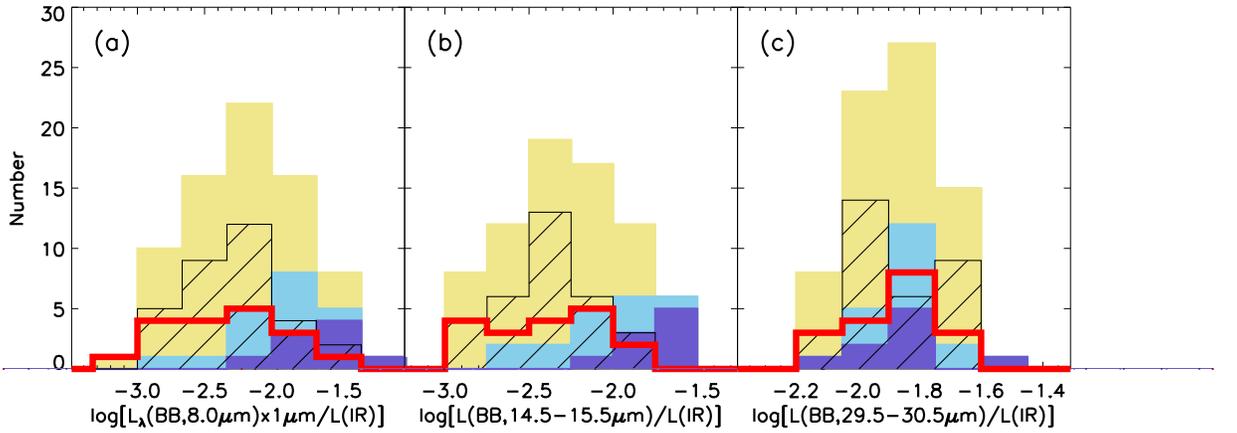}
\caption{ Distributions of the ratios of the monochromatic blackbody
  (i.e., excluding PAH emission) ($a$) 8, ($b$) 15, and ($c$)
  30 $\mu$m luminosities to the total infrared luminosities for 1-Jy
  ULIRGs of various optical spectral types (see the legend in
  Figure \ref{fig:bbtemp}). The ratios involving the 8 and 15 $\mu$m
  luminosities are distinctly larger among Seyfert ULIRGs than among
  HII-like / LINER ULIRGs.}
\label{fig:bblir}
\end{figure*}

The results from the SED decomposition analysis described in Section
5.2 are presented in Figures $\ref{fig:bbtemp}-\ref{fig:mirfir_v}$.
Figure \ref{fig:bbtemp} shows the distributions of temperatures for
the cold, warm, and hot blackbody components used in the fits. Note
that, as mentioned in Section 5.2, the temperatures of hot components
with $T \ga 1000$ K are not well constrained in the fits.  However, it
is clear that Seyfert ULIRGs, particularly Seyfert 1 ULIRGs, show a
tendency to have a warmer hot component than H~II-like and LINER
ULIRGs. This separation is not seen in the warm and cold components.

Figure \ref{fig:bblir} presents the distributions of observed
monochromatic 8-, 15-, and 30-$\mu$m blackbody to total infrared
luminosity ratios for all ULIRGs in the sample according to their
optical spectral types.  These blackbody luminosities represent the
sum of all blackbody components fitted to the IRS spectra of these
objects, uncorrected for extinction. K-S and Kuiper tests on these
figures confirm the stronger MIR (8 and 15 $\mu$m but not 30
$\mu$m) continuum emission in Seyfert ULIRGs, particularly Seyfert 1s,
than in H~II-like or LINER ULIRGs.  A similar result is found in
Figure \ref{fig:mirfir_v}$a$, where the 5.4 - 25 $\mu$m ``pure''
(PAH-free and silicate-free) blackbody emission is compared to the
FIR emission. This ratio is very strongly correlated with
$f_{30}/f_{15}$ (Figure \ref{fig:mirfir_v}$d$). Table \ref{tab:cont}
lists for each ULIRG the observed PAH- and silicate-free 5 -- 25
$\mu$m luminosities as a (logarithmic) fraction of the FIR luminosity.

One can safely assume that the continuum emission from the atmospheres
of the young stars in ULIRGs, and the very hot (10$^3$ K) small grain
NIR dust emission component inferred in ISOPHOT spectra of normal
galaxies (Lu et al.\ 2003) and presumed to also exist in these
objects, do not contribute significantly to the observed continuum
above $\sim$ 5 $\mu$m.  Consequently, the results in Figures
$\ref{fig:bbtemp}-\ref{fig:mirfir_v}$ most likely reflect an elevated
AGN contribution to the MIR emission of Seyfert 1 and 2 ULIRGs
relative to that of H~II-like or LINER ULIRGs.  This is discussed more
quantitatively in Section 7.1.

\subsection{Absorption Features}

Table \ref{tab:abs} lists the effective 9.7 $\mu$m silicate optical
depth, $\tau^{\rm eff}_{9.7}$, defined as
\begin{equation}
I^{\rm eff} {\rm exp}(-\tau^{\rm eff}_{9.7}) = \sum I_i {\rm exp}
[-\tau^i_{9.7}],
\end{equation}
where $I_{\rm eff} = \sum I_i$ and the sum is over the blackbody
components $i$. Note that the silicate feature is in emission in four
Seyfert 1 ULIRGs (Mrk~1014, F07598$+$6508, 3C~273, and 21219$-$1757)
and all QSOs (Paper III).  Also note that our fits were kept simple
and neglected possible variations in the 18/10 $\mu$m absorption
ratios in ULIRGs, so we cannot constrain the geometry of the dust
distribution in these objects in detail (e.g., Sirocky et al.\ 2008;
Li et al.\ 2008). For the same reason, we do not attempt to constrain
the fraction of silicate absorption that is from crystalline silicates
rather than amorphous silicates (e.g., Spoon et al.\ 2006). Finally,
it is important to point out that the effective silicate optical depth
discussed here is a true optical depth, defined with respect to the
unextincted blackbody flux level derived from our fits. It is
therefore different from those published in earlier studies (e.g.,
Brandl et al.\ 2006; Spoon et al.\ 2007; Armus et al. 2007; Imanishi
et al. 2007), where the depth of this feature is measured empirically
with respect to the observed (extincted) continuum. It would be the
same if the continuum and silicates were equally extincted but
unfortunately that is not generally the case. A comparison between our
measurements and those published in Armus et al.\ (10 objects)
indicate that $\tau^{\rm eff}_{9.7} \sim (2.5 \pm 0.8) \tau_{9.7}({\rm
  Armus})$ with a median ratio of 2.8.

Figure \ref{fig:tauhist} shows the distribution of $\tau^{\rm
  eff}_{9.7}$ {\em vs.\/} the optical spectral types of ULIRGs. The
  broad distribution of silicate strength in ULIRGs is well known from
  previous studies (e.g., Hao et al.\ 2007; Spoon et al. 2007). K-S
  and Kuiper tests indicate that LINER and H~II-like ULIRGs have
  significantly larger $\tau^{\rm eff}_{9.7}$ than Seyfert ULIRGs on
  average, in general agreement with Spoon et al.\ (2007) and Sirocky
  et al.\ (2008). H~II-like ULIRGs are statistically indistinguishable
  from LINER ULIRGs and the same is true between Seyfert 1 and Seyfert
  2 ULIRGs.  Combining these results with those in Section 6.1, we find
  that all of the Seyfert 1 ULIRGs and most of the Seyfert 2 ULIRGs
  cluster in the lower-left portion of the $\tau^{\rm eff}_{9.7}$ {\em
  vs.\/} $L$(8 $\mu$m)/$L$(IR), $L$(15 $\mu$m)/$L$(IR),
  $f_{30}/f_{6}$, and $f_{30}/f_{15}$ diagrams
  (Figure \ref{fig:tau_v}). No clear trend is seen between $\tau^{\rm
  eff}_{9.7}$ and $f_{30}/f_{15}$ among HII-like and LINER ULIRGs,
  contrary to the optically-selected starburst galaxies of Brandl et
  al.\ (2006), where objects with strong silicate absorption tend to
  have a steeper MIR continuum.

Also listed in Table \ref{tab:abs} are the equivalent widths of the
sum of the H$_2$0 ice (5.7-7.8 $\mu$m) and aliphatic hydrocarbons
(6.85 + 7.25 $\mu$m) features,
and individual equivalent widths for C$_2$H$_2$ 13.7 $\mu$m and HCN 14
$\mu$m. [The HCO$^+$ 12.1 $\mu$m and HNC 21.7 $\mu$m features, whose
millimetric transitions are important diagnostics of radiative pumping
and possibly the presence of AGN (e.g., Imanishi et al.\ 2006b;
Gu\'elin et al.\ 2007), were not detected in any individual
object. Upper limits of 5 $\times$ 10$^{-5}$ $\mu$m and 2 $\times$
10$^{-3}$ $\mu$m were measured for the equivalent widths of HCO$^+$
12.1 $\mu$m and HNC 21.7 $\mu$m, respectively, in the average spectrum
of ULIRGs with $\tau^{\rm eff}_{9.7}$ $>$ 3.86, the median $\tau^{\rm
  eff}_{9.7}$]. The equivalent widths of C$_2$H$_2$ and HCN were
measured directly, using SMART, only in objects with obvious
detections (26 and 20 objects, respectively, or 35\% and 27\% of all
ULIRGs). In contrast, the equivalent width of the H$_2$0 ice +
hydrocarbons feature was derived from the SED fit of each object. This
equivalent width is calculated with respect to the blackbody continuum
only (+ silicate continuum in four Seyfert 1 ULIRGs) i.e., PAH
emission is not counted as continuum.  Upper limits were set as
follows. Each spectrum was inspected visually to determine whether or
not the H$_2$O absorption fit was robust.  For those judged
questionable, the measured equivalent width was set as an upper limit.
For those objects with no H$_2$O absorption (often because we fixed it
that way) and that have a significant PAH contribution (which turn out
to be HII galaxies, LINERs, or Seyfert 2s), EW(H$_2$O + HC) was
assigned a limit of $<$ 0.1 $\mu$m.  This is obviously uncertain, but
implies a somewhat reasonable 5\% sensitivity to absorption if the PAH
contributes half the emission at these wavelengths.  For the Seyfert
1s with limits only, the upper limit was set equal to that of the
lowest Seyfert 1 measurement.  Firm measurements exist for 46 objects
(62\% of all ULIRGs) and upper limits on all the others.

Interestingly, objects with the strongest C$_2$H$_2$ 13.7 $\mu$m and
HCN 14 $\mu$m absorption features are not necessarily those with the
strongest silicate and H$_2$O ice features. Indeed, Figure
\ref{fig:abs} shows that the equivalent widths of H$_2$O ice +
hydrocarbons, C$_2$H$_2$, and HCN correlate only loosely with
$\tau^{\rm eff}_{9.7}$ and between each other. This implies
significant variations in composition of the dense absorbing material
from one ULIRG to the next. The strongest correlation is found between
C$_2$H$_2$ and HCN, which {\em a posteriori} is not surprising since both
features are believed to be tracers of high-density ($>$ 10$^8$
cm$^{-3}$), high-temperature chemistry (e.g., in Young Stellar
Objects; Lahuis \& van Dishoeck 2000; Lahuis et al.\ 2006, 2007).

\subsection{PAHs}

The PAH 6.2 and 7.7 $\mu$m equivalent widths and the total PAH to
infrared and FIR luminosity ratios are listed in Table
\ref{tab:pahs}.  The PAH luminosities are taken from our fits, and are
corrected for extinction in the four sources with fitted PAH
extinction (see Section 5.2 for more details).  The equivalent widths
are computed by dividing the PAH luminosity by the observed
(extincted) continuum fluxes at 6.22 and 7.9\micron.  Without proper
fits, the PAH 7.7$\mu$m equivalent width measurements are subject to
errors in the silicate 9.7$\mu$m absorption correction.  However, our
fits to the entire 5-30$\mu$m IRS spectra take this effect into
account in a robust manner.  In what follows, we use the 7.7\micron\
feature exclusively, though the 6.2\micron\ feature gives identical
results.

The results of our fits (Section 5.2) suggest that the detected PAHs
in our sources are lightly extincted ($A_V < 10$).  This means that
the extinction towards the observable PAH-emitting regions in ULIRGs
is small compared to the sometimes heavily extincted
blackbody-emitting regions.  This is consistent with the detection of
spatially extended PAH emission in compact U/LIRGs by Soifer et
al. (2002).

That said, we cannot rule out heavily extincted PAHs in the cores of
ULIRGs.  We can set limits on the contribution of such heavily
extincted components to the total, unextincted PAH emission.  First we
assume that any heavily obscured PAH emission is extincted to the same
degree as the continuum.  Then, for the median $\tau_{9.7}^{eff}$ in
our sample (3.9), any heavily obscured PAH emission must constitute
less than about a third of the total unextincted PAH emission for it
to not significantly alter the fit.  This obscured component could
rise to half of the total unextincted emission if $\tau_{9.7}^{eff}$
was about twice the median ($6-7$).  The four sources where PAH
extinction is detected (Section 5.2) may be cases where obscured PAH
emission starts to dominate.

Both PAH (e.g., F\"{o}rster-Schreiber et al. 2004; Peeters et
al. 2004; Calzetti et al. 2007) and FIR emission (e.g., Kennicutt
1998) are tracers of star formation in quiescent and actively star
forming galaxies.  We also argue in Papers I and II that PAH and FIR
emission in PG QSOs are produced by star formation.  We find a fairly
tight distribution of $L$(PAH)/$L$(FIR) in ULIRGs
(Figure \ref{fig:pahfir}$a$ and $b$), consistent with previous studies
(Peeters et al. 2004) as well as the notion that both trace star
formation.  [$L$(PAH) is the total PAH flux in the $5-30$\micron\
range.]  K-S and Kuiper tests indicate no significant trend with
optical spectral type.  The mean and standard deviation are
log~$L$(PAH)/$L$(FIR) $\simeq -1.71 \pm 0.3$.  The same conclusions
apply if $L$(IR) is substituted for $L$(FIR), and we measure
log~$L$(PAH)/$L$(IR) $\simeq -1.96 \pm 0.3$.  If the 7.7\micron\ PAH
luminosity is substituted for the total luminosity, these PAH ratios
are lower by 0.4 dex.  Thus, our results are consistent with $L$(PAH,
7.7\micron)/$L$(FIR) for PG~QSOs ($-2.0 \pm 0.3$; Paper I).

However, we do find that galaxies with stronger than average silicate
absorption have smaller $L$(PAH)/$L$(FIR) ratios by a factor of 2 than
galaxies with weaker than average absorption, as verified with K-S and
Kuiper tests (Figure \ref{fig:pahfir}$c$).  In fact, the PAH-to-FIR
ratio {\it anticorrelates} with effective silicate optical depth, such
that larger extinction corresponds to smaller $L$(PAH)/$L$(FIR)
(Figure \ref{fig:pahfir}$d$).  This effect is most pronounced in the HII
and LINER ULIRGs.  As we note above, $\sim$ half of the intrinsic PAH
emission may be completely buried in the most heavily obscured
sources.  A factor-of-two correction could close at least some of the
discrepancy between PAH-to-FIR ratios in heavily obscured and lightly
obscured ULIRGs, and further tighten the distribution of
$L$(PAH)/$L$(FIR).  However, we argue below that the differences we
observe are more likely due to a real suppression of PAH emission.

In Figure \ref{fig:weqpah7}$a$, we show the distribution of
7.7\micron\ equivalent widths, which is quite broad and shows
significant optical spectral type dependence.  Seyfert 1 ULIRGs have
much smaller PAH equivalent widths on average than H~II ULIRGs (--0.78
{\em vs} 0.13), while the PAH equivalent widths of Seyfert 2 (--0.29)
and LINER (--0.05) ULIRGs fall in between these values, confirming
earlier {\em ISO} results (e.g., Genzel et al.\ 1998; Lutz et al.\
1999) as well as recent {\em Spitzer} results (e.g., Desai et al. 2007;
Spoon et al. 2007).  PG QSOs overlap with Seyferts.

A weak luminosity dependence is also present.  ULIRGs with
log[$L$(IR)$/L_\odot$] $\ge$ 12.4 have slightly smaller 7.7\micron\
PAH equivalent widths than lower luminosity objects (--0.36 $\pm$ 0.09
{\em vs.\/} 0.04 $\pm$ 0.08; Figure \ref{fig:weqpah7}$b$), in agreement
with earlier {\em ISO} results (e.g., Lutz et al.\ 1998b; Tran et al.\
2001).  The PAH equivalent widths of PG~QSOs are similar to those of
Seyfert 1 ULIRGs, but they do not follow the trend with infrared
luminosity of the ULIRGs (this is not surprising since PG~QSOs were
not selected through infrared methods like the ULIRGs). The slight IR
luminosity dependence of EW(PAH) among ULIRGs coincides with the
excess of Seyfert 1 ULIRGs (7 out of 9) and deficit of HII ULIRGs (2
out of 18) in the high-luminosity bin of our sample.  In other words,
it parallels the well-known infrared luminosity dependence of the
optical spectral types of ULIRGs (Veilleux et al.\ 1995, 1999a).

Given the trend with optical spectral type, it is not surprising to
find that ULIRGs with warmer quasar-like MIR continua exhibit smaller
PAH equivalent widths than cooler systems (Figure \ref{fig:weqpah7}$c$).
The $f_{30}/f_{15}$ ratio is particularly efficient at separating
objects, including PG~QSOs, according to their PAH equivalent widths.
Arguably it is even better at it than the optical spectra type, since
there is a correlation between $f_{30}/f_{15}$ and EW(PAH) among
galaxies of a given spectral type.  Our results also indicate that
$f_{30}/f_{15}$ is a better proxy for EW(PAH) than $f_{30}/f_{6}$ and
even $f_{15}/f_{6}$, the continuum color diagnostic used by Laurent et
al.\ (2000).  We return to this point in Section 7.1.

As with the PAH-to-FIR ratios (Figure \ref{fig:weqpah7}$d$), the
7.7\micron\ equivalent width correlates strongly with extinction in
HII/LINER ULIRGs (Figure \ref{fig:weqpah7}$d$).  What is the origin of
these dependences?  For the PAH-to-FIR ratio, we cannot rule out
extinction effects.  We can for EW(PAH), as long as the continuum and
any unobserved, heavily obscured PAHs are extincted to roughly the
same degree.  The equivalent width is, however, affected by a strongly
varying amount of warm continuum.  We observe a broad distribution of
8\micron-to-IR ratios (Figure \ref{fig:bblir}$a$), suggesting that the
8\micron\ continuum plays an important role in regulating EW(PAH).
The anticorrelation of both PAH-to-FIR ratio and EW(PAH) with
extinction also points to the presence of PAH suppression at high
extinction / low EW(PAH).  This supression may be due to effects of
high density in the cores of ULIRGs, or to destruction of PAHs in the
harsh radiation field of AGN (whose importance increases with
increasing optical depth in HII/LINER ULIRGs; \S7.3).  In our recipe
for computing AGN contribution from EW(PAH), we assume that (a) PAH
emission is due to star formtion and that (b) an AGN causes both an
increase in the 8\micron\ continuum and PAH suppression (\S7.1).

A qualitatively similar result was found by Desai et al.\ (2007) and
Spoon et al.\ (2007) using large samples of starbursts, AGN, and
ULIRGs (some of these are also part of our sample).  The HII and LINER
ULIRGs in our sample populate a diagonal sequence joining the highly
absorbed, weak-PAH ULIRGs with the unabsorbed, PAH-dominated systems;
this sequence coincides with the ``diagonal branch'' of Spoon et al.
On the other hand, all Seyfert 1 ULIRGs and many, but not all, Seyfert
2 ULIRGs in our sample have both weak silicate absorption {\em and}
weak PAHs; they populate what Spoon et al.\ call the ``horizontal
branch''.  The existence of these two branches may reflect true
intrinsic differences in the power source and/or nuclear dust
distribution between galaxies on the two branches (e.g., Levenson et
al.\ 2007; Spoon et al.\ 2007; Sirocky et al.\ 2008). We return to
this point in Section 7.1 and Section 7.3.

\subsection{Fine Structure Lines}

In this section we use the strengths of the fine structure lines to
constrain the properties of the warm ionized gas near the central
energy source of our sample galaxies.  We first discuss the
low-excitation features that are commonly detected in star-forming
galaxies before discussing the high-excitation lines, direct probes of
the AGN phenomenon.

Figure \ref{fig:linelum_v_fir}$a$ compares the luminosity of the
[Ne~II] 12.8 $\mu$m line in the ULIRGs and QSOs of our sample with the
FIR luminosity.  The average and median values of the [Ne~II]/FIR
luminosity ratios are remarkably similar regardless of the optical
spectral type, including the Seyfert 1 ULIRGs and the QSOs:
log~$L$([Ne~II])/$L$(FIR) = --3.35 $\pm$ 0.10. The similarity of this
ratio for ULIRGs and QSOs was first pointed out in Paper I, where this
result in combination with the similar PAH-to-FIR luminosity ratio (\S
6.3) was used to argue that the bulk of the FIR luminosity in QSOs is
produced via obscured star formation rather than the AGN. Not
surprisingly, the less obscured optically-selected {\em ISO}
starbursts (Verma et al.\ 2003) and Seyfert galaxies (Sturm et al.\
2002) plotted in Figure \ref{fig:linelum_v_fir}$a$ have noticeably
larger [Ne~II]/FIR ratios (by a factor of $\sim$ 2 and 4,
respectively).

The ([Ne~III] 15.5 $\mu$m)/([Ne~II] 12.8 $\mu$m) line ratio is
commonly used to diagnose the excitation properties of star-forming
galaxies (e.g., Thornley et al.\ 2000; Verma et al.\ 2003; Brandl et
al.\ 2006).  Since the ionization potentials needed to produce
Ne$^{+}$ and Ne$^{++}$ are 21.6 and 41.0 eV, respectively, the
[Ne~III]/[Ne~II] ratio is sensitive to the hardness of the ionizing
radiation and therefore to the (effective temperature of the) most
massive stars in a starburst or to the presence of an AGN, if
applicable. Figure \ref{fig:ne3ne2} shows this ratio as a function of
the infrared and FIR luminosities and the MIR continuum colors,
$f_{25}/f_{60}$ and $f_{30}/f_{15}$.  No obvious trend with the F/IR
luminosities is seen among ULIRGs.  However, a clear dependence is
seen with optical spectral type and MIR continuum colors, confirming
earlier studies (e.g., Dale et al. 2006; Farrah et al. 2007). Larger
[Ne~III]/[Ne~II] ratios go hand-in-hand with warmer MIR continuum.
QSOs have larger [Ne~III]/[Ne~II] ratios on average than Seyfert
ULIRGs, and Seyfert ULIRGs have larger ratios on average than HII-like
and LINER ULIRGs. K-S and Kuiper tests indicate that the
[Ne~III]/[Ne~II] ratios of LINER ULIRGs are statistically
indistinguishable from those of HII-like ULIRGs, and the same
statement also applies when comparing Seyfert 1 ULIRGs with Seyfert 2
ULIRGs. The two Seyfert 1 ULIRGs with [Ne~III] upper limits are
F07598+6508 and F13218+0552. Both of them have unusually small optical
narrow-line [OIII] $\lambda$5007/H$\beta$ ratios, consistent with
low-luminosity high-excitation regions (e.g., Kim et al.\ 1998). The
dependence of [Ne~III]/[Ne~II] on spectral type and MIR colors induces
a slight trend between this ratio and EW(PAH 7.7)
(Figure \ref{fig:ne3ne2}$e$). No obvious trend between [Ne~III]/[Ne~II]
and EW(PAH 7.7) is seen within HII-like and LINER ULIRGs, a result
that is consistent with that found for optically-selected starburst
galaxies (Brandl et al.\ 2006). Given the relatively high metallicity of
ULIRGs (Section 6.6 in this paper and Rupke et al. 2008), the large
[Ne~III]/[Ne~II] ratios seen in Seyfert ULIRGs cannot be explained by
star formation alone as in the case of low-metallicity dwarf galaxies.

Other low ionization fine structure lines such as [Fe~II] 25.99
$\mu$m, [S~III] 33.48 $\mu$m, and [Si~II] 34.82 $\mu$m have been found
to be useful diagnostics of activity in galactic nuclei (e.g., Lutz et
al. 2003; Sturm et al. 2005; Dale et al. 2006). Unfortunately, these
lines are often redshifted out of the wavelength range of our data so
they cannot be used for any kind of statistical analysis. We do not
discuss these lines any further in this paper.

The ionizing spectra of all but the hottest O stars cut off near the
He II edge (54.4 eV), so the detection of [O~IV] 25.9 $\mu$m from
three-times ionized oxygen with ionization energy $\chi$ $\sim$ 55 eV
is potentially a good indicator of AGN activity.  This line is
detected in 30/34 QSOs, 3/9 Seyfert 1 ULIRGs, 8/13 Seyfert 2 ULIRGs,
6/28 LINER ULIRGs, and only one of the 18 HII-like ULIRGs
(F21208$-$0519:N) with high-resolution spectra. In Figure
\ref{fig:linelum_v_fir}$b$, we plot the (upper limits on the) [O~IV]
25.9 $\mu$m luminosity {\em versus} the FIR luminosity of ULIRGs and
QSOs.  The solid line is a fit to the data of HII ULIRGs, so it is
formally only an upper limit, as indicated by the arrows.  This upper
limit is above, therefore consistent with, the measured values in {\em
ISO} starbursts. The PG~QSOs and the few Seyfert 1 ULIRGs with [O~IV]
detections lie on average $\sim$1.2 dex above that line. All Seyfert 2
ULIRGs with [O~IV] detections lie $\sim$ 0.8 dex above that line,
while the upper limits on [O~IV] derived for the other Seyfert ULIRGs
are consistent with those for the HII ULIRGs and reflect the flux
detection threshold across the sample. Interestingly, the
optically-selected ISO Seyfert galaxies have [O~IV]/FIR luminosity
ratios that are similar to those of the PG~QSOs.

As in the case of [Ne~III]/[Ne~II], there is no strong trend between
the [O~IV]/[Ne~II] ratios and M/IR luminosities of ULIRGs, but a
strong dependence with optical spectral type and MIR continuum colors
is detected (Figure \ref{fig:o4ne2}). These results are similar to those
found by Farrah et al.\ (2007) on a different but overlapping sample
of ULIRGs.  The lack of an obvious luminosity dependence among ULIRGs
may be surprising in the light of the optical and {\em ISO} results
which suggest a larger AGN contribution to the bolometric luminosity
among the more luminous ULIRGs (e.g., Veilleux et al.\ 1995, 1999a;
Lutz et al.\ 1998b; Tran et al.\ 2001).  Possible explanations for
this apparent discrepancy include: (1) the optical spectral
classification is affected by dust obscuration so it is not reliable;
(2) [O~IV] 25.9 $\mu$m is not as good an AGN indicator as the PAH
equivalent width (e.g., contaminating [O~IV] emission from WR stars
and ionizing shocks, Lutz et al.\ 1998a; Abel \& Satyapal 2008); (3)
given typical ULIRG redshifts and actual IRS sensitivity at $\ga$ 30
$\mu$m, detecting [O~IV] is difficult. The number of ULIRGs with
actual [O~IV] detection is small, especially among HII-like and LINER
ULIRGs, so small number statistics mask the correlation. We favor this
last possibility. Explanation \#1 can be rejected outright since we
detect in the present paper (as in Lutz et al.\ 1999) clear
correlations of the MIR parameters with optical spectral types. If the
optical classification were unreliable, these relations would be
erased. Scenario \#2 seems unlikely since starbursts with large
[O~IV]/[Ne~II] are (low metallicity) dwarfs; at the relatively high
metallicity of ULIRGs, this ratio is expected to be small. Note,
however, that several Seyfert 1 ULIRGs have no detected [O~IV]
emission. This is not purely a sensitivity effect. As pointed out in
our discussion of the [Ne~III]/[Ne~II] ratios in Seyfert 1 ULIRGs,
there is strong corraborating evidence at optical wavelengths that
many of these objects have high-excitation regions of unexpectedly low
luminosity. The exact cause of this effect is unclear. With these
caveats in mind, we will use the [O~IV]/[Ne~II] ratio in Section 7.1
as a diagnostic of nuclear activity in ULIRGs.

Despite the fact that [Ne~V] 14.3\micron\ is fainter in these sources
than [O~IV], the better S/N ratio of the IRS data at shorter
wavelengths has allowed us to put similar constraints on the [Ne~V]
and [O~IV] lines.  ([Ne~V] 24.3\micron\ was also detected in many
sources with [Ne~V] 14.3\micron\ emission, but at a lower rate than
[Ne~V] 14.3 \micron).  [Ne~V] 14.3\micron\ was detected in 25/34 QSOs,
4/9 Seyfert 1 ULIRGs, 9/13 Seyfert 2 ULIRGs, 4/28 LINER ULIRGs
(F04103$-$2838, UGC~5101, F13335$-$2612, NGC~6240), and even one of
the 18 HII ULIRGs with high-resolution spectra (F20414$-$1651).  The
relatively small redshifts of three of the five detected HII/LINER
ULIRGs makes it apparent that sensitivity plays a role in the
detectability of these lines in sources where they are intrinsically
weak. Once again, the relatively modest number of detections in
Seyfert 1 ULIRGs point to intrinsically weak high-excitation regions
in some of these objects.

The very high ionization potential of Ne$^{4+}$, $\chi$ = 97.1 eV,
makes this line an unambiguous signature of nuclear activity. Indeed
the separation with spectral type and MIR continuum colors previously
seen in [O~IV]/(F/IR) and [O~IV]/[Ne~II] is clearer when [Ne~V] is
substituted for [O~IV] (Figure \ref{fig:linelum_v_fir}$c$ and
\ref{fig:ne5ne2}). This separation is further emphasized in Figure
\ref{fig:o4ne2_v_ne5ne2}, where we compare the values of
[Ne~V]/[Ne~II] with [O~IV]/[Ne~II] measured in our sample of QSOs and
ULIRGs.  The solid diagonal line in these diagrams is a line of
constant [Ne~V]/[O~IV] and changing [Ne~II].  This line may be
interpreted as a mixing line if [Ne~V] and [O~IV] are only produced by
an AGN and [Ne~II] by starburst activity.  The tickmarks along the
line indicate the percentage contribution of the starburst to [Ne~II]
(from 0 to 99\%). Optically-selected ISO Seyfert galaxies lie in the
same region as the PG~QSOs in all these diagrams (Figures
\ref{fig:linelum_v_fir}, \ref{fig:ne5ne2}, and
\ref{fig:o4ne2_v_ne5ne2}). We will return to this last figure in our
discussion of the energy source in ULIRGs and QSOs (Section 7.1).

In the ULIRGs and PG~QSOs with high-S/N SL spectra, we also searched
for redshifted [Ne~VI] 7.65 $\mu$m ($\chi$ = 126 eV), another powerful
AGN indicator.  This line was unambiguously detected in 6 QSOs, 2
Seyfert 1 ULIRGs, and 5 Seyfert 2 ULIRGs, but in none of the LINER and
HII-like ULIRGs.  Confusion between [Ne~VI] and PAH substructure can
mask weak [Ne~VI] emission in the latter objects.  By and large, [Ne~VI]
follows the same trends with spectral type and MIR continuum colors as
[Ne~V] (Figures \ref{fig:linelum_v_fir}$d$ and \ref{fig:ne6ne2}).

The [Ne~VI]/[O~IV] {\em vs.\/} [Ne~VI]/[Ne~II] diagram from Sturm et
al.\ (2002) is reproduced in Figure \ref{fig:highionlines}$a$.  Both
axes scale with the AGN excitation, but [Ne~VI]/[Ne~II] can also be
influenced by contributions from star forming regions to the [Ne~II]
line.  As shown by the grid of AGN models from Groves et al.\ (2004),
pure AGN are expected to lie roughly along a diagonal line in this
diagram (both ratios increase with increasing hardness of the
radiation).  Composite sources, however, have stronger [Ne~II] lines
than pure AGN, so they are expected to lie to the left of the pure
AGN sources.  Because Seyfert ULIRGs are composite sources, with
significant starburst contribution to [Ne~II]
(Figure \ref{fig:o4ne2_v_ne5ne2}), they lie leftward of most of the
comparison Seyferts and PG~QSOs.  The comparison Seyferts are in
decent agreement with the models, though some may suffer minor
starburst contamination to [Ne~II].  The [Ne~VI]/[Ne~II] ratios of
most of the PG~QSOs are too high by factors of $2-3$.  There is some
disagreement in [Ne~VI]/[O~IV], as well, which may be due to incorrect
Ne/O abundance ratios in the models (\S6.6).

A much better match with the models is found when considering the
[Ne~VI]/[Ne~III] {\em vs.\/} [Ne~VI]/[Ne~V] diagnostic diagram
(Figure \ref{fig:highionlines}$b$), which is independent of relative
metal abundance effects. This suggests that the bulk of the [Ne~VI],
[Ne~V], and possibly even [Ne~III] emission in PG~QSOs, Seyfert
ULIRGs, and {\em ISO} Seyferts is produced by the AGN.

\subsection{Molecular Hydrogen Lines} 

Three lines from rotational transitions of warm H$_2$ were regularly
detected in the spectra of ULIRGs and QSOs: $v = 0$ $J = 3 - 1$ S(1)
17.04 $\mu$m, $v = 0$ $J = 4 - 2$ S(2) 12.28 $\mu$m, and $v = 0$ $J =
5 - 3$ S(3) 9.67 $\mu$m.  The $v = 0$ $J = 2 - 0$ S(0) 28.22 $\mu$m
transition was detected in 10 objects (F09039+0503, UGC~5101,
F12112+0305, F13335$-$2612, Mrk~273, F14248$-$1447, F21208$-$0519:N,
PG~1211+143, PG~1440+356, and B2~2201+31A). Higher-level transitions
were also detected in a few objects [S(4), S(5), S(6), and S(7) in
F09039+0503 and F15130$-$1958, and S(5) in UGC~5101, F12112+0305,
F17208$-$0014, and PG~1700+518].

To first order, the strengths of the H$_2$ lines scale linearly with
the star formation rate indicators of ULIRGs: F/IR, [Ne~II], and PAH
luminosities (Figure \ref{fig:h2s1lum_v}). This is true in detail for
the HII-like ULIRGs, but strong departures from the linear relation
are seen among the other ULIRGs and the QSOs.  The H$_2$ line emission
in these latter objects tends to be overluminous for a given F/IR or
PAH luminosity. Similar departures from the linear relation were seen
among the SINGS LINER/Seyfert targets (Roussel et al.\ 2007) and
attributed to shock heating. The fact that the departures are
strongest among the QSOs of our sample suggests that heating by the
AGN is important in these objects.  Rigopoulou et al.\ (2002) came to
a similar conclusion based on the H$_2$/PAH ratio of a sample of 9
optically-selected Seyfert galaxies.

The temperature-sensitive H$_2$ S(2)/S(1), S(3)/S(1), and S(3)/S(2)
ratios may be used to shed some light on the possible role of the AGN.
Their distributions are shown in Figure \ref{fig:h2rat}. Warm Seyfert
ULIRGs tend to have larger [smaller] S(3)/S(2) [S(2)/S(1)] ratios than
cool HII-like/LINER ULIRGs, while no statistically significant trend
is seen in the S(3)/S(1) ratio distribution. The K-S and Kuiper
probabilities that the distributions of S(3)/S(2) and S(2)/S(1)
ratios, when put in two bins above and below the median
$f_{25}/f_{60}$, arise from the same parent distribution are $<$2\%,
confirming the apparent trends. The number of QSOs with reliable H$_2$
line ratios is too small to be able to detect statistically
significant trends.

We used these ratios, when available, to construct an excitation
diagram for each object, assuming LTE, an ortho-to-para ratio of 3,
and no extinction. Figure \ref{fig:h2tex} shows a few examples. A
straight line in these diagrams indicates that a single temperature
applies to all transitions, with the excitation temperature ($T_{ex}$)
being the reciprocal of the slope. For an ortho-to-para ratio of 3,
temperatures derived from adjacent lines should increase with J. This
is illustrated in Figure \ref{fig:h2texdiff}, where we compare
$T_{ex}$(J = 4 - 3) and $T_{ex}$(J = 5 - 4), the excitation
temperatures derived from the S(2)/S(1) and S(3)/S(2) ratios,
respectively, as a function of $f_{25}/f_{60}$ and the effective
silicate optical depth, $\tau_{9.7}^{\rm effective}$. A temperature
difference that is negative implies that extinction and/or ortho-para
effects are at play, as illustrated by the downward arrows on the
right in each diagram (see caption to this figure for an explanation
of the arrows).

The small number of Seyfert 1s and QSOs for which both $T_{ex}$(J =
4-3) and $T_{ex}$(J = 5 - 4) are available prevents us from detecting
any obvious dependence on optical spectral type.  A visual inspection
of Figure \ref{fig:h2texdiff}$a$ suggests a possible trend of
increasing $T_{ex}$(J = 5 - 4) - $T_{ex}$(J = 4-3) with increasing
$f_{25}/f_{60}$, which would support the role of AGN in heating the
molecular gas in these objects.  In fact, the K-S and Kuiper
probabilities that the distributions of temperature differences binned
according to $f_{25}/f_{60}$ (above and below the median
$f_{25}/f_{60}$) arise from the same parent distribution are 0.6\% and
6\%, respectively, so there is a statistically significant trend.
However, much of this trend may be due primarily to extinction, as
shown in Figure \ref{fig:h2texdiff}$b$. The K-S and Kuiper
probabilities that the distributions of temperature differences binned
according to $\tau_{9.7}^{\rm effective}$ (above and below the median
$\tau_{9.7}^{\rm effective}$) arise from the same parent distribution
are only 0.8\% and 7\%, respectively.  Thus, an important conclusion
of this discussion is that any trends with $f_{25}/f_{60}$ could be
masked by extinction and ortho-para effects (c.f.\ Higdon et al.\
2006).

Table \ref{tab:h2} lists the warm H$_2$ masses derived from the
strength of the S(1) line and the average $T_{ex}$ for each
object. Values range from 0.5 to 20 $\times$ 10$^8$ M$_\odot$ with an
average (median) of $\sim$ 3.8 (3.3) and 3.6 (3.2) $\times$ 10$^8$
M$_\odot$ for the ULIRGs and QSOs, respectively. These values are
slightly larger on average than those of Higdon et al.\ (2006) and
imply that the warm gas mass is typically a few percent of the cold
gas mass derived from $^{12}$CO observations (0.4 -- 1.5 $\times$
10$^{10}$ M$_\odot$; e.g., Solomon et al.\ 1997; Downes \& Solomon
1998; Evans et al. 2001, 2002, 2006; Scoville et al. 2003).

\subsection{Metal Abundance}

In principle, one can use the strengths of the fine structure lines
relative to the hydrogen recombination lines in our data to derive the
metallicity of the gas producing these emission features.  In
practice, the only hydrogen line within the wavelength range of our
data is the very faint Hu$\alpha$ 12.4 $\mu$m (H 7-6), so the S/N of
our data only allow us to marginally detect or put upper limits on the
strength of this line in individual objects. However, Hu$\alpha$ is
detected at S/N$\sim6$ in the average spectrum of 27 PAH-dominated
ULIRGs (Figure\ 36$a$), so we can use the strength of this line to
derive an average metallicity in these systems. We follow the methods
of Verma et al.\ (2003), using the ratios [Ne~II] 12.8
$\mu$m/Hu$\alpha$ = 62 and [Ne~III] 15.5 $\mu$m/Hu$\alpha$ = 18
measured from the average spectrum to derive the abundance of neon in
these systems (Figure\ 36$b$). The recombination line Hu$\alpha$,
[Ne~II] 12.8 $\mu$m which is tracing the dominant singly ionized state
of neon, and [Ne~III] 15.5 $\mu$m tracing doubly ionized neon are
found at similar MIR wavelengths. They can be used to reach optically
obscured regions and obtain a metallicity measurement that is much
less sensitive to extinction effects than results obtained in
combination of MIR lines with NIR recombination lines. We
did not apply an extinction correction to the observed MIR line
ratios. Adopting an electron temperature of 5000\,K appropriate for
dusty starbursts (e.g. Puxley et al.\ 1989), Hu$\alpha$ emissivity
from Storey \& Hummer (1995) and neon collision strengths from Saraph
\& Tully (1994) and Butler \& Zeippen (1994), we find a neon abundance
12+log(Ne/H) = 8.30.  As discussed in the last paragraph of the
present section, the value of the solar neon abundance is currently the
subject of a heated debate. If we adopt the revised solar photospheric
neon abundance of Asplund et al.\ (2004), our neon abundance is $\sim$
2.9 $\times$ solar.

An underabundance compared to local luminosity-metallicity and
mass-metallicity relations of galaxies was recently reported in the
optical study of 100 star-forming LIRGs and ULIRGs by Rupke et al.\
(2008), who attributed it to a combination of two effects: a decrease
of abundance with increasing radius in the progenitor galaxies and
strong, interaction- or merger-induced gas inflow into the galaxy
nucleus. Thirteen objects from the sample of Rupke et al.\ (2008) are
in common with the current Spitzer sample. The oxygen abundance, 12 +
log(O/H), of the ULIRGs in common with both samples ranges from 8.43
to 9.04. Using the value from Asplund et al.\ (2004) for the
solar oxygen abundance, 12 + log(O/H)$_\odot$ = 8.66, these numbers
translate into 0.6 -- 2.4 $\times$ solar.  Given this relatively
narrow range of abundance and small number of objects, it is perhaps
not surprising that no trend was found in our sample between the
oxygen abundances of Rupke et al.\ (2008) and any {\em Spitzer}-derived
continuum or line ratios.  In particular, we note that the oxygen
abundance of these ULIRGs is well above the threshold abundance, 12 +
log(O/H) = 8.1 or $\sim$ 0.3 solar, below which PAH emission is
apparently suppressed (e.g., Engelbracht et al.\ 2005; O'Halloran et
al.\ 2006; Smith et al.\ 2007).

Supersolar neon abundance derived from the MIR spectra (if the Asplund
et al.\ 2004 value of the solar neon abundance is correct) and close
to solar oxygen abundance derived from the optical spectra may trace
different layers of the ULIRGs, in both extinction and abundance. In
the picture outlined by Rupke et al.\ (2008), it is plausible that
less obscured regions are dominated by lower metallicity gas
transported in from the outskirts of the galaxies. In contrast, the
dusty inner regions may be dominated by more pre-enriched gas from the
inner regions of the progenitor galaxies, compressed to the immediate
circumnuclear region during the merger process and enriched further by
the intense circumnuclear star formation.  However, the excellent
overall agreement reported in Section 7.1 between optical and MIR
diagnostics of nuclear activity do not seem to favor this
picture. These results imply that the optical line spectrum in most
cases traces gas that ``knows'' what the true power source is and
therefore should also trace gas that fairly samples the
metallicity. An alternative explanation for the higher neon abundance
-- and the one we favor -- is that it reflects {\em in-situ}
enrichment in the most heavily obscured (densest) star-forming
regions, but these regions are distributed throughout the ULIRG rather
than preferentially near the center.

An important caveat in comparing the optical and MIR metallicity
measurements is the assumed solar Ne/O ratio. The exact value of this
ratio has been the subject of a heated debate in recent years, some
groups arguing that it is considerably higher than the standard value
(e.g., Drake \& Testa 2005; Wang \& Liu 2008, although see Schmelz et
al. 2005 for a counterexample). A larger solar Ne/O ratio would bring
our {\em Spitzer} measurements in closer agreement with the optical
results.

\section{Discussion}

\subsection{Energy Source: Starburst {\em  vs.\/}  AGN}

In this section, we use the data presented in Section 6 to estimate
the fractional contribution of nuclear activity to the bolometric
luminosity of the ULIRGs and PG~QSOs in our sample (herafter called
the ``AGN contribution'' for short). We use six different methods
based on (1) the [O~IV] 25.9 $\mu$m/[Ne~II] 12.8 $\mu$m ratio, (2) the
[Ne~V] 14.3 $\mu$m/[Ne~II] 12.8 $\mu$m ratio, (3) the equivalent width
of PAH 7.7 $\mu$m, (4) the PAH (5.9 - 6.8 $\mu$m) to continuum (5.1 --
6.8 $\mu$m) flux ratio combined with the continuum (14 -- 15 $\mu$m) /
(5.1 -- 5.8 $\mu$m) flux ratio (see Figure 34), (5) the MIR blackbody
to FIR flux ratio, and (6) the $f_{30}/f_{15}$ continuum flux
ratio. These methods are described in detail in Appendix A. The zero
points, bolometric corrections, and basic results from each method are
listed in Tables $\ref{tab:agnfrac_zp}-\ref{tab:agnfrac_indiv}$. We
compare the results from the various methods and look for trends with
optical and infrared parameters in Section 7.1.1 and Section 7.1.2,
respectively.

\subsubsection{Comparisons of Results from Different Methods}

Tables $\ref{tab:agnfrac_avg}-\ref{tab:agnfrac_indiv}$ and Figure
\ref{fig:agnfrac_comp} indicate a remarkably good agreement between
the AGN fractional contributions to the bolometric luminosities of
ULIRGs and PG~QSOs derived from the various methods. {\em The mean
ULIRG AGN contribution is $\sim$ 38.8 $\pm$ 21.1\% averaged over all
ULIRGs and all methods} (in Table \ref{tab:agnfrac_avg}, the
average-of-averages and standard errors are calculated by first
averaging over all methods for individual objects, then averaging over
objects). {\em This mean ULIRG AGN contribution is in agreement with,
and refines the results of, Genzel et al.\ (1998): ULIRGs are
composite objects, but on average powered mostly by star formation.}
The various methods give average AGN contributions that are within
$\sim$ $\pm$10$-$15\% of each other, taking into account the range of
AGN fractional contributions derived from the fine structure line
ratios measured from average spectra (see discussion in Appendix
A). These small differences between the various methods can easily be
explained by uncertainties on the pure-starburst zero points (see
discussion in Appendix A; the pure-AGN zero points are considered more
robust since they are based on the FIR-undetected PG~QSOs) and modest
differential extinction ($A_V$ $\la$ 10 mags) between the inner
line-emitting region (where the bulk of the [O~IV] and [Ne~V] emission
is produced on average) and outer line-emitting region (where the bulk
of the [Ne~II] and PAH emission is produced on average). The good
agreement between the various methods is not in contradiction with the
results of Armus et al.\ (2007) since here we compare AGN fractional
contributions to the bolometric luminosities, while Armus et al.\ did
not apply bolometric corrections to their numbers so they were
comparing AGN fractional contributions to the [Ne~II] and MIR
luminosities and found them to be different.

Note that there is systematic uncertainty associated with the choice
of what defines an AGN or starburst (or HII region / PDR in the case
of the Laurent method; see Appendix A for detailed discussion on the
choices of zero points and bolometric correction factors for each
method).  For instance, there may be a range of possible emission-line
ratios or continua that define a ``pure'' AGN or starburst.
Experiments show that reasonable changes in zero-point values do not
significantly change the results for a given method.  Nonetheless,
this uncertainty may contribute to the scatter observed when comparing
differing diagnostics.  To smooth over these possible systematics, in
what follows we compute the average AGN contribution over all methods
for each object.  This minimizes the chance that a stronger systematic
effect in any one method will affect the results.

Furthermore, we cannot rule out the possibility of a third class of
physically distinct systems.  In other words, a pure starburst or pure
AGN may not describe all of parameter space.  For instance, one
possibility is that heavily obscured systems host unique physical
conditions in high-density cores that do not replicate starbursting or
AGN ULIRG environments.  To examine this possibility, we highlight
systems with effective silicate optical depths above the median in
Figure \ref{fig:agnfrac_comp}.  It is evident from this figure that
HII and LINER ULIRGs with higher obscuration tend to have higher AGN
contribution.  Whether this is due to fundamentally different physics,
or simply an obscured AGN, is unclear from this diagram.  We return to
this issue in Section 7.1.2 and Section 7.3, where we uncover smooth trends
between AGN contribution, obscuration, and merger phase which are
difficult to explain if fundamentally different physics were at play.

Finally, we cannot formally rule out the possibility of deeply buried
AGN invisible at MIR wavelengths but contributing significantly to the
FIR emission in some of these objects. However, it is now considered a
highly contrived scenario given the good agreement between the variety
of methods used to evaluate the AGN contribution to the bolometric
luminosity.  As described in Appendix A, these methods use the full
gamut of diagnostic tools available at 6 -- 30 um. The diagnostic
features are produced under different conditions (density, dust
content) and over a range of distances from the center. They also
cover a broad range in wavelength and therefore dust optical depth.
If obscured AGN are contributing significantly to the FIR emission of
several of these sources, one would expect diagnostics that use
long-wavelength emission and probe deep into the cores (e.g.,
$f_{30}/f_{15}$ ratio) to give systematically different results than
the others. This is not seen in our data.

\subsubsection{Trends with Optical Spectral Types, $f_{25}/f_{60}$
  ratios, Infrared Luminosities, and Extinctions}

We detect strong correlations between {\em Spitzer}-derived AGN
contributions on the one hand and optical spectral types and
$f_{25}/f_{60}$ ratios on the other (Figure \ref{fig:agnfrac_v}$a$ and
$c$).  These results confirm and expand on earlier results. The AGN
contribution ranges from $\sim$$15-35$\% among HII and LINER ULIRGs
(taking into account the range of AGN fractional contributions derived
from the fine structure line ratios measured from average spectra [see
discussion in Appendix A]) to $\sim$ 50 and 75\% among Seyfert 2 and
Seyfert 1 ULIRGs, respectively. The presence of a dominant AGN in
Seyfert 1 ULIRGs was first deduced from the strengths of the
optical/NIR broad lines in a few objects (Figures 4 and 5 of Veilleux et
al.\ 1997 and 1999b, respectively); the new {\em Spitzer} results now
show that this statement applies to Seyfert 1 ULIRGs in general. The
excellent correlation between optical spectral types and 7.7 $\mu$m
PAH-derived AGN contribution was first pointed out by Taniguchi et
al.\ (1999) and Lutz et al.\ (1999) using {\em ISO} data, but we have
now quantified this correlation and detected similar ones when using
the fine-structure line and continuum slope methods. The correlation
between AGN contributions and $f_{25}/f_{60}$ ratios is equally strong
and quantitatively confirms the qualitative statement made more than
twenty years ago by de Grijp et al.\ (1985) that this ratio is an
excellent indicator of AGN activity. The AGN contribution among cool
ULIRGs ($f_{25}/f_{60}$ $<$ 0.2) is $\sim$ 30\% on average compared
with $\sim$ 60\% among warm ULIRGs.

Figure \ref{fig:agnfrac_v} also displays the AGN contributions of the
PG~QSOs. These fall right along the extrapolation of the spectral type
and $f_{25}/f_{60}$ sequences, with AGN contributions typically larger
than $\sim$ 80\% among the QSOs. (Recall that only 8 PG~QSOs -- only
those that are FIR-undetected -- were used to set the pure-AGN zero
points so this last statement is not circular.)  These results bring
support to the concept of an excitation sequence between the cool,
HII/LINER ULIRGs, the warm Seyfert-like ULIRGs, and the PG~QSOs. They
are also consistent with the evolution scenario proposed by Sanders et
al.\ (1988a, 1988b), if the excitation sequence is also a merger
sequence. This question is examined in Section 7.3 below.

A weaker correlation is seen between the AGN contributions and
infrared luminosities of ULIRGs (Figure \ref{fig:agnfrac_v}$b$).  We
observe average AGN contributions of $\sim$ 34\% and $\sim$ 48\% for
ULIRGs with log[$L$(IR)$/L_\odot$] below and above 12.4, respectively.
This general trend with infrared luminosity is consistent with the
optical results of Veilleux et al.\ (1995, 1999a) and the {\em ISO}
results of Lutz et al.\ (1999) and Tran et al.\ (2001). The PG~QSOs
are distinctly less infrared luminous than the ULIRGs, yet they have
larger AGN contributions. If the evolution scenario of Sanders et al.\
(1988a, 1988b) is to apply to PG~QSOs, the infrared-luminous starburst
in these objects must have subsided from its peak activity during the
ULIRG phase.

Another prediction of this evolutionary scenario is that the AGN
eventually emerges out of its dusty cocoon. The only diagnostic tool
at our disposal to estimate the amount of dust in these systems is the
effective silicate optical depth (Section 6.2). We return to Figure
\ref{fig:weqpah7}$d$, this time considering the AGN contribution
rather than simply the optical spectral types. The results are shown
in Figure \ref{fig:weqpah7_v_siltau_v}$a$ and summarized in Table
\ref{tab:weqpah7_v_tau}.  {\em We find a remarkably strong trend in
  AGN contribution, leading from the lower right through the upper
  region and ending in the lower left (we have labeled these regions
  R1, R2, and R3 for convenience).  All of the objects in R1 are
  starburst-dominated.  In R2, the objects have larger AGN
  contributions, but are still mostly starburst-dominated.  In R3, the
  objects are either AGN dominated or show a balance between starburst
  and AGN. These results are consistent with the evolution scenario if
  the objects on the Spoon et al.\ diagonal branch (regions R1 and R2)
  are in an earlier phase of ULIRG evolution than objects on the left
  tip of the horizontal branch (region R3).  Differences between
  ULIRGs populating R1 and R2 may also be explained in the context of
  the evolution scenario if extinction increases during the
  intermediate stages of merger evolution (from R1 to R2) before dust
  is destroyed or blown away by the AGN (R3). We explore this
  possibility in Section 7.3.}

\subsection{Black Hole Growth Rate}

Here we calculate the Eddington ratio, $\eta$, i.e., the ratio
of AGN bolometric luminosity to the Eddington luminosity, $L$(Edd) $=
3.3 \times 10^4~(M_{\rm BH}/M_\odot)$~$L_\odot$, for each system.  The
results are shown in Tables
$\ref{tab:eddrat_phot_avg}-\ref{tab:eddrat_dyn_avg}$ and Figure
\ref{fig:eddrat_v}. Two methods were used to estimate the black hole
masses in these systems: (1) ``dynamical'' black hole masses based on
the stellar velocity dispersion of the spheroidal component in these
objects from Dasyra et al. (2006a, 2006b, 2007) and the stellar
velocity dispersion $-$ black hole mass relation of Tremaine et
al. (2002), (2) ``photometric'' black hole masses based on
measurements of the H-band luminosity of the spheroidal component in
these systems (free of the central point source) from Veilleux et
al. (2002, 2006, 2009) and the H-band spheroid luminosity $-$ black
hole mass relation of Marconi \& Hunt (2003).

Photometric black hole masses are available for all ULIRGs and PG~QSOs
in the {\em Spitzer} sample, while dynamical estimates are available for
only a third of the sample.  Note also that the dynamical black hole
mass measurements of ULIRGs and PG~QSOs are smaller on average than
the photometric estimates, hence the Eddington ratios derived from the
dynamical black hole masses are larger on average than those based on
the photometric method: for the ULIRGs in our sample, log($\eta$) =
$-1.08 \pm 0.40$ and $-0.35 \pm 0.63$ (Tables
$\ref{tab:eddrat_phot_avg}-\ref{tab:eddrat_dyn_avg}$; in these tables,
the average-of-averages and standard errors are calculated by first
averaging over all methods for individual objects, then averaging over
objects).  A similar discrepancy is found among the PG~QSOs when
comparing the dynamical estimates with those from reverberation
mapping or the virial method. A detailed comparison of the various
black mass estimates in ULIRGs and PG~QSOs is beyond the scope of the
present paper (interested readers should refer to Veilleux et al.\
2009 for a more detailed discussion and a table of the black hole
masses from the various methods).  Suffice it to say that the absolute
values of all Eddington ratios are quite uncertain so the present
discussion focuses on overall {\em relative} trends, which should be
much more robust.

Figure \ref{fig:eddrat_v} shows the distribution of the
photometrically-derived Eddington ratios as a function of spectral
types, infrared luminosities, $f_{25}/f_{60}$ ratios, nuclear
separations, and interaction classes (Table \ref{tab:eddrat_phot_avg}
summarizes the results).  No obvious correlations exist between the
Eddington ratios and any of these parameters. Weak trends may be
present with the morphological quantities: Eddington ratio appears to
be larger at the smallest nuclear separations and latest interaction
classes (see Section 7.3 for description of interaction classes). In both
cases, the addition of PG~QSOs seems to either extend or enhance these
trends. However, a rigorous statistical analysis of these data cannot
confirm the trends involving the ULIRGs. Similarly, the number of
objects with dynamical Eddington ratios (Table
\ref{tab:eddrat_dyn_avg}) is generally too small to allow us to detect
any significant trends involving this quantity.

\subsection{Merger Evolution}

Virtually all 1-Jy ULIRGs and most PG~QSOs show clear signs of strong
tidal interaction/merger. ULIRGs are on-going mergers that sample the
Toomre merger sequence beyond the first peri-passage (Veilleux et al.\
2002, 2006), while many PG~QSOs are advanced mergers where the nuclei
of the progenitor galaxies have apparently coalesced (e.g., Surace et
al. 2001; Guyon et al. 2006; Veilleux et al.\ 2009). It is therefore
natural to ask whether the excitation sequence we see in Figure
\ref{fig:agnfrac_v} in fact corresponds to the final stages of the
evolution sequence first suggested by Sanders et al.\ (1988a,
1988b). In this section we examine this question by using a number of
morphological indicators of merger phase: apparent (projected) nuclear
separation, lengths of tidal tails, compactness of merger remnant, and
strength of tidally-induced morphological anamolies in coalesced
systems. We also use the interaction classes of 1-Jy ULIRGs and
PG~QSOs derived by Veilleux et al.\ (2002, 2006, 2009) and based on
the classification scheme of Surace (1998). This scheme combines all
morphological indicators of merger phase and compares the results with
published numerical simulations of mergers (which we describe in more
detail below).  In brief, Classes I through V correspond to first
approach, first contact, pre-merger [subdivided into $a$ and $b$ for
wide ($>$ 10 kpc) and close ($\le$ 10 kpc) pairs], merger (subdivided
into $a$ and $b$ for diffuse and compact systems), and old merger,
respectively. The results are summarized in Table
\ref{tab:agnfrac_avg} and presented in Figure \ref{fig:agnfrac_v}$d$
and $e$.

A simple comparison between binary- and single-nucleus ULIRGs reveals
a slight difference between their respective AGN fractional
contribution to the bolometric luminosity (32\% for the binaries {\em
versus} 46\% for the singles; Table \ref{tab:agnfrac_avg}). A closer
look at this result indicates that the increase in AGN contribution
generally takes place when the apparent nuclear separation is less
than $\sim$1 kpc (Figure \ref{fig:agnfrac_v}$d$). This trend with
nuclear separation seems driven primarily by the large number of
Seyfert 1 ULIRG among merged systems. Projection effects undoubtedly
add scatter to the data. This is also illustrated in Figure
\ref{fig:agnfrac_v}$e$, where we substituted the more physically
meaningful interaction class for the apparent nuclear separation. By
and large, the results on the ULIRGs are consistent with the evolution
scenario of an increasingly more dominant AGN among late mergers,
although with considerable scatter. Some of this scatter may be due to
multiple episodes of AGN dominance during the merger process.  Most
wide ($NS > $ 6 kpc) binaries do not have confirmed redshifts for both
pair members so the slightly larger AGN contributions among this class
of objects may be due to misidentifications. Our results on the
PG~QSOs bring additional support to the evolution scenario: the high
AGN contribution and unresolved morphologies of the PG~QSOS falls
right along the trends observed among ULIRGs
(Figure \ref{fig:agnfrac_v}$d$ and $e$).

Figure \ref{fig:eddrat_v}$d$ and $e$ suggest that pre-merger ULIRGs
are indeed on average less actively accreting matter onto the black
holes than late mergers, based on their Eddington ratios, with the
PG~QSOs nicely falling along these trends. However, as mentioned in \S
7.2, these trends are statistically not very significant.

In Section 7.1, we found smaller dust obscuration in AGN-dominated systems
and wondered if it was an evolutionary effect. We revisit Figure
\ref{fig:weqpah7}$d$, taking into account the interaction classes and
nuclear separations. {\em In Figure \ref{fig:weqpah7_v_siltau_v}$b$
and $c$, we do find trends moving from regions R1 thru R2 to R3 (see
summary in Table \ref{tab:weqpah7_v_tau}): (a) close pairs or singles
(separation $<$ 1~kpc) occupy 36/47/63\% of the total in R1/R2/R3; (b)
middle interaction stages (IIIa/b) decrease from R1-R3, occupying
68/43/29\% of the total; and (c) late interaction stages (IVa/b)
increase from 26/39/54\% of the total in R1/R2/R3.  These trends with
morphology suggest that the importance of dust extinction generally
peaks during the intermediate stages of merger evolution (IIIb/IVab)
before dust gets destroyed or blown away during the last phase of the
merger.} A similar trend was found by Rossa et al.\ (2007) in the
(smaller) sample of galaxies of the Toomre sequence.

Even in the absence of projection effects and misidentifications,
there are theoretical grounds for significant scatter in the simple
evolutionary picture outlined above.  Numerical simulations have been
used extensively to study the dynamical evolution of merging galaxies
and the associated inflow of gas to the central regions (e.g., Barnes
\& Hernquist 1996; Mihos \& Hernquist 1996; Springel et al. 2005; Cox
et al. 2006; Naab et al. 2006). These simulations capture star
formation and AGN fueling via ``sub-resolution physics'':
phenomenological models which tie the star formation and AGN accretion
-- and their subsequent feedback on the gas -- to the physical
properties of the gas on $\sim$100 pc scales. Because the physical
scales for accretion onto AGN are orders of magnitude smaller, these
simulations can only provide a broad brush picture of the evolution of
activity in merging galaxies. Nonetheless, even with these limitations
they provide a plausible framework for discussing both the trends and
the scatter in this evolutionary picture.

The relative strengths of the first and second inflow phases depend on
a wide variety of factors, which will lead to significant scatter in
the evolutionary paths of mergers.  As the early inflow is moderated
by dynamical instabilities in the host galaxies' disks, this phase is
very sensitive to the intrinsic properties of the host, such as the
presence or absence of central bulges to stabilize the disks (Mihos \&
Hernquist 1996), the disk surface mass density (Mihos et al. 1997), or
the gas fraction of the disk (Springel et al.  2005).  Galaxies more
susceptible to these disk instabilities are more likely to suffer
early inflow and onset of AGN activity, increasing the scatter in AGN
properties in the early interaction stage.  Other factors playing into
the merger evolution include the orbital geometry (e.g., prograde
versus retrograde encounters) and mass ratio of the encounter.
However, simulations show that these factors are secondary to the
structural properties of the galaxies, as long as we consider major
mergers like ULIRGs (e.g., Mihos \& Hernquist 1996; Springel et
al. 2005; Younger et al. 2008).

The strengths of the starburst and AGN activity during different
stages may also couple via the induced activity.  The early fueling of
starbursts and AGN can both deplete and heat the gas, potentially
limiting the ability to form powerful starbursts or AGN late in the
merging process.  Simulations which include both AGN and starburst
heating (e.g., Springel et al. 2005) suggest this is a small effect
early in the encounter when there is ample fuel supply, but that at
later stages the AGN heating is sufficient to cease further starburst
activity (DiMatteo et al. 2005).  The fact that we see examples of
post-mergers with significant starburst activity is somewhat
problematic for these models, and may indicate that the AGN feedback
models may be overly efficient in these simulations.  Weaker feedback
could halt accretion near the AGN without terminating star formation
in the more extended distribution of gas.

The fact that there are multiple inflow epochs along the merging
sequence implies that a scenario in which AGN only turn on at the
final stages of coalescence is oversimplified.  Indeed, depending on
the complex interplay of these factors, multiple bursts of strong
starburst or AGN activity throughout the process are not completely
unexpected.  In a probabilistic sense, the likelihood of strong AGN
fueling is highest as the galaxy nuclei coalesce, due to the rapidly
varying gravitational potential that drives high inflow rates.
However, the {\it potential} for AGN fueling exists throughout the
interaction process.  In a large sample like ours, this trend can be
seen in the data, where later interaction classes and single nucleus
objects do indeed have higher AGN fractions and larger Eddington
ratios on average.  However, the large scatter also indicates that AGN
activity can in some cases dominate the galaxy's radiative energy
output even at intermediate merger stages.

Numerical simulations plausibly show that the evolution of starburst
and AGN activity among mergers can vary significantly from merger to
merger, due to variations in the global properties of the encounter
and the progenitor galaxies.  However, there may be scatter in
evolutionary paths that is unresolved by the simulations.  Ultimately
(and unfortunately) the detailed predictions for star formation and
AGN accretion in the models depend critically on the sub-resolution
physics.  Varying the hydrodynamical equation of state or the
prescriptions for star formation or AGN activity can significantly
change the detailed results (e.g., Barnes 2004; Cox et al. 2006).
Furthermore, the sub-resolution density structure of the gas is
critical for driving continued inflow from the $\sim$100 pc resolution
limit of the models down to the accretion scale of the central AGN.
Given these uncertainties and limitations inherent to the simulations,
it is likely that the simulations merely give a time averaged
expectation for AGN and starburst activity, averaged over the
dynamical timescale for the inner few hundred pc.  The instantaneous
rate of activity (as measured by our observational data set) may show
significant time variation -- not seen in simulations -- due to
stochastic, small scale physical processes.  This may also account for
a significant amount of the scatter seen in the evolutionary trends
shown in Figures 36 and 38. This random component of accretion is
known to be important among local AGN (e.g., Davies et al.\ 2007).

{\em In summary, we see trends in AGN fraction, nuclear
  obscuration, and possibly Eddington ratio as a function of
  interaction class and nuclear separation, but with considerable
  scatter along the merger sequence.  As shown in Figure
  \ref{fig:weqpah7_v_siltau_v}$b$ and $c$, roughly half of fully
  merged ULIRGs have not (yet) succeeded in producing AGN-dominated
  systems and some pre-merger ULIRGs are already AGN-dominated.  Some
  part of this scatter may be attributed to projection effects and
  misidentifications (that is, observational effects).  However, an
  equal or greater portion of the scatter is probably due to the
  physics identified by numerical simulations: the varying initial
  conditions among interactions and the fact that starburst and AGN
  activity can peak locally in intensity prior to final coalesence.
  Further scatter may arise due to small-scale stochastic processes
  that are presently unresolved by these simulations.

  A revision of the evolution scenario of Sanders et al.\ (1988a,
  1988b) is needed to explain all of these results.  The ``softer''
  version we propose requires the presence of multiple evolutionary
  paths in the phase space of AGN contribution, Eddington ratio, and
  dust obscuration {\it versus} merger phase.  For AGN contribution
  and Eddington ratio, this almost certainly includes paths that are
  not monotonically increasing with time.}

\section{Summary}

We have carried out a detailed {\em Spitzer} IRS study of the
MIR continuum, absorption, and emission line properties of a
carefully selected sample of 74 ULIRGs and 34 PG~QSOs within $z \sim
0.3$. For the first time in ULIRGs, the continuum and dust features
were modeled using a combination of PAH templates, blackbodies
punctuated by deep extinction and absorption features, and silicate
emission features, when necessary. The main observational results are
the followings:

\begin{enumerate}

\item We find that the $f_{30}/f_{15}$ and (PAH-free) MIR/FIR flux
  ratios are powerful continuum diagnostics of AGN activity among
  ULIRGs and QSOs.

\item We confirm the broad range of silicate obscuration among ULIRGs,
  with the optically classified Seyfert ULIRGs being less obscured on
  average than the HII-like and LINER ULIRGs. The loose correlations
  seen between silicate, H$_2$O ice + hydrocarbons, C$_2$H$_2$, and
  HCN absorption features imply significant variations in composition
  of the dense absorbing material from one ULIRG to the next. The
  HCO$^+$ 12.1 $\mu$m and HNC 21.7 $\mu$m features were not detected
  in any individual object or the average spectrum of the more
  obscured ULIRGs.

\item The average PAH-to-FIR flux ratio of ULIRGs is remarkably
  similar to that of PG~QSOs.  No obvious trend is seen with optical
  spectral type, but both $L$(PAH)/$L$(FIR) and the 7.7\micron\ PAH
  equivalent width decrease with increasing extinction in HII/LINER
  ULIRGs.  We confirm the strong correlation of EW(7.7\micron\ PAH)
  with optical spectral types and find a similarly strong correlation
  with the $f_{30}/f_{15}$ ratio.

\item Our analysis of the fine structure lines in ULIRGs and QSOs
  reveals a continuous excitation sequence with the cool ($f_{25
    \mu{\rm m}}/f_{60 \mu{\rm m}} \la $ 0.1) optically classified
  HII-like and LINER ULIRGs at the low-excitation end of the sequence,
  the PG~QSOs at the high-excitation end, and the warm optically
  classified Seyfert ULIRGs in between.

\item Warm H$_2$ masses range from $\sim$ 0.5 to 20 $\times$ 10$^8$
  M$_\odot$ with an average (median) of $\sim$ 3.8 (3.3) and 3.6 (3.2)
  $\times$ 10$^8$ M$_\odot$ for the ULIRGs and QSOs,
  respectively. These masses are typically a few percent of the cold
  gas mass derived from $^{12}$CO observations. The
  temperature-sensitive H$_2$ S(2)/S(1), S(3)/S(1), and S(3)/S(2) flux
  ratios suggest possible heating by the AGN in Seyfert ULIRGs and
  PG~QSOs and shock excitation in LINER ULIRGs. However, dust
  extinction and/or variations in the ortho-to-para ratio make the
  results inconclusive.

\item The average MIR spectrum of PAH-dominated ULIRGs suggests
  supersolar neon abundance while optical spectra indicate roughly
  solar oxygen abundance. Uncertainties on the exact value of the
  solar Ne/O ratio may (partly) erase this discrepancy. However, if
  confirmed, this discrepancy may imply that these two methods trace
  different layers or star-forming regions of the ULIRGs, different in
  both extinction and abundance.  No trend is seen in the sample
  galaxies between optically-derived metallicity and emission-line,
  absorption-line, and continuum properties. This result is not
  surprising given the relatively narrow range of optical metallicity
  (0.6 -- 2.4 $\times$ solar) covered by our sample and the small
  number (13) of objects for which we have reliable metallicity
  measurements.

\item The contribution of an AGN to the bolometric luminosity in these
  systems is quantified using six different methods based on (1) the
  [O~IV] 25.9 $\mu$m/[Ne~II] 12.8 $\mu$m ratio, (2) the [Ne~V] 14.3
  $\mu$m/[Ne~II] 12.8 $\mu$m ratio, (3) the equivalent width of PAH
  7.7 $\mu$m, (4) the PAH (5.9 - 6.8 $\mu$m) to continuum (5.1 -- 6.8
  $\mu$m) flux ratio combined with the continuum (14 -- 15 $\mu$m) /
  (5.1 -- 5.8 $\mu$m) flux ratio, (5) the MIR blackbody to FIR flux
  ratio, and (6) the $f_{30}/f_{15}$ continuum flux ratio.  Good
  agreement to within $\sim$ $\pm$10$-$15\% on average is seen amongst
  the various methods.  This agreement rules out the possibility that
  a MIR-buried but FIR-bright AGN is present in many of these objects.

\end{enumerate}

\noindent{\em From these results we draw the following three main
  conclusions: }

\begin{enumerate}

\item {\em The average AGN contribution in ULIRGs is $\sim$ 35$-$40\%,
    in agreement with previous {\em ISO} studies.  Strong correlations
    exist between AGN contributions, optical spectral types, and
    $f_{25}/f_{60}$ ratios.  The AGN contributions range from
    $\sim$$15-35$\% among cool HII and LINER ULIRGs to $\sim$50\% and
    75\% among warm Seyfert 2 and Seyfert 1 ULIRGs, respectively.  The
    PG~QSOs fall along the extrapolation of these trends, with AGN
    contributions typically larger than $\sim$ 80\%.  The largest AGN
    contributions are also observed at the smallest nuclear
    separations and latest interaction classes.}

\item {\em All ULIRGs in our sample fall in three distinct AGN
    classes: (1) Objects with small extinctions and large PAH
    equivalent widths are highly starburst-dominated, (2) Systems with
    large extinctions and modest PAH equivalent widths have larger AGN
    contributions, but are still mostly starburst-dominated, (3)
    ULIRGs with both small extinctions and PAH equivalent widths are
    either AGN dominated or show a balance between starburst and
    AGN. The AGN contributions in highly obscured, class 2 ULIRGs are
    necessarily more uncertain than in the other objects, and we
    cannot formally rule out the possibility that these objects
    represent a physically distinct type of ULIRGs. However, a weak
    trend is seen toward smaller nuclear separations and later merger
    stages along the sequence (1) -- (2) -- (3). These results suggest
    that dust extinction generally peaks during the intermediate
    stages of merger evolution, before the dust gets destroyed or
    blown away during the late-merger phase.}

\item {\em A ``softer'' version of the standard ULIRG -- QSO evolution
  scenario is needed to explain the scatter in trends of AGN
  contribution, Eddington ratio, and dust obscuration with merger
  stage.  With our large sample size we are able to discern the
  average trends discussed above.  However, roughly half of fully
  merged ULIRGs have not (yet) succeeded in producing AGN-dominated
  systems or blown away their obscuring dust screen, and some
  pre-merger ULIRGs are already AGN-dominated.  Our revised
  evolutionary picture permits multiple paths that are not necessarily
  monotonic in quantities like AGN contribution and Eddington ratio.
  Such a scenario is consistent with numerical simulations of
  merger-induced starburst and AGN activity.  These simulations show
  the highest inflow rates when the galaxies coalesce, but allow for
  significant episodes of inflow and nuclear activity throughout a
  major galaxy merger.  The strength and timing of these episodes will
  vary depending on the initial conditions of the interaction, and
  quite possibly on stochastic processes presently unresolved by
  simulations.}

\end{enumerate}

Finally, we point out that the continuum-based methods used here to
quantify the power source in these local ULIRGs and QSOs are ideally
suited to the study of faint high-$z$ systems.  In particular, the
calibration of the $f_{30}/f_{15}$ method we derived from our local
sample may be used in the future to quantify the AGN contribution in
U/LIRGs at $z \sim 1 - 1.5$ using the MIPS 70/24 $\mu$m flux ratio
(e.g., Sajina et al. 2007) and at higher redshifts with {\em
  Herschel}.

\acknowledgements This work is based on observations carried out with
the {\em Spitzer Space Telescope}, which is operated by the Jet
Propulsion Laboratory, California Institute of Technology, under NASA
contact 1407.  Support for this work was provided by NASA through
contracts 1263752 (S.V., D.S.N.R., and D.C.K.), 101185-07.E.7991.01.08
(J.M., S.L.), 1264025 (J.C.M.), and 1264791 (D.B.S. A.S., R.D.J.,
J.E.B.)  issued by JPL/Caltech. S. V. acknowledges support from a
Senior Award from the Alexander von Humboldt Foundation and thanks the
host institution, MPE Garching, where some of this work was performed.
We thank the anonymous referee for a critical reading of the
manuscript and acknowledge useful conversations with G. Share and
J. Graci\'a Carpio.  This work has made use of NASA's Astrophysics
Data System Abstract Service and the NASA/IPAC Extragalactic Database
(NED), which is operated by the Jet Propulsion Laboratory, California
Institute of Technology, under contract with the National Aeronautics
and Space Administration.  The SMART software package was developed by
the IRS team at Cornell University and is available through the {\it
  Spitzer} Science Center at Caltech.

\centerline{{\bf APPENDIX A.}~~~SIX METHODS TO DERIVE AGN CONTRIBUTIONS} 

The six methods used to calculate the contribution of the AGN to the
bolometric luminosity of the ULIRG or PG~QSO are described briefly in
the notes to Table \ref{tab:agnfrac_avg}. Here we first discuss the
assumptions that apply to all of the methods, and then describe each
method individually with their respective strengths and weaknesses.

For each of these methods, we compare the observed quantities derived
from our data with pure-AGN and pure-starburst zero points [for method
\#3, we also compare the data with a pure-PDR (photodissociation
region) zero point].  The pure-AGN zero point is set to the average
value of the FIR-undetected (unobscured) PG~QSOs in our sample, while
the pure-starburst zero point is set to the average value of the most
starburst-like ULIRGs in our sample.  The results of these comparisons
provide AGN fractional contributions to the [Ne~II], 8 $\mu$m
continuum, 5.3 -- 5.8 $\mu$m continuum, FIR, and 15-$\mu$m
luminosities, respectively. Next, we apply correction factors to
transform these various luminosities into bolometric
luminosities. These correction factors are listed in Table
\ref{tab:agnfrac_bolcor} for starburst (average of HII-like ULIRGs)
and AGN (average of FIR-undetected PG~QSOs).  The method used to
calculate these correction factors is described in the note to that
table.  For this and the other methods, we assumed $L(bol) =
1.15~L$(IR) for all ULIRGs (Kim \& Sanders 1998) and $L(bol) =
7~L$(5100 \AA)$+L$(IR) for all PG QSOs.  The latter includes both
contributions from the 'intrinsic' AGN luminosity (Paper II), as well
as AGN and starburst luminosity reprocessed by dust.

We make no attempts to correct our data for extinction. The impact of
dust extinction on the various features (continuum, fine structure
lines, PAH features) depends greatly on the distribution of the dust
relative to the sources of emission (e.g., dust screen {\em versus}
mixed distribution) and we have very little information on this issue
(except for the fact that it is almost certainly lower for PAHs than
for the continuum and that foreground dust screens fit the continuum
better than mixed dust screens; Sections 5.2 and 6.3). So, rather than
making {\em ad hoc} assumptions on the dust geometry and running the
risk of producing unphysical results (e.g., absurdly high PAH or MIR
luminosities), we did not apply any dust extinction correction to the
measured quantities. A comparison of our (extincted) measurements with
those of (unobscured) FIR-undetected PG~QSOs may therefore
underestimate the AGN contribution in our objects.

Dust extinction is not an issue when comparing our data with the
pure-starburst zero point since this latter is based on the observed
values of the most starburst-like ULIRGs. Starburst galaxies of lower
infrared luminosities (lower star formation rates and/or extinction)
were not used to set this zero point or for the starburst bolometric
corrections. Indeed, as pointed out by a number of studies, the
unusual conditions (higher density, more intense radiation field) in
ULIRGs produce systematic shifts in the relations between FIR, [Ne~II]
and PAH emission in such a way that we cannot use the relations
derived from normal starburst galaxies to quantify ULIRGs. The
competition of the dust with the gas for absorption of the ionizing
photons becomes increasingly more effective as the density of the
star-forming regions increases, so the [Ne~II]/FIR ratio is reduced in
ULIRGs (e.g., Rigby \& Rieke 2004; Dopita et al.\ 2006; Calzetti et
al.\ 2007). The more intense radiation field in ULIRGs induces a
greater ionization or dehydrogenation of the PAHs, so the PAH/FIR
ratio is also reduced (e.g., Tielens et al. 1999; Helou et
al. 2001). Pure-starburst zero points and bolometric corrections based
on starburst galaxies of lower infrared luminosities would therefore
underestimate the true contribution of star formation in
ULIRGs.$\footnote{Another concern about the use of published
  [Ne~II]/FIR and PAH/FIR ratios of normal starburst galaxies is
  aperture effects: the [Ne~II] and PAH features are derived from {\em
    Spitzer} spectra with entrance apertures that are much smaller
  than {\em IRAS}, from which the FIR fluxes are derived. This effect
  will underestimate the actual [Ne~II]/FIR and PAH/FIR ratios of
  normal starburst galaxies. In contrast, these aperture effects do
  not affect the ratios of ULIRGs significantly because most of the
  FIR, [Ne~II], and PAH emission is produced within the central kpc of
  these objects, so is contained well within the {\em Spitzer}
  apertures.}$

{\bf Method \#1 ([O~IV]/[Ne~II] ratio) and Method \#2 ([Ne~V]/[Ne~II]
  ratio).} The large number of ULIRGs without firm [O~IV] and/or
[Ne~V] detection makes the results based on the individual
[O~IV]/[Ne~II] and [Ne~V]/[Ne~II] ratios subject to potentially large
systematic uncertainties. We treated upper limits as detections in our
analysis so the AGN contribution as derived with this method should be
considered upper limits as well. To further constrain these numbers,
we measured the line ratios from average spectra, produced by
normalizing individual spectra in each category listed in Table
$\ref{tab:agnfrac_avg}$ (e.g, spectral type, $f_{25}/f_{60}$, infrared
luminosity, morphological classes) to the same [Ne~II] or FIR
flux. The resulting average AGN fractional contributions, while
affected by systematic uncertainties arising during the averaging
procedure, are roughly consistent with the numbers in Table
$\ref{tab:agnfrac_avg}$. The average spectra suggest that the actual
average AGN contributions are generally not lower by more than about
5$-$10\% from the limits themselves, although for a few categories
(e.g., H~II regions), the actual values may be lower by a larger
factor (10$-$20\%).  The pure-AGN zero point of the [O~IV]/[Ne~II]
method is set at log([O~IV]/[Ne~II]) = 0.6, corresponding to the
average value of the FIR-undetected PG~QSOs in our sample, while the
pure-starburst ratio [O~IV]/[Ne~II] is set at zero (as described in
Table \ref{tab:agnfrac_zp}, the non-zero ratios actually measured in
starbursts are negligible for the purposes of computing AGN
contributions). The assumption here is that [Ne~II] emission from a
pure starburst lowers [O~IV]/[Ne~II] below the pure AGN value, as
shown by the diagonal line in Figure \ref{fig:o4ne2_v_ne5ne2}.  This
figure also shows the calibration based on the [Ne~V]/[Ne~II] ratio.
The pure-AGN log([Ne~V]/[Ne~II]) = 0.10 and is based once again on the
average value of FIR-undetected PG~QSOs. The correction factors used
to transform [Ne~II] luminosities into bolometric luminosities are
listed in Table \ref{tab:agnfrac_bolcor}. Note that Armus et al.\
(2007) did not apply this last correction to their numbers so their
AGN fractional contributions relate to the [Ne~II] luminosities, not
the bolometric luminosities, and are considerably smaller than the
numbers presented here.

{\bf Method \#3 (PAH 7.7 $\mu$m equivalent width).} To facilitate
comparisons with most of the published {\em ISO} results (e.g., Genzel
et al.\ 1998; Lutz et al.\ 1999; Rigopoulou et al. 1999; Tran et al.\
2001), we used the PAH 7.7 $\mu$m equivalent widths to quantify the
role of AGN in the 1-Jy ULIRGs. Recall that our fits allow only a
small range ($\sim$ 0.13 dex) in PAH 6.2/7.7 and 11.3/7.7 ratios so
the conclusions based on the 7.7 $\mu$m feature also apply to first
order to the other PAH features (see description of Method \#4
below). The pure-starburst zero point of this method, log[EW(PAH 7.7
$\mu$m)] = 0.75, is near the maximum value observed in our sample. The
AGN will reduce this quantity by contributing to the continuum
emission at this wavelength and destroying the PAH molecules (e.g.,
Voit 1992).  We assume EW(PAH 7.7 $\mu$m) = 0 for a pure AGN and PAH
destruction due to AGN radiation that is proportional to the AGN
fractional contribution. The results do not depend sensitively on this
last assumption. The correction factors used to transform 8 $\mu$m
continuum luminosities into bolometric luminosities are listed in
Table \ref{tab:agnfrac_bolcor}.

{\bf Method \#4 (modified Laurent et al.\ method).} This method is
inspired by Laurent et al.\ (2000), but uses the modifications of
Armus et al.\ (2007) to avoid contamination by PAHs in the continuum
fluxes (5.3 -- 5.8 $\mu$m instead of 5.1 -- 6.8 $\mu$m).  Figure
\ref{fig:laurent} shows the results for our sample. The zero points
for the pure starburst and PDR are from Armus et al.\ (2007) and the
zero point for the pure AGN corresponds to the average value for the
FIR-undetected PG~QSOs to reduce possible starburst contributions to
the continuum emission (Paper II).  We have moved the Armus et
al.\ pure PDR point to the right by 0.3~dex to encompass the group of
points that would otherwise fall outside the mixing region.  Note that
there is also uncertainty in the ``pure AGN'' point, since this is an
average spectrum and the PAH luminosity is only an upper limit.  The
correction factors to transform 5.3 -- 5.8 $\mu$m continuum
luminosities into bolometric luminosities are listed in Table
\ref{tab:agnfrac_bolcor}. 

\begin{figure*}[ht]
\epsscale{1.0}
\plotone{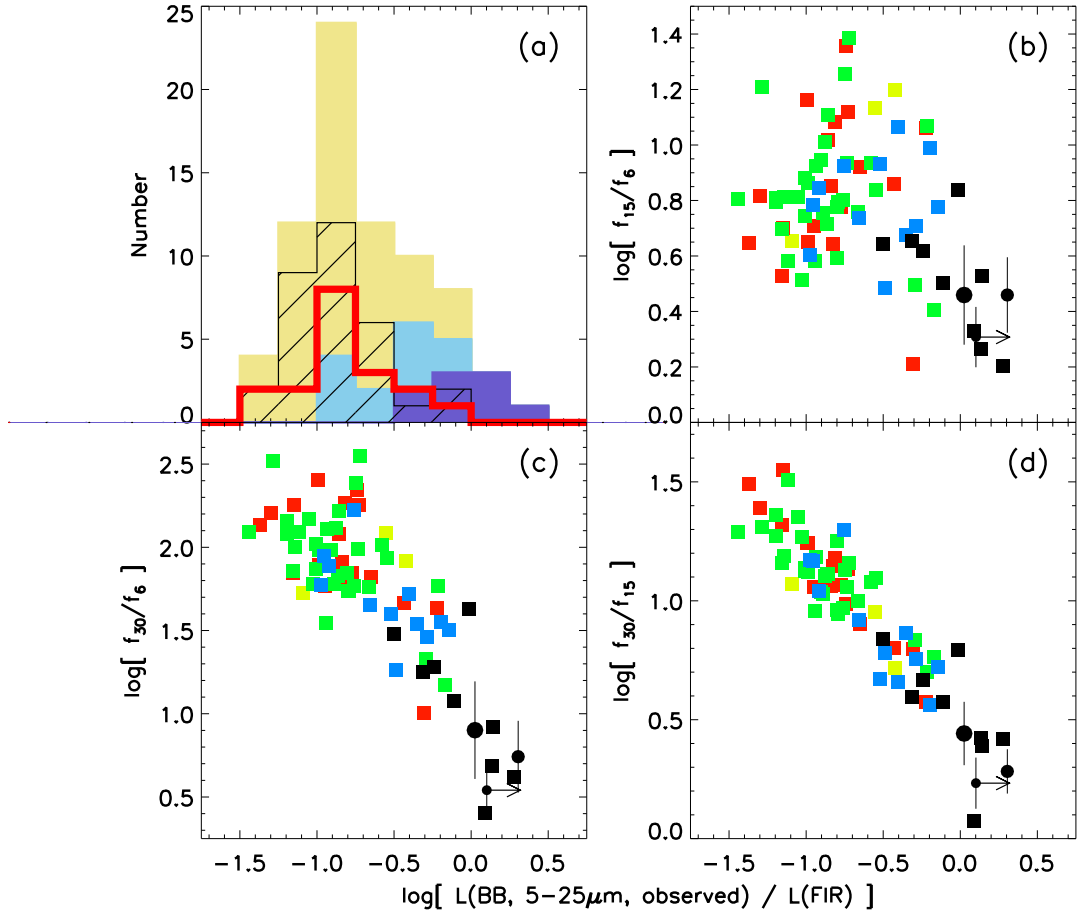}
\caption{ ($a$) Distributions of the ratios of 5 -- 25 $\mu$m
  blackbody (i.e., excluding PAH emission) luminosities to the
  FIR luminosities for 1-Jy ULIRGs of various optical
  spectral types (see the legend in Figure \ref{fig:bbtemp}). This ratio
  is distinctly larger among Seyfert ULIRGs than among HII-like /
  LINER ULIRGs. ($b$, $c$, $d$) Ratios of 5 -- 25 $\mu$m blackbody
  (i.e., excluding PAH emission) luminosities to the
  FIR luminosities for 1-Jy ULIRGs of various optical
  spectral types {\em vs. \/} 15-to-6 $\mu$m flux ratios ($b$),
  30-to-6 $\mu$m flux ratios ($c$), and 30-to-15 $\mu$m flux ratios
  ($d$). The meaning of the symbols is the same as
  Figure \ref{fig:fluxcomp}. In addition, the small, medium-size, and
  large black circles correspond to the FIR-undetected, FIR-faint, and
  FIR-bright PG~QSOs, respectively, as defined in Paper II.  The
  strongest correlation is seen in ($d$).}
\label{fig:mirfir_v}
\end{figure*}

\begin{figure*}[ht]
\epsscale{0.5}
\plotone{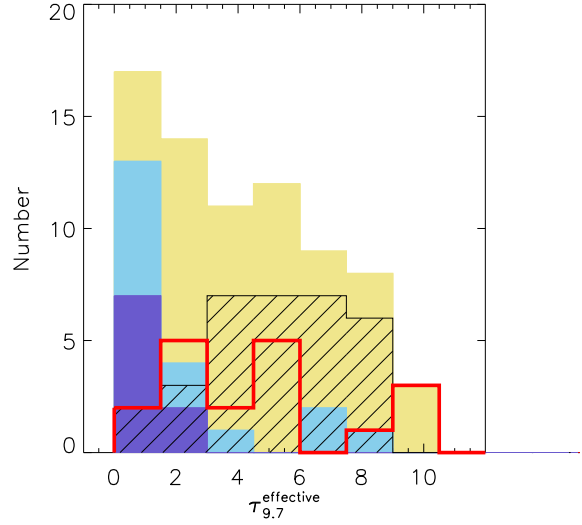}
\caption{ Distributions of the effective 9.7 $\mu$m silicate optical
  depth for 1-Jy ULIRGs of various optical spectral types (see the
  legend in Figure \ref{fig:bbtemp}). The four Seyfert 1 ULIRGs with
  silicate in emission are not shown in this figure. Optically
  classified Seyferts generally have smaller optical depths than
  HII-like/LINER ULIRGs.  }
\label{fig:tauhist}
\end{figure*}

\begin{figure*}[ht]
\epsscale{1.0}
\plotone{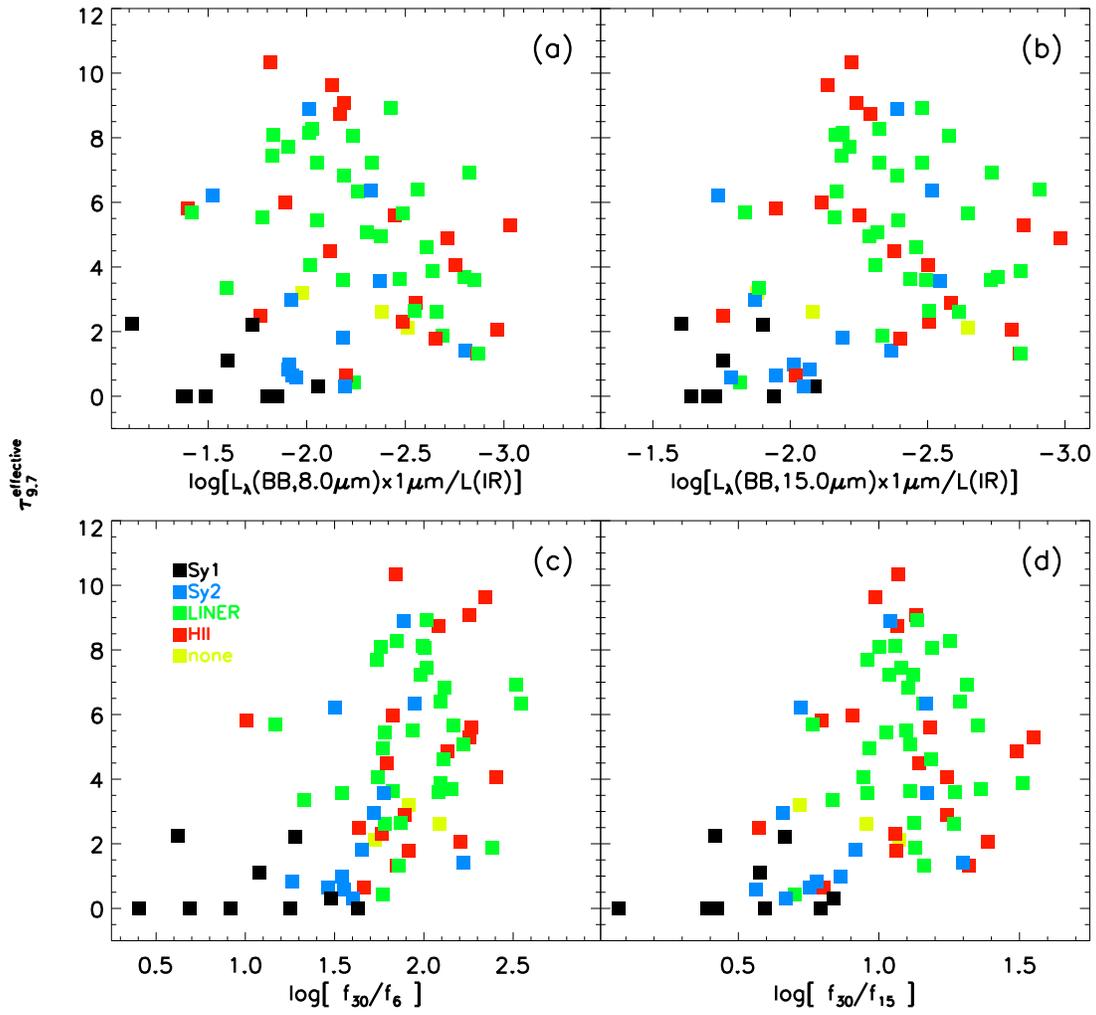}
\caption{ Effective silicate optical depths {\em vs.\/} ($a$) ratios
  of monochromatic blackbody 8 $\mu$m luminosities to total infrared
  luminosities, ($b$) ratios of monochromatic blackbody 15 $\mu$m
  luminosities to total infrared luminosities, ($c$) 30-to-6 $\mu$m
  flux ratios, and ($d$) 30-to-15 $\mu$m flux ratios for the 1-Jy
  ULIRGs of various optical spectral types. The meaning of the symbols
  is the same as in Figure \ref{fig:fluxcomp}. Seyfert ULIRGs generally
  populate the lower left portion of these diagrams. }
\label{fig:tau_v}
\end{figure*}

\begin{figure*}[ht]
\epsscale{1.0}
\plotone{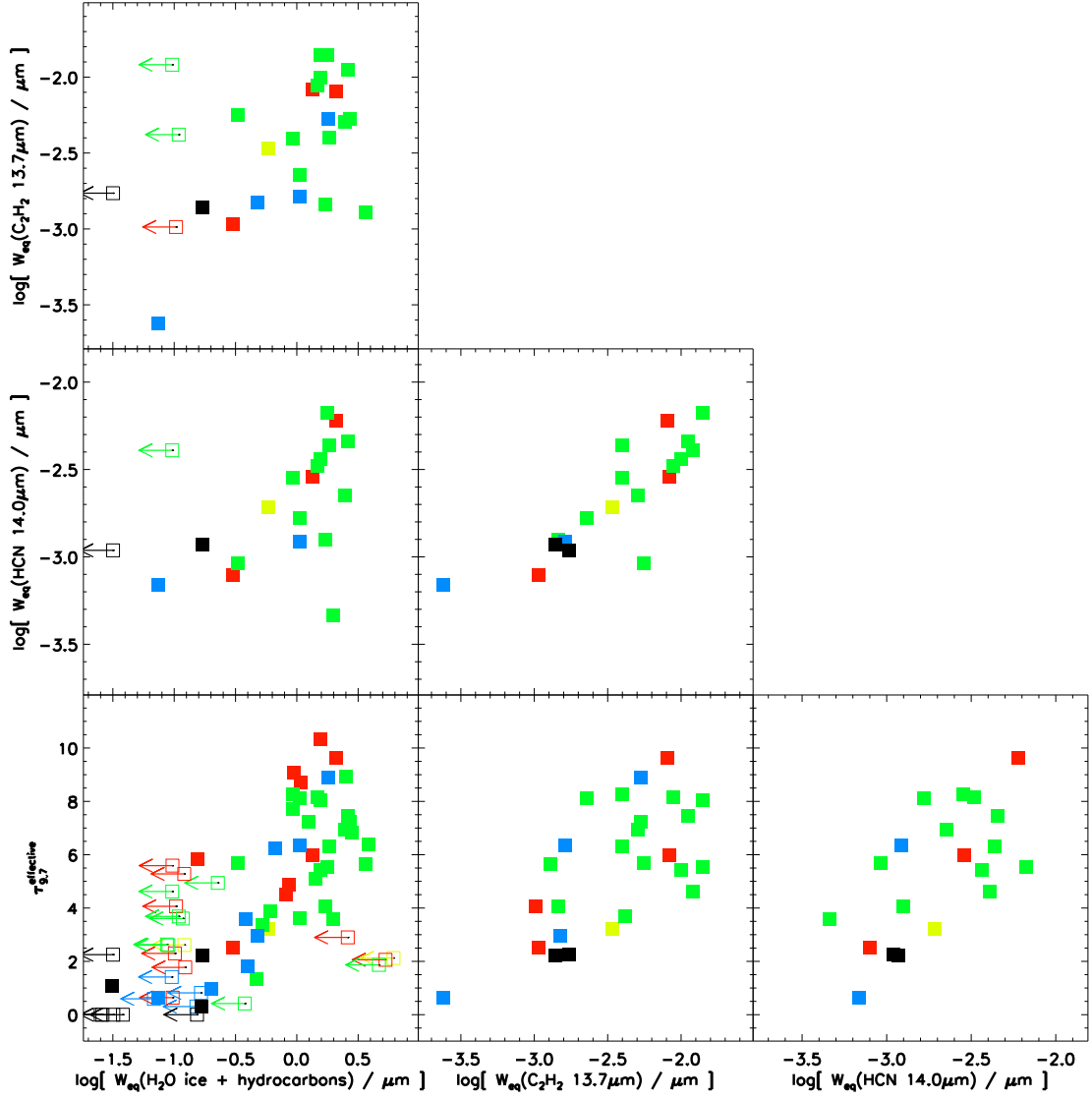}
\caption{ Comparisons of absorption features: Effective optical depths
  of 9.7 $\mu$m silicate, equivalent widths of 14 $\mu$m HCN;
  equivalent widths of H$_2$O ice + hydrocarbons; equivalent widths of
  13.7 $\mu$m C$_2$H$_2$, and equivalent widths of 14 $\mu$m HCN. The
  meaning of the symbols is the same as in
  Figure \ref{fig:fluxcomp}. The strengths of these features are only
  loosely correlated, implying significant variations in the
  composition of the dense absorbing material from one ULIRG to the
  next. }
\label{fig:abs}
\end{figure*}

\begin{figure*}[ht]
\epsscale{0.95}
\plotone{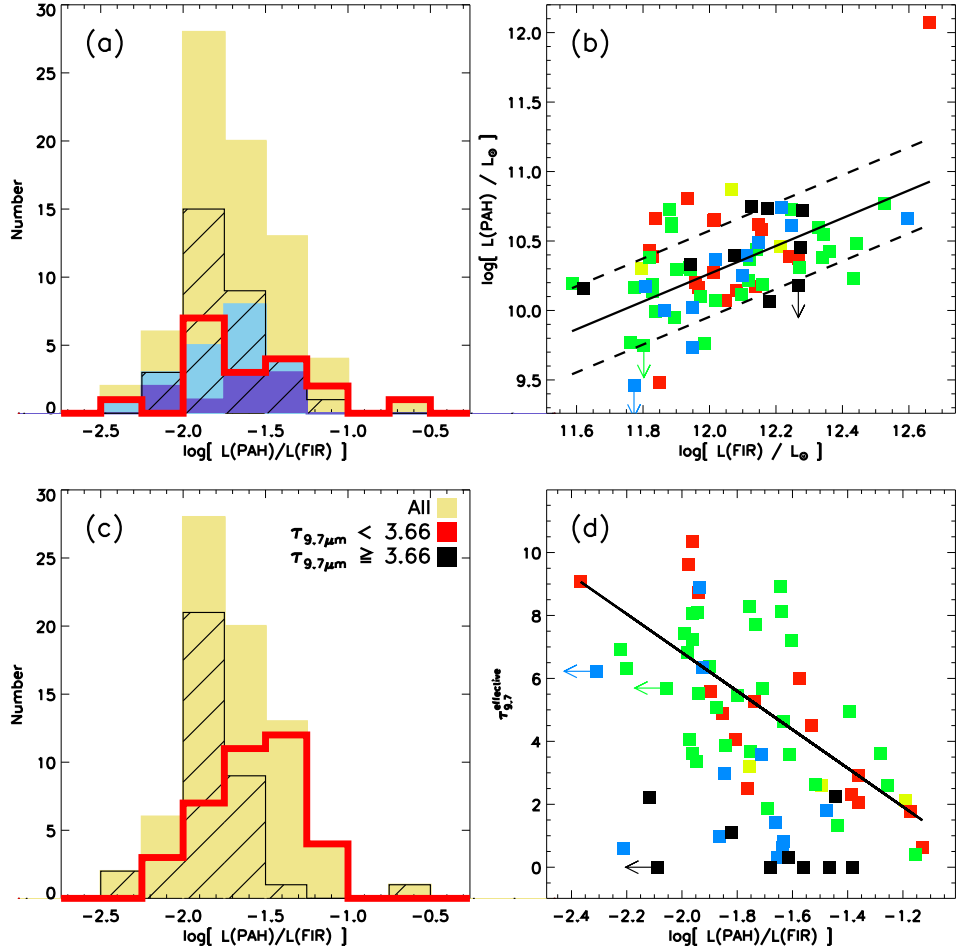}
\caption{ ($a$),($c$) Distributions of the ratio of the total PAH
  luminosities to FIR luminosities of the 1-Jy ULIRGS
  according to optical spectral types (see the legend in
  Figure \ref{fig:bbtemp}) and effective silicate optical depth
  (relative to the median value).  There is a tight distribution
  (standard deviation 0.3 dex), as also illustrated in panel ($b$),
  where the PAH and FIR luminosities are plotted against each other.
  The meaning of the symbols is the same as in
  Figure \ref{fig:fluxcomp}, the solid line is the mean ratio assuming a
  slope of unity, and the dashed lines show the standard deviation
  from the mean.  (The HII-like ULIRG that lies well above the best
  linear PAH-FIR fit is F00397$-$1312; the PAHs in this system are
  corrected for significant extinction.) This ratio is very similar to
  that of the PG~QSOs (Paper I).  ($d$) $L$(PAH)/$L$(FIR)
  vs. effective silicate optical depth.  The line is a fit to
  HII/LINER ULIRGs.  (F00397$-$1312, with the highest observed
  $L$(PAH)/$L$(FIR) value, is not visible in this plot).  An
  anticorrelation exists between PAH/FIR ratio and silicate optical
  depth.}
\label{fig:pahfir}
\end{figure*}

\begin{figure*}[ht]
\epsscale{1.0}
\plotone{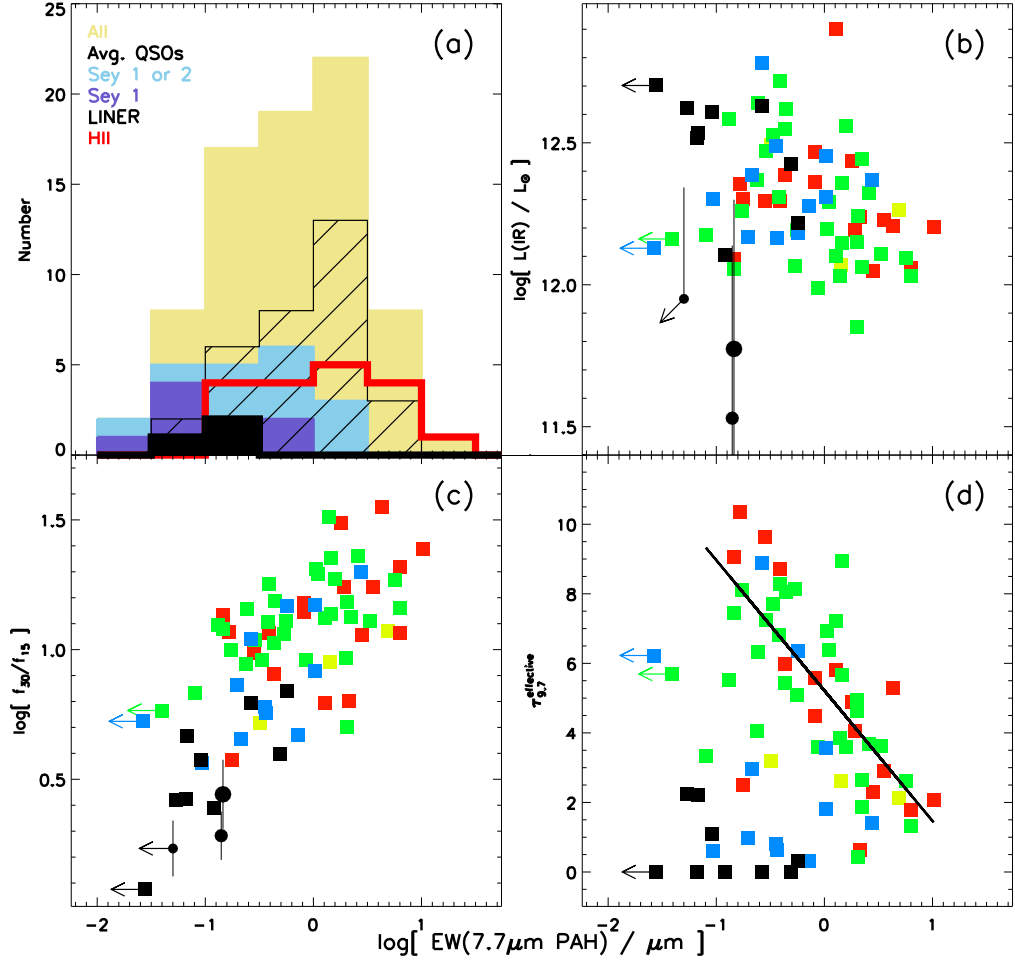}
\caption{($a$) Distribution of equivalent widths of the 7.7 $\mu$m PAH
  feature for 1-Jy ULIRGs and average FIR-bright, FIR-faint, and
  FIR-undetected PG~QSOs (see the legend in Figure \ref{fig:bbtemp}).
  Note the similarity between Seyfert 1 ULIRGs and QSOs, and the
  higher EW(PAH) values in HII galaxies and LINERs than in Seyferts.
  In the other panels, the 7.7 $\mu$m PAH equivalent width is plotted
  as a function of ($b$) the infrared luminosity, ($c$) the 30-to-15
  $\mu$m flux ratio, and ($d$) the silicate optical depth.  The
  meaning of the square (circle) symbols is the same as in
  Figure \ref{fig:fluxcomp} (\ref{fig:mirfir_v}). These figures show
  that EW(PAH) is weakly correlated with infrared luminosity, strongly
  correlated with $f_{30}/f_{15}$ (more so than with optical spectral
  type), and strongly correlated with silicate optical depth for
  HII/LINER galaxies.  The line in ($d$) is a fit to the HII/LINER
  detections.  Seyferts deviate from the HII/LINER correlation and
  populate the lower left-hand corner of the plot.}
\label{fig:weqpah7}
\end{figure*}

\begin{figure*}[ht]
\epsscale{1.1}
\plotone{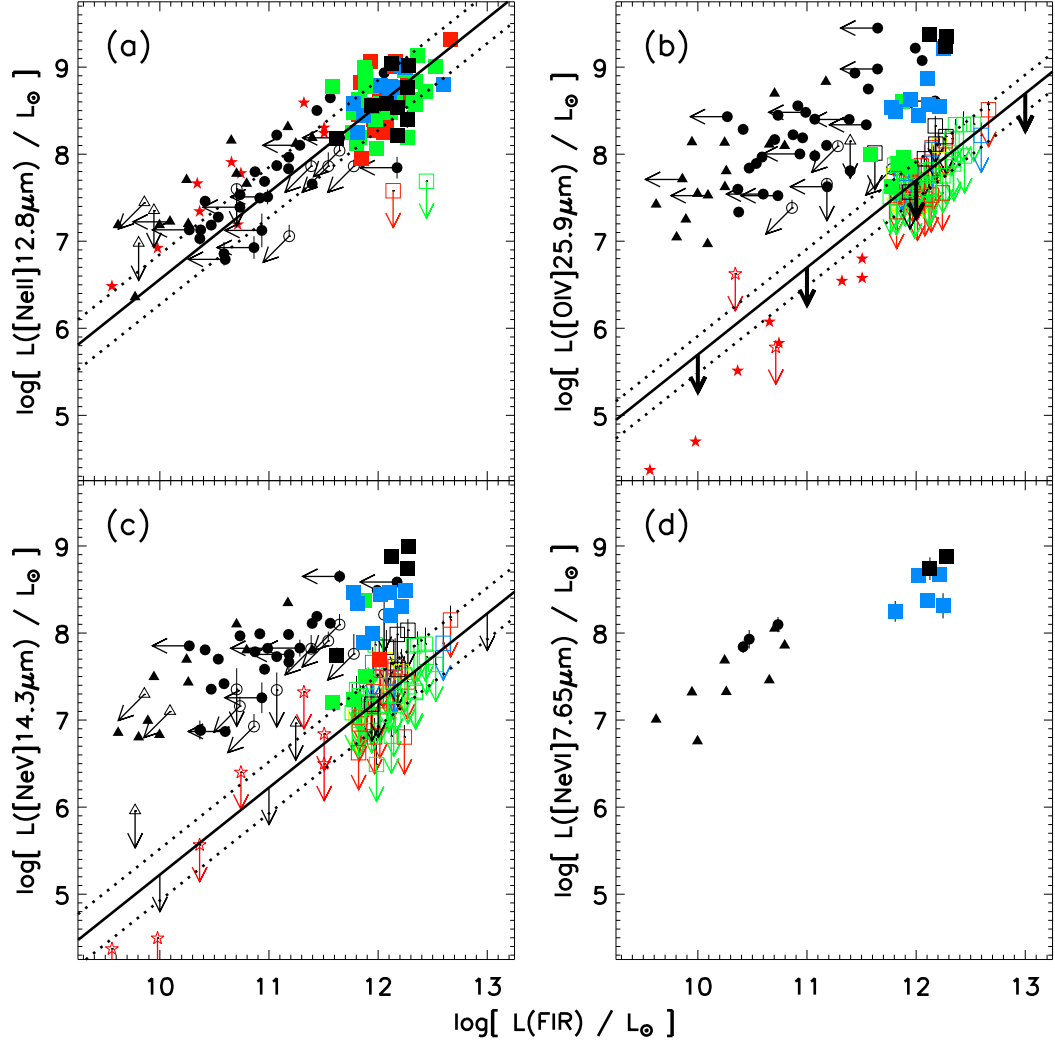}
\caption{ ($a$) [Ne II] 12.8 $\mu$m luminosity, ($b$) [O~IV] 25.9
  $\mu$m luminosity, ($c$) [Ne~V] 14.3 $\mu$m luminosity, and ($d$)
  [Ne~VI] 7.65 $\mu$m luminosity {\em vs.\/} FIR luminosity
  of the 1-Jy ULIRGs and PG~QSOs in the {\em Spitzer} sample as well as
  some optically-selected starbursts and Seyfert 2 galaxies observed
  with {\em ISO}.  The meaning of the square and circle (other) symbols is
  the same as in Figure \ref{fig:fluxcomp} (\ref{fig:mircolor}). The
  solid (dotted) line represents the mean (standard deviation) of the
  ratio $L$(emission line)$/L$(FIR) for HII ULIRGs.  All ULIRGs lie
  close to the fitted [Ne~II]-to-FIR ratio of $10^{-3.4}$, but QSOs
  and Seyfert ULIRGs are increasingly above the values of the [O~IV]-
  and [Ne~V]-to-FIR luminosity ratios of HII ULIRGs. This result is
  readily explained if the AGN contributes increasingly to the
  fine-structure line emission at higher ionization levels.}
\label{fig:linelum_v_fir}
\end{figure*}

\begin{figure*}[ht]
\epsscale{1.1}
\plotone{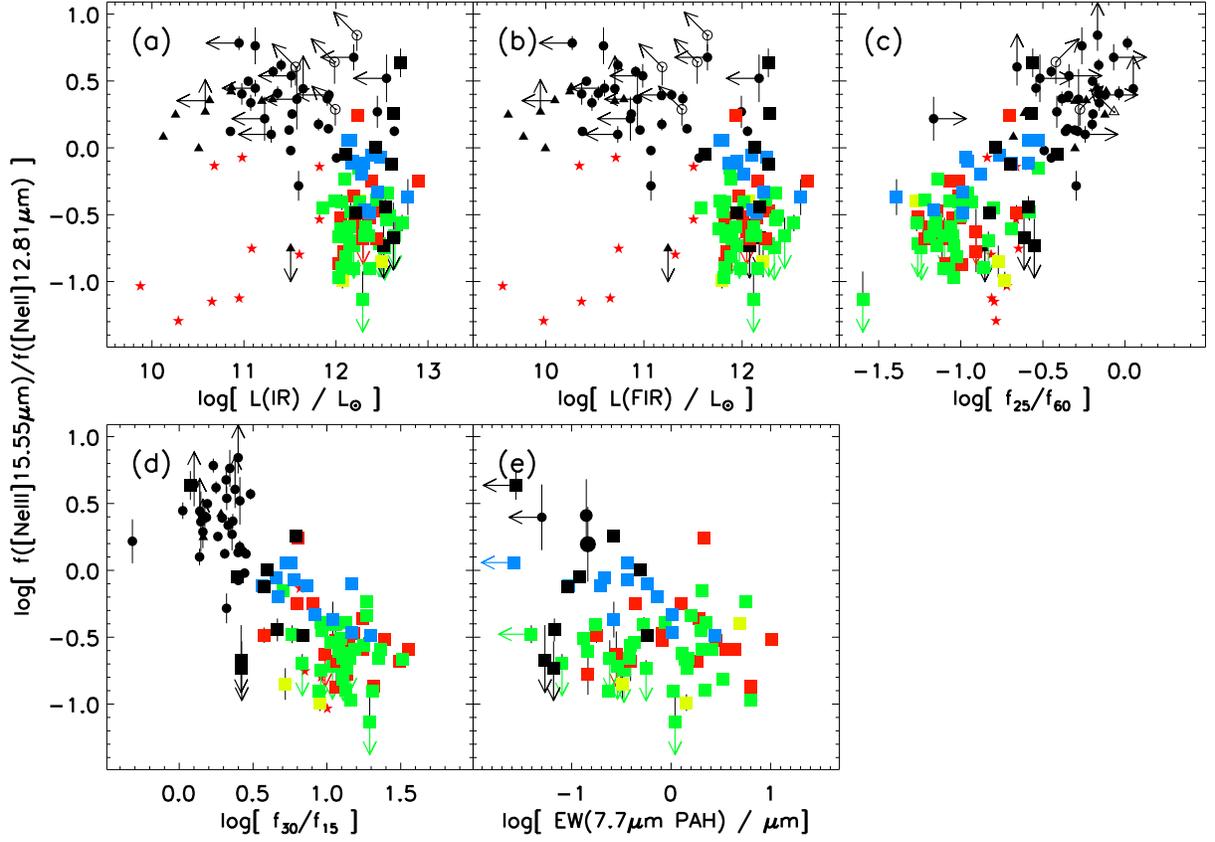}
\caption{ [Ne~III] 15.5 $\mu$m/[Ne~II] 12.8 $\mu$m flux ratios {\em
    vs.\/} ($a$) total infrared luminosity, ($b$) FIR
    luminosity, ($c$) 25-to-60 $\mu$m flux ratio, ($d$) 30-to-15
    $\mu$m flux ratio, and ($e$) 7.7 $\mu$m PAH equivalent width.  The
    meaning of the small square and circle (other) symbols is the same
    as in Figure \ref{fig:fluxcomp} (\ref{fig:mircolor} and
    \ref{fig:mirfir_v}).  In addition, the small, medium-size, and
    large black circles in panel ($e$) correspond to the average
    FIR-undetected, FIR-faint, and FIR-bright PG~QSOs from Paper II.
    No obvious trend is seen with FIR or total infrared luminosity
    among ULIRGs, but a clear dependence is seen on optical spectral
    type and IR continuum colors. }
\label{fig:ne3ne2}
\end{figure*}

\begin{figure*}[ht]
\epsscale{1.1}
\plotone{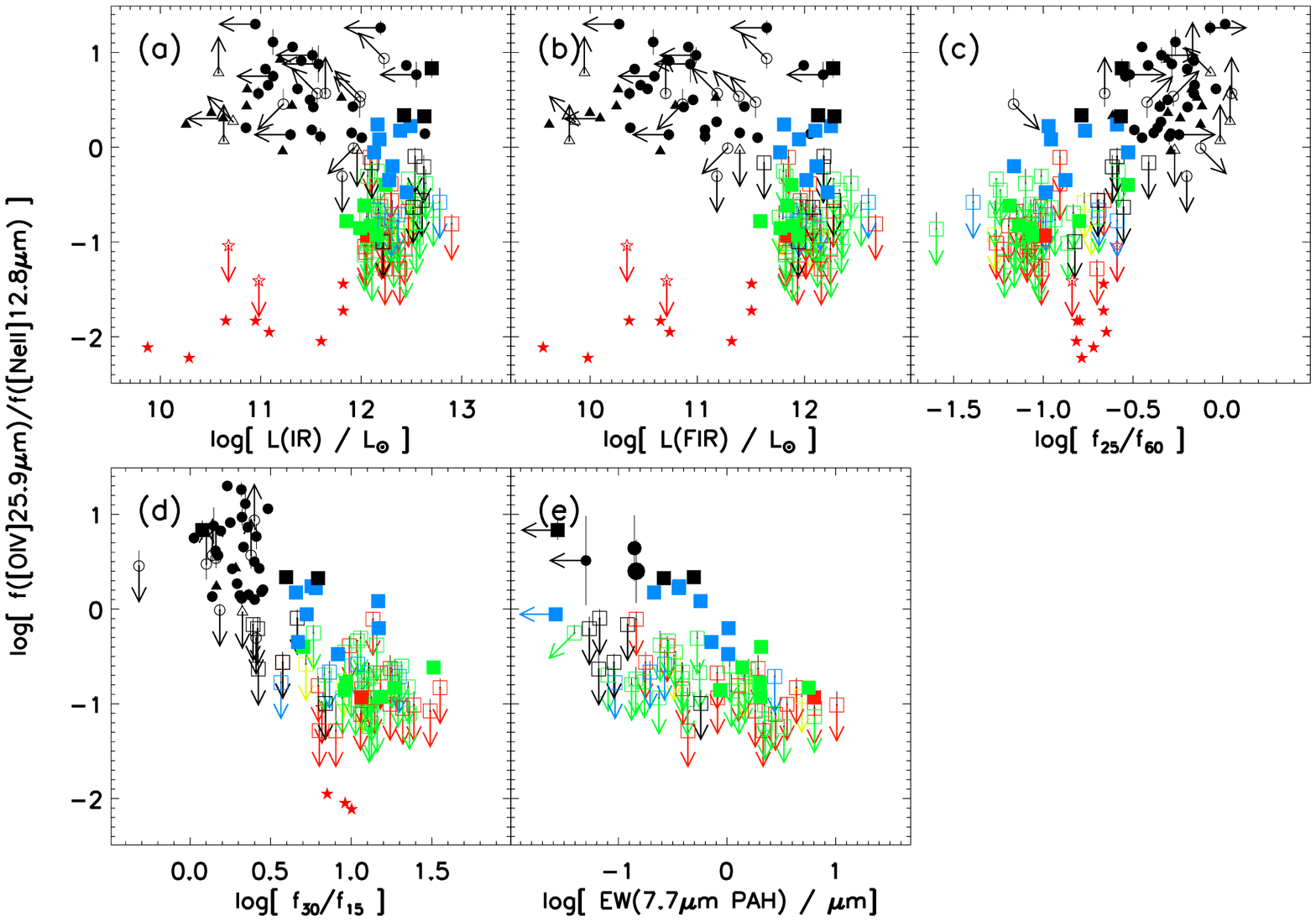}
\caption{ [O~IV] 25.9 $\mu$m/[Ne~II] 12.8 $\mu$m flux ratios {\em
    vs.\/} ($a$) total infrared luminosity, ($b$) FIR
    luminosity, ($c$) 25-to-60 $\mu$m flux ratio, ($d$) 30-to-15
    $\mu$m flux ratio, and ($e$) 7.7 $\mu$m PAH equivalent width.  The
    meaning of the small square and circle (other) symbols is the same
    as in Figure \ref{fig:fluxcomp} (\ref{fig:mircolor} and
    \ref{fig:mirfir_v}). This ratio among ULIRGs is slightly larger on
    average among ULIRGs with larger far- or total infrared
    luminosity. [O~IV]/[Ne~II] is more clearly larger among warmer,
    Seyfert ULIRGs. PG~QSOs nicely fit along this excitation sequence.
    }
\label{fig:o4ne2}
\end{figure*}

\begin{figure*}[ht]
\epsscale{1.1}
\plotone{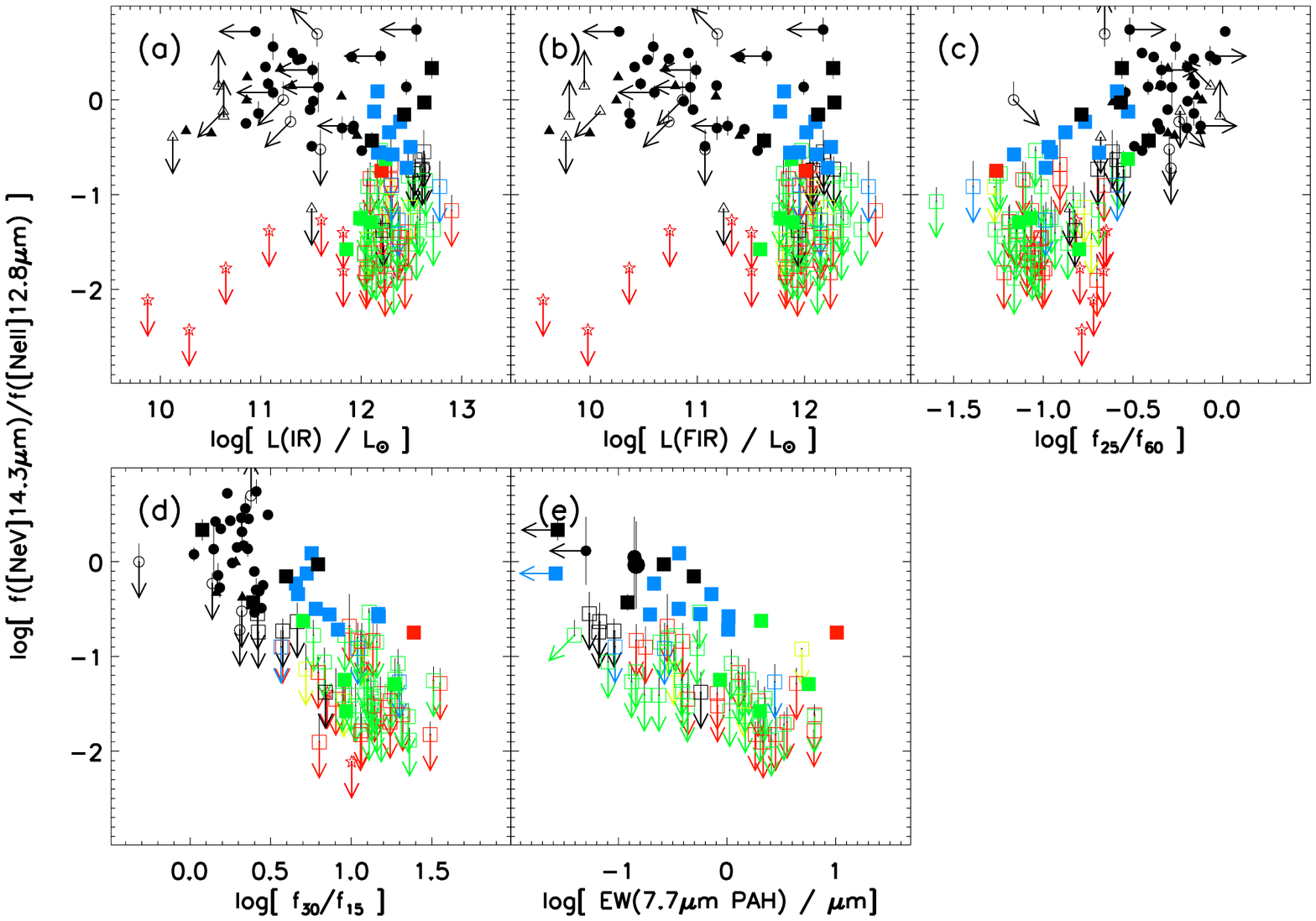}
\caption{ [Ne~V] 14.3 $\mu$m/[Ne~II] 12.8 $\mu$m flux ratios {\em
    vs.\/} ($a$) total infrared luminosity, ($b$) FIR
    luminosity, ($c$) 25-to-60 $\mu$m flux ratio, ($d$) 30-to-15 $mu$m
    flux ratio, and ($e$) 7.7 $\mu$m PAH equivalent width.  The
    meaning of the small square and circle (other) symbols is the same
    as in Figure \ref{fig:fluxcomp} (\ref{fig:mircolor} and
    \ref{fig:mirfir_v}).  This ratio among ULIRGs is slightly larger
    on average among ULIRGs with larger FIR or total infrared
    luminosity. [Ne~V]/[Ne~II] is more clearly larger among warmer,
    Seyfert ULIRGs. PG~QSOs nicely fit along this excitation
    sequence. }
\label{fig:ne5ne2}
\end{figure*}

\begin{figure*}[ht]
\epsscale{0.8}
\plotone{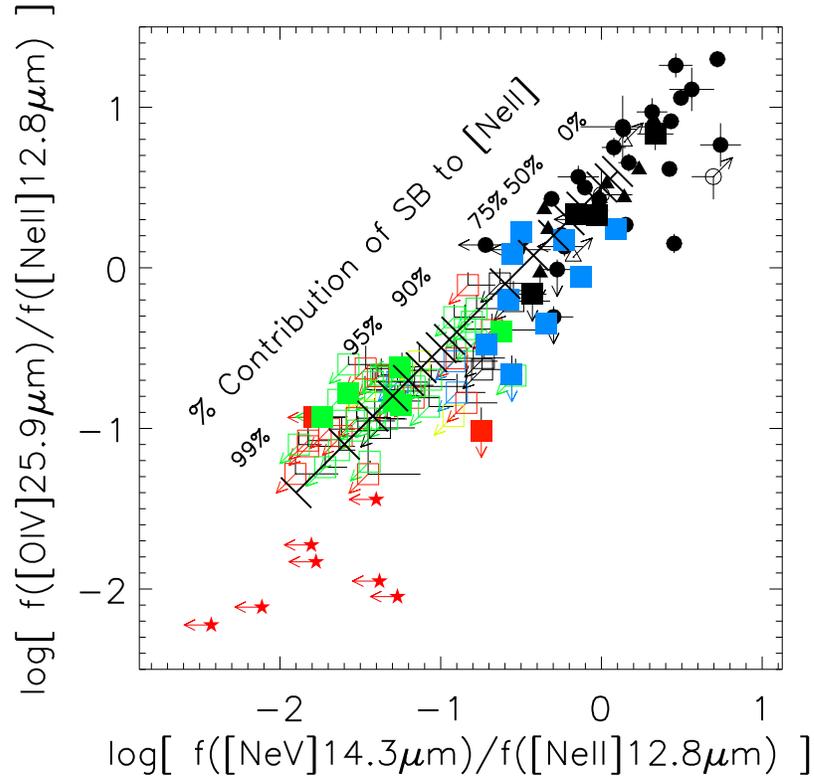}
\caption{ [Ne~V] 14.3 $\mu$m/[Ne~II] 12.8 $\mu$m {\em vs.\/} [O~IV]
  25.9 $\mu$m/[Ne~II] 12.8 $\mu$m for ULIRGs and PG~QSOs in our
  sample.  The meaning of the square and circle (other)
  symbols is the same as in Figure \ref{fig:fluxcomp}
  (\ref{fig:mircolor}).  A clear positive correlation is observed,
  representing an excitation sequence anchored with the HII-like/LINER
  ULIRGs, moving up in excitation level to the Seyfert 2 and Seyfert 1
  ULIRGs, and ending with the PG~QSOs.
  The solid diagonal line in these diagrams indicates constant
  [Ne~V]/[O~IV] and changing [Ne~II].  This line may be interpreted as
  a mixing line if [Ne~V] and [O~IV] are only produced by an AGN and
  [Ne~II] by starburst activity.  The tickmarks along the line
  indicate the percent contribution of the starburst to [Ne~II].  The
  anchor point of this line has some systematic uncertainty due to
  variable AGN physical conditions.}
\label{fig:o4ne2_v_ne5ne2}
\end{figure*}

\begin{figure*}[ht]
\epsscale{1.1}
\plotone{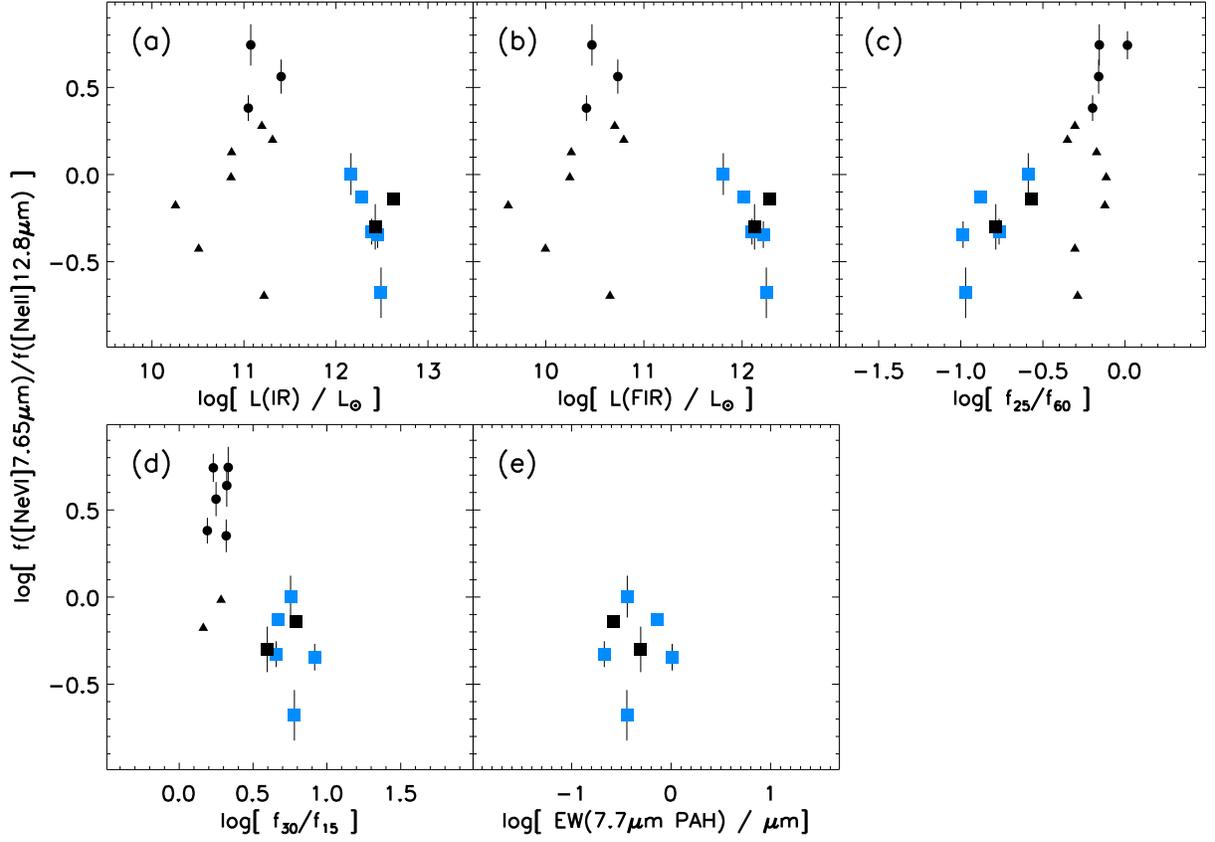}
\caption{ [Ne~VI] 7.65 $\mu$m/[Ne~II] 12.8 $\mu$m flux ratios {\em
    vs.\/} ($a$) total infrared luminosity, ($b$) FIR
  luminosity, ($c$) 25-to-60 $\mu$m flux ratio, ($d$) 30-to-15 $\mu$m
  flux ratio, and ($e$) 7.7 $\mu$m PAH equivalent width.  The meaning
  of the square and circle (triangle) symbols is the same as in
  Figure \ref{fig:fluxcomp} (\ref{fig:mircolor}).  [Ne~VI] is detected
  only in Seyfert ULIRGs and QSOs.  The small number of detections
  prevent us from looking for statistically significant correlations
  with infrared or FIR luminosity, optical spectral type, or
  continuum colors.  }
\label{fig:ne6ne2}
\end{figure*}

\begin{figure*}[ht]
\epsscale{1.1}
\plotone{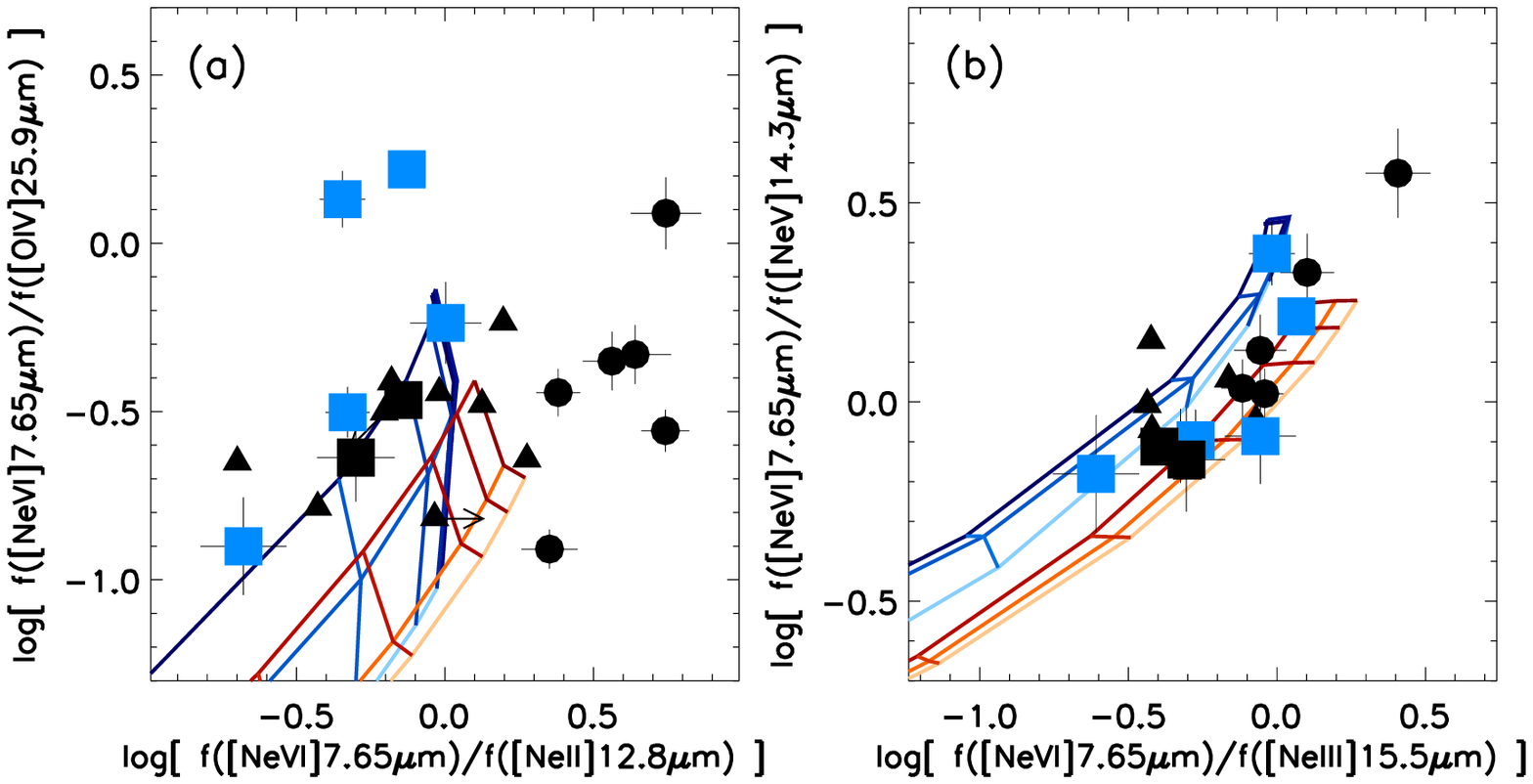}
\caption{($a$) [Ne~VI] 7.65 $\mu$m/[Ne~II] 12.8 $\mu$m {\em vs.\/}
  [Ne~VI] 7.65 $\mu$m/[O~IV] 25.9 $\mu$m and ($b$) [Ne~VI] 7.65
  $\mu$m/[Ne~III] 15.5 $\mu$m {\em vs.\/} [Ne~VI] 7.65 $\mu$m/[Ne~V]
  14.3 $\mu$m for ULIRGs and PG~QSOs in our sample.  The meaning of
  the square and circle (triangle) symbols is the same as in
  Figure \ref{fig:fluxcomp} (\ref{fig:mircolor}).  The grids show the
  predictions from the dusty, solar-metallicty, radiation
  pressure-dominated photoionization models of Groves et al. (2004)
  for two different power-law slopes [$f_\nu \propto \nu^\alpha$,
  $\alpha = -1.2$ (blue) and $\alpha = -2.0$ (red)]. The two
  parameters that vary across the grids are the ionization parameter
  (log $U$ = $-$4 -- 0) and the hydrogen density (log $n_H$ = 2 -- 4).
  The color hues correspond to low values (light color) and high
  values (dark color); e.g., dark blue means high density or high
  ionization parameter.  Poor agreement is seen in ($a$) due to
  starburst contamination of [Ne~II] or problems with the model
  (highlighted by disagreements in both axes).  Panel ($b$) is more
  useful, since it uses only high ionization lines of a single element
  (Ne).  The good agreement means that AGN may power not only [Ne~VI]
  and [Ne~V], but also [Ne~III], in these systems.}
\label{fig:highionlines}
\end{figure*}

\begin{figure*}[ht]
\epsscale{1.1}
\plotone{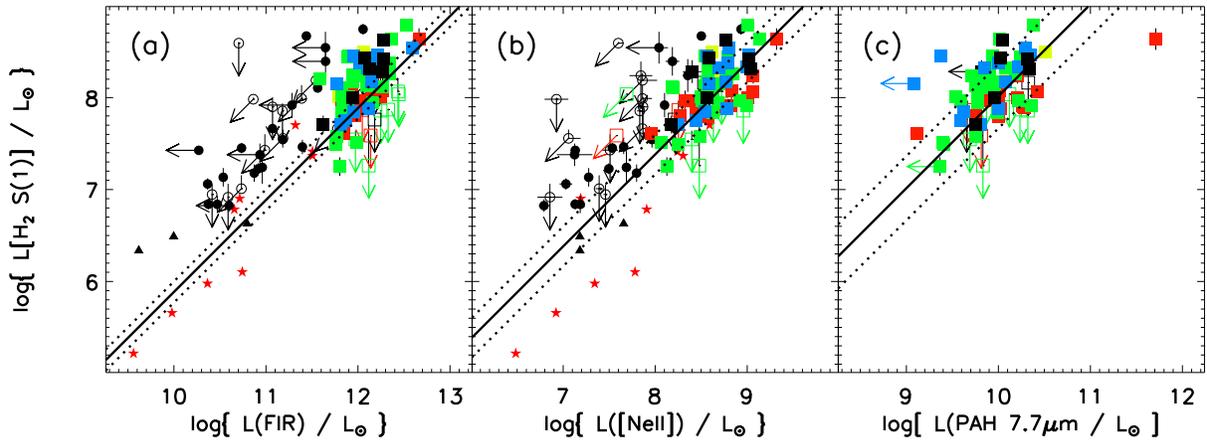}
\caption{Luminosity of H$_2$ S(1) 17.04 $\mu$m {\em vs.\/} various
  star formation indicators: ($a$) FIR luminosity, ($b$)
  luminosity of [Ne~II] 12.8 $\mu$m, and ($c$) luminosity of PAH 7.7
  $\mu$m.  The meaning of the square and circle (other) symbols is the
  same as in Figure \ref{fig:fluxcomp} (\ref{fig:mircolor}).  The solid
  and dotted lines show the mean and standard deviation of $L$[H$_2$
  S(1)]/$L$(X) (where X $=$ FIR, etc.)  for HII ULIRGs.  A positive
  correlation is seen between these quantities for both ULIRGs and
  PG~QSOs. However, Seyfert ULIRGs and especially PG~QSOs have larger
  H$_2$-to-FIR, H$_2$-to-[Ne~II], and H$_2$-to-PAH ratios than HII
  ULIRGs, suggesting that H$_2$ emission is influenced by the presence
  of the AGN in these objects.  The LINERs also have significantly
  larger ratios and dispersions in panels ($b$) and ($c$).}
\label{fig:h2s1lum_v}
\end{figure*}

\begin{figure*}[ht]
\epsscale{1.1}
\plotone{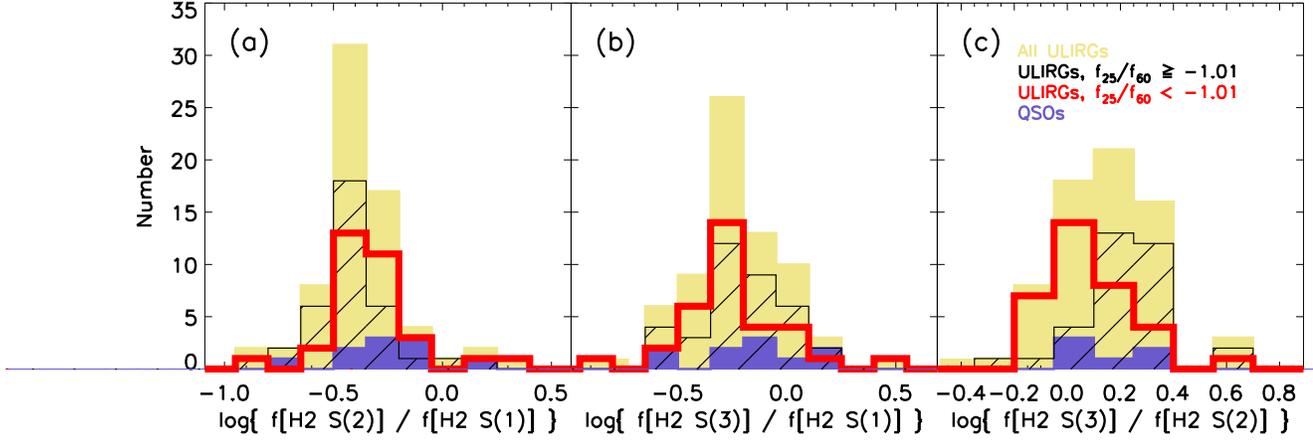}
\caption{ Histograms of H$_2$ ($a$) S(2)/S(1), ($b$) S(3)/S(1), and
  ($c$) S(3)/S(2) flux ratios (uncorrected for extinction) as a
  function of $f_{25}/f_{60}$ in ULIRGs and QSOs.  The ULIRG
  distributions are divided at the median value of $f_{25}/f_{60}$.
  These ratios are indicative of the excitation temperature of H$_2$
  gas, but can also be affected by extinction and variable
  ortho-to-para ratios.  Warm ULIRGs have smaller S(2)/S(1) ratios on
  average than cool ULIRGs by 0.1 dex, with a K-S (Kuiper) probability
  of arising from the same distribution of $<$0.1\% (1\%).  Warm
  ULIRGs have larger S(3)/S(2) ratios on average than cool ULIRGs by
  0.15 dex, with a K-S (Kuiper) probability of arising from the same
  distribution of $<$0.1\% (2\%).  Warm and cool ULIRGs are
  statistically indistinguishable in S(3)/S(1).}
\label{fig:h2rat}
\end{figure*}

\begin{figure*}[ht]
\epsscale{0.9}
\plotone{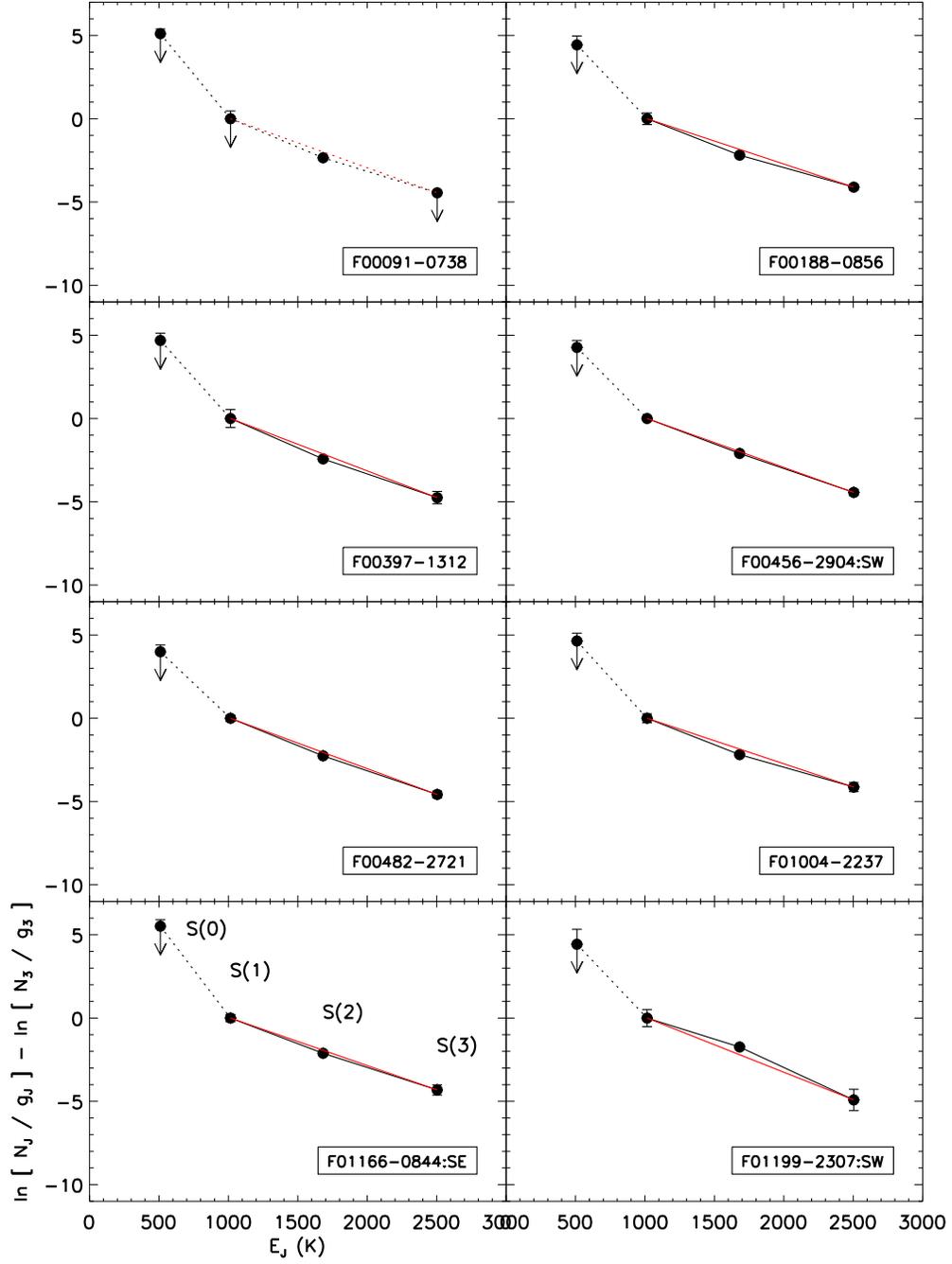}
\caption{ H$_2$ excitation diagram for a few ULIRGs in our sample. The
  excitation temperature ($T_{ex}$) is the reciprocal of the slope
  between any two data points.  Solid lines join points where both
  lines are detected, and dashed lines join points where at least one
  line is an upper limit.  Black lines join immediately adjacent
  lines, and red lines join the the S(1) and S(3) transitions.  Error
  bars are 2$\sigma$.}
\label{fig:h2tex}
\end{figure*}

\begin{figure*}[ht]
\epsscale{1.0}
\plotone{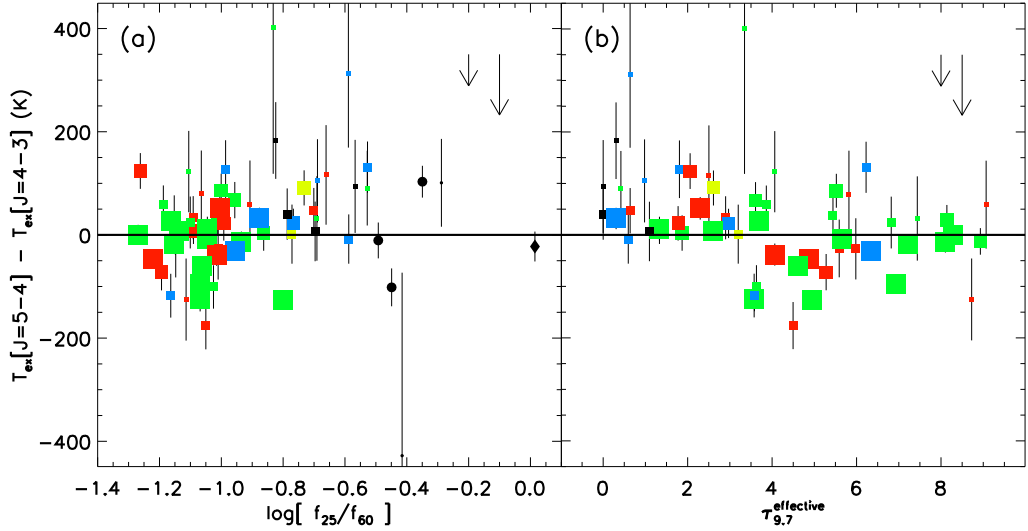}
\caption{ Difference between the excitation temperatures derived from
  the H$_2$ S(3)/S(2) and S(2)/S(1) flux ratios {\em vs.\/} ($a$)
  $f_{25}/f_{60}$ and ($b$) the 9.7 $\mu$m silicate effective optical
  depth.  The meaning of the symbols is the same as in
  Figure \ref{fig:fluxcomp}.  The size of the symbol reflects the
  relative uncertainties on each data point, where the quartile of
  most certain points are the largest and the quartile of least
  certain points are the smallest.  All objects should be above the
  solid line, unless extinction and/or ortho-para effects are at
  play. The small downward arrow on the right in each diagram reflects
  the effect of changing the ortho-to-para ratio from 3 to 2, while
  the long arrow reflects the effect of an extinction $A_V = 10$.
  Significant trends among ULIRGs are seen, with decreasing
  temperature difference with increasing silicate optical depth and
  decreasing $f_{25}/f_{60}$.  ULIRGs with extinction greater than the
  median (and $f_{25}/f_{60}$ lower than the median) have a lower
  temperature difference by $60-70$ K, with a K-S (Kuiper)
  significance of $<$0.1\% (6$-$7\%).  The trend with extinction
  implies extinction of the molecular lines.  When corrected for this
  extinction, the trend with $f_{25}/f_{60}$ will lessen or
  disappear.}
\label{fig:h2texdiff}
\end{figure*}

{\bf Method \#5 (PAH-free 5 -- 25 $\mu$m to FIR continuum ratio).} In
Section 6.1.3, we showed that the PAH-free, silicate-free MIR(5 -- 25
$\mu$)-to-FIR ratio derived from our fits was an excellent probe of
nuclear activity in the 1-Jy ULIRGs (Figure  \ref{fig:mirfir_v}).  We
adopt log[L(MIR)/L(FIR)] = 0.35 and --1.25 as the zero points for pure
AGN and starburst, respectively. The AGN contribution is calculated
from a linear interpolation between these two extremes. The zero point
for the pure AGN corresponds to the average MIR/FIR ratio of
FIR-undetected PG~QSOs.  The zero point for the pure starbursts is
calculated from the ten ULIRGs with the lowest MIR/FIR ratios.  The
correction factors to transform FIR luminosities into bolometric
luminosities are listed in Table \ref{tab:agnfrac_bolcor}.

{\bf Method \#6 ($f_{30}/f_{15}$ continuum ratio).} The application of
method \#5 to a large number of ULIRGs is time consuming since it
involves detailed template fitting of the MIR SED. A more
straightforward method, method \#6, is based on the $f_{30}/f_{15}$
continuum ratio, which was found to be more tightly correlated with
the PAH-free, silicate-free MIR/FIR ratio than any other MIR continuum
ratio at our disposal (Figure \ref{fig:mirfir_v}). This method is
roughly equivalent to using the MIPS 70/24 $\mu$m flux ratios to
search for AGN activity in $z \sim 1$ U/LIRGs. Here we use
$f_{30}/f_{15}$ as a surrogate of the PAH-free, silicate-free MIR/FIR
ratio and adopt log($f_{30}/f_{15}$) = 0 and 1.35 as the zero points
for pure AGN and starburst, respectively. The AGN contribution is
calculated from a linear interpolation between these two extremes. The
zero point for the pure AGN corresponds to the average $f_{30}/f_{15}$
ratio of FIR-undetected PG~QSOs, while the zero point for the pure
starbursts is calculated from the ten ULIRGs with the largest
$f_{30}/f_{15}$ ratios.  The correction factors to transform 15-$\mu$m
luminosities into bolometric luminosities are listed in Table
\ref{tab:agnfrac_bolcor}.  

\clearpage

\begin{deluxetable}{clcccrc}
\tablecolumns{7}
\tabletypesize{\scriptsize}
\tablecaption{Sample}
\tablewidth{0pt}
\tablehead{
\colhead{Galaxy} & \colhead{$z$} & \colhead{log($L(bol)/L_\sun$)} & \colhead{Type} & \colhead{IC} & \colhead{NS} & \colhead{Ref} \\
\colhead{(1)} & \colhead{(2)} & \colhead{(3)} & \colhead{(4)} & \colhead{(5)} & \colhead{(6)} & \colhead{(7)}
}
\startdata
\cutinhead{ULIRGs}
                 F00091$-$0738  & 0.118  & 12.36  &  HII  &   IIIb  &     2.31  &      1 \\
                 F00188$-$0856  & 0.128  & 12.43  &    L  &      V  &$<$  0.34  &      2 \\
                 F00397$-$1312  & 0.262  & 12.96  &  HII  &      V  &$<$  0.61  &      2 \\
              F00456$-$2904:SW  & 0.110  & 12.29  &  HII  &   IIIa  &    22.80  &      2 \\
                 F00482$-$2721  & 0.129  & 12.09  &    L  &   IIIb  &     7.39  &      1 \\
\enddata
\tablerefs{1 = Sakamoto et al. 1999; 2 = Scoville et al. 2000; 3 = Beswick et al. 2001; 4 = Kim et al. 2002; 5 = Veilleux et al. 2006; 6 = Veilleux et al. 2009}
\tablecomments{
Col.(1): Galaxy name.  Coordinate-based names beginning with "F" are sources in the IRAS Faint Source Catalog.  Col.(2): Redshift.  Col.(3): Bolometric luminosity.  For ULIRGs, we assume $L(bol) = 1.15 L(IR)$.  For PG QSOs, we assume $L(bol) = 7 L(5100~\AA) + L(IR)$ (Netzer et al. 2007).  Col.(4): Optical spectral type, from Veilleux et al. (1995, 1999a) and Rupke et al. (2005a).  Col.(5): Interaction class, from Veilleux et al. 2009, Veilleux et al. 2006, or Veilleux et al. 2002 (in order of preference).  Col.(6): Nuclear separation, in kpc.  Col.(7): Reference for nuclear separation.
}
\label{tab:sample}
\end{deluxetable}
 
\begin{deluxetable}{ccrrrrr}
\tablecolumns{7}
\tabletypesize{\scriptsize}
\tablecaption{Observations}
\tablewidth{0pt}
\tablehead{
\colhead{} & \colhead{} & \multicolumn{5}{c}{Exposure Time} \\
\colhead{Galaxy} & \colhead{PID} & \colhead{SL2} & \colhead{SL1} & \colhead{SH} & \colhead{LH} & \colhead{LL} \\
\colhead{(1)} & \colhead{(2)} & \colhead{(3)} & \colhead{(4)} & \colhead{(5)} & \colhead{(6)} & \colhead{(7)}
}
\startdata
\cutinhead{ULIRGs}
    F00091$-$0738  &      3187 &     240 &     240 &     480 &     960 & \nodata \\
    F00188$-$0856  &       105 &     240 &     240 &     720 &     480 & \nodata \\
    F00397$-$1312  &       105 &     240 &     240 &     720 &     480 & \nodata \\
 F00456$-$2904:SW  &      3187 &     240 &     240 &     960 &     720 & \nodata \\
    F00482$-$2721  &      3187 &     240 &     240 &     720 &    1440 & \nodata \\
\enddata
\tablecomments{
Col.(1): Galaxy name.  Col.(2): {\it Spitzer} proposal ID(s) under which data was taken.  Col.(3-7): Exposure times for each IRS module, in seconds.  The LL exposure time is listed only when these data were used in the fit.
}
\label{tab:obs}
\end{deluxetable}

\begin{deluxetable}{clllllllll}
\tablecolumns{10}
\tabletypesize{\scriptsize}
\tablecaption{Emission Line Fluxes}
\tablewidth{0pt}
\tablehead{
\colhead{Galaxy} & \colhead{[\ion{Ne}{6}]7.65} & \colhead{H$_2$~S(3)~9.66} & \colhead{[\ion{S}{4}]10.51} & \colhead{H$_2$~S(2)~12.28} & \colhead{Hu$\alpha$~12.37} & \colhead{[\ion{Ne}{2}]12.81} & \colhead{[\ion{Ne}{5}]14.32} & \colhead{[\ion{Ne}{3}]15.55} & \colhead{H$_2$~S(1)~17.03} \\
\colhead{(1)} & \colhead{(2)} & \colhead{(3)} & \colhead{(4)} & \colhead{(5)} & \colhead{(6)} & \colhead{(7)} & \colhead{(8)} & \colhead{(9)} & \colhead{(10)}
}
\startdata
\cutinhead{ULIRGs}
                 F00091$-$0738 &               \nodata &       $<$5.18E-22    &        $<$3.76E-22    &  \phm{$<$}2.55E-22(59) &       $<$3.15E-22    &  \phm{$<$}1.88E-21( 6) &       $<$3.92E-22    &        $<$4.45E-22    &        $<$7.71E-22   \\ 
                 F00188$-$0856 &               \nodata & \phm{$<$}7.27E-22( 8) &       $<$8.96E-23    &  \phm{$<$}2.99E-22(27) &       $<$8.99E-23    &  \phm{$<$}4.27E-21( 1) &       $<$1.66E-22    &  \phm{$<$}5.30E-22(10) & \phm{$<$}7.71E-22(17)\\
                 F00397$-$1312 &               \nodata & \phm{$<$}3.89E-22(18) & \phm{$<$}2.66E-22(27) & \phm{$<$}2.39E-22(38) &       $<$1.43E-22    &  \phm{$<$}3.78E-21( 1) &       $<$2.56E-22    &  \phm{$<$}2.14E-21( 9) & \phm{$<$}7.86E-22(27)\\
              F00456$-$2904:SW &               \nodata & \phm{$<$}6.62E-22( 9) &       $<$1.24E-22    &  \phm{$<$}4.14E-22(18) &       $<$9.59E-23    &  \phm{$<$}5.88E-21( 0) &       $<$1.19E-22    &  \phm{$<$}1.50E-21( 2) & \phm{$<$}9.77E-22( 4)\\
                 F00482$-$2721 &               \nodata & \phm{$<$}4.73E-22(10) &       $<$1.90E-22    &  \phm{$<$}2.90E-22(15) &       $<$1.34E-22    &  \phm{$<$}2.79E-21( 1) &       $<$1.55E-22    &  \phm{$<$}6.04E-22( 6) & \phm{$<$}8.03E-22( 9)\\
\enddata
\tablecomments{
Col.(1): Galaxy name.  Col.(2$-$10): Atomic fine structure and H$_2$ rotational emission line fluxes, in W~cm$^{-2}$.  Percent errors are given in parentheses.  Upper limits are 3$\sigma$.
}
\label{tab:eml1}
\end{deluxetable}

\begin{deluxetable}{cllllllll}
\tablecolumns{9}
\tabletypesize{\scriptsize}
\tablecaption{Emission Line Fluxes}
\tablewidth{0pt}
\tablehead{
\colhead{Galaxy} & \colhead{[\ion{Fe}{2}]17.94} & \colhead{[\ion{S}{3}]18.71} & \colhead{[\ion{Ne}{5}]24.32} & \colhead{[\ion{O}{4}]25.89} & \colhead{[\ion{Fe}{2}]25.99} & \colhead{H$_2$~S(0)~28.22} & \colhead{[\ion{S}{3}33.48]} & \colhead{[\ion{Si}{2}]34.81} \\
\colhead{(1)} & \colhead{(2)} & \colhead{(3)} & \colhead{(4)} & \colhead{(5)} & \colhead{(6)} & \colhead{(7)} & \colhead{(8)} & \colhead{(9)}
}
\startdata
\cutinhead{ULIRGs}
                 F00091$-$0738 &       $<$7.71E-22    &        $<$1.03E-21    &        $<$9.27E-22    &        $<$7.71E-22    &        $<$7.71E-22    &        $<$1.13E-21    &                \nodata &              \nodata \\
                 F00188$-$0856 &       $<$4.94E-22    &        $<$3.09E-22    &        $<$4.68E-22    &        $<$4.92E-22    &        $<$4.94E-22    &        $<$5.76E-22    &                \nodata &              \nodata \\
                 F00397$-$1312 &       $<$5.89E-22    &  \phm{$<$}1.29E-21(13) &       $<$5.52E-22    &        $<$5.87E-22    &        $<$5.89E-22    &        $<$7.58E-22    &                \nodata &              \nodata \\
              F00456$-$2904:SW &       $<$5.82E-22    &  \phm{$<$}3.40E-21( 4) &       $<$5.75E-22    &        $<$5.82E-22    &        $<$5.82E-22    &        $<$6.21E-22    &                \nodata &              \nodata \\
                 F00482$-$2721 &       $<$1.69E-22    &  \phm{$<$}1.05E-21( 7) &       $<$2.75E-22    &  \phm{$<$}6.75E-22(10) &       $<$1.69E-22    &        $<$3.87E-22    &                \nodata &              \nodata \\
\enddata
\tablecomments{
Col.(1): Galaxy name.  Col.(2$-$9): Atomic fine structure and H$_2$ rotational emission line fluxes, in W~cm$^{-2}$.  Percent errors are given in parentheses.  Upper limits are 3$\sigma$.
}
\label{tab:eml2}
\end{deluxetable}

\begin{deluxetable}{crrrrrrrr}
\tablecolumns{9}
\tablecaption{Continuum Measurements}
\tabletypesize{\scriptsize}
\tablewidth{0pt}
\tablehead{
\colhead{Galaxy} & \colhead{6~$\micron$} & \colhead{15~$\micron$} & \colhead{20~$\micron$} & \colhead{25~$\micron$} & \colhead{30~$\micron$} & \colhead{12~$\micron$ (IRAS)} & \colhead{25~$\micron$ (IRAS)} & \colhead{log[$L$(MIR)/$L$(FIR)]} \\
\colhead{(1)} & \colhead{(2)} & \colhead{(3)} & \colhead{(4)} & \colhead{(5)} & \colhead{(6)} & \colhead{(7)} & \colhead{(8)} & \colhead{(9)}
}
\startdata
\cutinhead{ULIRGs}
                 F00091$-$0738 &     5.2 &   117.7 &   178.2 &   546.1 &  1145.5 &    42.0 &   325.0 &     -0.73
                            \\
                 F00188$-$0856 &    12.6 &    78.3 &   125.0 &   327.8 &   690.8 &    26.5 &   203.2 &     -0.79
                            \\
                 F00397$-$1312 &    78.3 &   126.9 &   146.8 &   422.7 &   792.8 &    42.0 &   157.8 &     -0.30
                            \\
              F00456$-$2904:SW &     8.9 &    39.7 &   111.8 &   324.2 &   691.9 &    23.3 &   210.7 &     -0.98
                            \\
                 F00482$-$2721 &     2.4 &     9.3 &    42.0 &   129.9 &   300.5 &     5.2 &    73.6 &      -1.1
                            \\
\enddata
\tablecomments{
Col.(1): Galaxy name.  Col.(2-6): Rest-frame {\it Spitzer} flux densities, computed from the IRS spectra using a 3.3\%\ bandpass, in mJy.  Col.(7-8): Observed-frame {\it Spitzer} flux densities, computed using step function approximations to the {\it IRAS} 12 and 25 \micron\ system response functions.  The flux given is the average $f_\nu$ under the step function.  Col.(9): Extincted $5-25$ \micron\ luminosity minus PAH$+$silicate emission (i.e., blackbody only) as a (logarithmic) fraction of the far-infrared ($40-122$ \micron) luminosity.  For the PG QSOs, this latter quantity is available only for the three average spectra divided by $L$(FIR).
}
\label{tab:cont}
\end{deluxetable}

\begin{deluxetable}{crrrr}
\tablecolumns{5}
\tablecaption{Fit Results: Absorption Measurements}
\tabletypesize{\scriptsize}
\tablewidth{0pt}
\tablehead{
\colhead{Galaxy} & \colhead{$\tau_{9.7\micron}^{eff}$} & \colhead{log[$W_{eq}$(H$_2$O+HC)]} & \colhead{log[$W_{eq}$(C$_2$H$_2$)]} & \colhead{log[$W_{eq}$(HCN)]} \\
\colhead{(1)} & \colhead{(2)} & \colhead{(3)} & \colhead{(4)} & \colhead{(5)}
}
\startdata
                 F00091$-$0738 &\phm{$<$}   9.62 & \phm{$<$}   0.32 & \phm{$<$}  -2.09 & \phm{$<$}  -2.22 \\
                 F00188$-$0856 &\phm{$<$}   4.06 & \phm{$<$}   0.23 & \phm{$<$}  -2.84 & \phm{$<$}  -2.90 \\
                 F00397$-$1312 &\phm{$<$}   5.82 & \phm{$<$}  -0.81 &          \nodata &          \nodata \\
              F00456$-$2904:SW &\phm{$<$}   2.90 &       $<$   0.42 &          \nodata &          \nodata \\
                 F00482$-$2721 &\phm{$<$}   3.86 & \phm{$<$}  -0.21 &          \nodata &          \nodata \\
\enddata
\tablecomments{
Col.(1): Galaxy name.  Col.(2): Effective peak silicate optical depth, computed using the ratio of the total extincted flux to the total unextincted flux.  Col.(3-5): Rest-frame equivalent widths (in microns) of the water ice + hydrocarbon feature at $5-7$~\micron; the C$_2$H$_2$ 13.7~\micron\ absorption feature; and the HCN 14.0~\micron\ absorption feature, respectively.
}
\label{tab:abs}
\end{deluxetable}

\begin{deluxetable}{crrrr}
\tablecolumns{5}
\tablecaption{Fit Results: PAH Measurements}
\tabletypesize{\scriptsize}
\tablewidth{0pt}
\tablehead{
\colhead{Galaxy} & \colhead{log[$W_{eq}$(6$\micron$ PAH)]} & \colhead{log[$W_{eq}$(7$\micron$ PAH)]} & \colhead{log[$L$(PAH)/$L$(IR)]} & \colhead{log[$L$(PAH)/$L$(FIR)]} \\
\colhead{(1)} & \colhead{(2)} & \colhead{(3)} & \colhead{(4)} & \colhead{(5)}
}
\startdata
                 F00091$-$0738 &\phm{$<$}  -0.78 & \phm{$<$}  -0.55 & \phm{$<$}  -2.23 & \phm{$<$}  -1.98 \\
                 F00188$-$0856 &\phm{$<$}  -1.17 & \phm{$<$}  -0.63 & \phm{$<$}  -2.18 & \phm{$<$}  -1.97 \\
                 F00397$-$1312 &\phm{$<$}  -0.59 & \phm{$<$}   0.11 & \phm{$<$}  -0.83 & \phm{$<$}  -0.59 \\
              F00456$-$2904:SW &\phm{$<$}   0.14 & \phm{$<$}   0.55 & \phm{$<$}  -1.57 & \phm{$<$}  -1.36 \\
                 F00482$-$2721 &\phm{$<$}  -0.36 & \phm{$<$}   0.14 & \phm{$<$}  -2.04 & \phm{$<$}  -1.84 \\
\enddata
\tablecomments{
Col.(1): Galaxy name.  Col.(2-3): Logarithmic rest-frame equivalent widths (in microns) of the PAH 6.2~\micron\ and 7.7~\micron\ features.  Col.(4-5): Logarithmic ratios of the total PAH luminosity to the total infrared and far-infrared luminosities.
}
\label{tab:pahs}
\end{deluxetable}

\begin{deluxetable}{cllr}
\tablecolumns{4}
\tabletypesize{\scriptsize}
\tablecaption{H$_2$ Properties}
\tablewidth{0pt}
\tablehead{
\colhead{Galaxy} & \colhead{$T_{ex}(4-3)$} & \colhead{$T_{ex}(5-4)$} & \colhead{log[$M$(H$_2$)/$M_\sun$]} \\
\colhead{(1)} & \colhead{(2)} & \colhead{(3)} & \colhead{(4)}
}
\startdata
\cutinhead{ULIRGs}
                 F00091$-$0738 &      $>$  284.36    &       $>$  391.26    &      \nodata \\
                 F00188$-$0856 &\phm{$<$}  305.07(14) &\phm{$<$}  427.87(15) &    8.34(31) \\
                 F00397$-$1312 &\phm{$<$}  274.51(19) &\phm{$<$}  353.93(18) &    9.16(45) \\
              F00456$-$2904:SW &\phm{$<$}  317.97( 8) &\phm{$<$}  351.26( 8) &    8.37(19) \\
                 F00482$-$2721 &\phm{$<$}  295.44( 8) &\phm{$<$}  354.25( 8) &    8.46(18) \\
\enddata
\tablecomments{
Col.(1): Galaxy name.  Col.($2-3$): Molecular hydrogen excitation temperatures determined from the S(1)/S(2) and S(2)/S(3) fluxes, respectively.  Percent errors are given in parentheses.  Col.(4): Molecular hydrogen mass computed using the partition function from Herbst et al. (1996), the S(1) flux, and the average of the ($4-3$) and ($5-4$) excitation temperatures.  Percent errors are given in parentheses.
}
\label{tab:h2}
\end{deluxetable}

\begin{deluxetable}{crrr}
\tablecaption{Zero-Point Values for Computing AGN Contribution}
\tablewidth{0pt}
\tablehead{
\colhead{Quantity} & \colhead{AGN} & \colhead{SB/\ion{H}{2}} & \colhead{PDR} \\
\colhead{(1)} & \colhead{(2)} & \colhead{(3)} & \colhead{(4)}
}
\startdata
                           log($[$\ion{O}{4}]/[\ion{Ne}{2}])  &    0.60  & \nodata\tablenotemark{a}  & \nodata \\
                            log([\ion{Ne}{5}]/[\ion{Ne}{2}])  &    0.10  & \nodata\tablenotemark{a}  & \nodata \\
                               log[$W_{eq}$(PAH 7.7\micron)]  & \nodata  &    0.75  & \nodata \\
              log[$f$(PAH 6.2\micron)/$f$($5.3-5.8$\micron)]  &   -1.82  &   -0.40  &    0.62 \\
              log[$f$($14-16$\micron)/$f$($5.3-5.8$\micron)]  &    0.07  &    1.54  &   -0.30 \\
                                      log[$L$(MIR)/$L$(FIR)]  &    0.35  &   -1.25  & \nodata \\
                                        log[$f_{30}/f_{15}$]  &    0.20  &    1.35  & \nodata \\
\enddata
\tablecomments{
Col.(1): Physical quantity by which AGN contribution to the bolometric luminosity is computed.  Col.($2-4$): Values of this quantity for a "pure" AGN, starburst / \ion{H}{2} region, or photo-dissociation region (PDR).  The values for [\ion{O}{4}]/[\ion{Ne}{2}] and [\ion{Ne}{5}]/[\ion{Ne}{2}] are averages among the 8 PG QSOs undetected in the FIR (Netzer et al. 2007).  The $W_{eq}$(PAH 7.7\micron) value is the maximum value observed in our sample.  The $f$(PAH 6.2\micron)/$f$($5.3-5.8$\micron) and $f$($14-16$\micron)/$f$($5.3-5.8$\micron) values are taken from Armus et al. 2007 for the \ion{H}{2} region and PDR vertices, and from our FIR-undetected PG QSO subsample for the AGN vertex.  Finally, the $L$(MIR)/$L$(FIR) and $f_{30}/f_{15}$ values are estimated from our data, using the average of the FIR-undetected PG QSOs as a pure AGN.
}
\tablenotetext{a}{Technically, the [O~IV] and [Ne~V] emission in normal starbursts is non-zero (Lutz et al. 1998; Abel \& Satyapal 2008); e.g., log([O~IV]/[Ne~II]) $\sim$ -1.9 for the {\it ISO} starbursts with detected [O~IV] (Verma et al. 2003).  However, for the purposes of computing AGN contribution we can safely assume it is negligible.}
\label{tab:agnfrac_zp}
\end{deluxetable}

\begin{deluxetable}{crr}
\tablecaption{Bolometric Corrections for Computing AGN Contribution}
\tablewidth{0pt}
\tablehead{
\colhead{Quantity} & \colhead{AGN} & \colhead{SB} \\
\colhead{(1)} & \colhead{(2)} & \colhead{(3)}
}
\startdata
             log([\ion{Ne}{2}]/$L(bol)$)  &   -4.66  &   -3.71 \\
         log[$L(5.3-5.8\micron)/L(bol)$]  &   -1.92  &   -2.85 \\
       log[$L_\lambda(8\micron)/L(bol)$]  &   -1.93  &   -2.39 \\
          log[$L_\nu(15\micron)/L(bol)$]  &  -14.33  &  -14.56 \\
                  log[$L$(FIR)/$L(bol)$)  &   -1.05  &   -0.29 \\
\enddata
\tablecomments{
Col.($2-3$): Values of the quantity in column 1 for a "pure" AGN or starburst.  Starburst values are averages over \ion{H}{2} ULIRGs, and AGN values are averages over the 8 FIR-undetected PG QSOs from Netzer et al. 2007.  The bolometric luminosity is computed according to $L(bol) = 1.15~L(IR)$ for all ULIRGs and $L(bol) = 7~L$(5100 \AA)$+L$(IR) for all PG QSOs.
}
\label{tab:agnfrac_bolcor}
\end{deluxetable}

\begin{deluxetable}{ccrrrrrrr}
\tablecolumns{9}
\tablecaption{Binned AGN Contribution}
\tablewidth{0pt}
\tabletypesize{\footnotesize}
\tablehead{
\colhead{Bin} & \colhead{No.} & \colhead{[\ion{O}{4}]/[\ion{Ne}{2}]} & \colhead{[\ion{Ne}{5}]/[\ion{Ne}{2}]} & \colhead{$W_\mathrm{eq}$(PAH 7.7\micron)} & \colhead{Laurent} & \colhead{$L$(MIR)/$L$(FIR)} & \colhead{$f_{30}/f_{15}$} & \colhead{All}\\
\colhead{(1)} & \colhead{(2)} & \colhead{(3)} & \colhead{(4)} & \colhead{(5)} & \colhead{(6)} & \colhead{(7)} & \colhead{(9)}
}
\startdata
          All ULIRGs  &$72-74$  &38.5($\pm$22.5) &42.4($\pm$26.3) &40.0($\pm$23.2) &37.4($\pm$26.5) &31.5($\pm$25.9) &43.3($\pm$27.1) &38.8($\pm$21.1)\\
\cutinhead{Spectral Type}
          \ion{H}{2}  &$17-18$  &$<$26.7($\pm$15.2) &$<$28.8($\pm$19.6) &29.1($\pm$23.1) &20.4($\pm$16.8) &22.5($\pm$19.0) &33.1($\pm$24.9) &27.1($\pm$15.0)\\
               LINER  &$31-32$  &$<$30.8($\pm$12.9) &$<$31.1($\pm$17.4) &36.9($\pm$21.0) &35.8($\pm$23.8) &20.4($\pm$18.0) &32.2($\pm$19.6) &31.2($\pm$14.6)\\
                Sey2  &     12  &58.5($\pm$22.9) &70.2($\pm$20.5) &48.8($\pm$18.8) &44.8($\pm$18.0) &40.7($\pm$21.3) &55.7($\pm$24.9) &53.1($\pm$16.2)\\
                Sey1  &      9  &63.7($\pm$27.9) &72.2($\pm$23.6) &65.2($\pm$14.1) &75.8($\pm$22.5) &76.0($\pm$17.4) &83.3($\pm$12.3) &72.7($\pm$14.4)\\
\cutinhead{log($f_{25}/f_{60}$)}
       $X \leq -1.2$  &$ 9-10$  &$<$29.4($\pm$14.4) &$<$40.4($\pm$18.2) &32.9($\pm$21.6) &28.3($\pm$17.2) & 9.5($\pm$ 8.7) &25.0($\pm$18.5) &27.3($\pm$12.5)\\
$-1.2 < X \leq -1.0$  &$29-30$  &$<$27.4($\pm$12.5) &$<$27.1($\pm$17.0) &30.4($\pm$19.2) &28.2($\pm$19.3) &16.4($\pm$12.3) &27.7($\pm$17.8) &26.4($\pm$12.1)\\
$-1.0 < X \leq -0.8$  &     13  &   43.8($\pm$22.8) &   44.8($\pm$24.6) &39.4($\pm$21.6) &37.2($\pm$26.6) &30.1($\pm$15.1) &43.6($\pm$21.8) &39.8($\pm$16.2)\\
          $-0.8 < X$  &     21  &   54.5($\pm$26.0) &   62.7($\pm$28.0) &57.4($\pm$21.1) &55.0($\pm$31.2) &64.2($\pm$18.6) &74.3($\pm$15.5) &61.3($\pm$18.8)\\
\cutinhead{log($L(IR)/L_\sun$)}
       $X \leq 12.2$  &     27  &$<$37.2($\pm$21.4) &$<$37.7($\pm$27.4) &34.7($\pm$25.5) &32.7($\pm$28.7) &27.8($\pm$24.3) &38.5($\pm$22.9) &34.8($\pm$21.9)\\
$12.2 < X \leq 12.4$  &$23-24$  &$<$31.1($\pm$18.0) &$<$39.5($\pm$24.2) &35.7($\pm$21.9) &30.3($\pm$19.4) &26.9($\pm$21.3) &40.2($\pm$28.8) &34.1($\pm$16.1)\\
$12.4 < X \leq 12.6$  &     14  &   43.7($\pm$24.3) &   45.3($\pm$24.8) &46.0($\pm$18.4) &44.5($\pm$24.9) &33.3($\pm$25.9) &49.0($\pm$27.4) &43.6($\pm$20.5)\\
          $12.6 < X$  &$ 8- 9$  &   55.4($\pm$27.1) &   61.0($\pm$27.3) &57.9($\pm$16.6) &58.9($\pm$28.8) &51.7($\pm$34.9) &57.3($\pm$31.8) &55.9($\pm$24.4)\\
\cutinhead{Number of Nuclei}
      Binary Systems  &     31  &$<$36.5($\pm$19.9) &$<$36.3($\pm$24.5) &35.3($\pm$25.6) &31.2($\pm$25.4) &23.2($\pm$21.3) &30.7($\pm$25.1) &32.2($\pm$20.3)\\
      Single Systems  &$34-36$  &   43.3($\pm$25.0) &   51.3($\pm$26.7) &45.3($\pm$21.0) &43.5($\pm$27.7) &40.4($\pm$28.5) &54.6($\pm$25.6) &46.2($\pm$20.9)\\
\cutinhead{Nuclear Separation (kpc)}
          $X \leq 1$  &     28  &$<$42.1($\pm$25.1) &$<$47.2($\pm$28.1) &44.0($\pm$22.2) &43.8($\pm$29.8) &41.2($\pm$30.2) &51.7($\pm$28.2) &45.0($\pm$23.1)\\
      $1 < X \leq 6$  &     14  &$<$34.9($\pm$18.9) &$<$31.1($\pm$22.9) &30.3($\pm$25.8) &28.1($\pm$25.5) &19.6($\pm$20.1) &25.9($\pm$25.3) &28.3($\pm$20.2)\\
             $X > 6$  &     17  &$<$36.4($\pm$22.0) &$<$40.1($\pm$26.1) &38.4($\pm$26.3) &33.1($\pm$26.2) &26.9($\pm$21.8) &36.7($\pm$23.8) &35.3($\pm$20.5)\\
\cutinhead{Interaction Class}
                IIIa  &      9  &$<$35.5($\pm$23.9) &$<$42.7($\pm$26.1) &37.4($\pm$25.1) &37.7($\pm$30.8) &26.1($\pm$17.8) &38.9($\pm$17.1) &36.4($\pm$18.9)\\
                IIIb  &     22  &$<$36.9($\pm$18.7) &$<$33.6($\pm$23.9) &34.4($\pm$26.4) &28.5($\pm$23.1) &22.0($\pm$22.9) &27.4($\pm$27.3) &30.5($\pm$21.0)\\
                 IVa  &$ 7- 8$  &$<$31.5($\pm$14.6) &$<$37.1($\pm$16.0) &36.5($\pm$18.4) &26.5($\pm$15.4) &15.1($\pm$13.1) &33.1($\pm$13.6) &29.7($\pm$ 8.2)\\
                 IVb  &$18-19$  &   48.9($\pm$30.2) &   58.6($\pm$30.7) &46.4($\pm$22.7) &47.8($\pm$29.4) &48.1($\pm$26.5) &60.4($\pm$26.3) &51.6($\pm$22.1)\\
                   V  &      9  &   41.5($\pm$17.3) &   47.7($\pm$20.6) &50.9($\pm$18.7) &49.5($\pm$29.0) &47.0($\pm$31.5) &61.2($\pm$23.5) &49.6($\pm$20.0)\\
\enddata
\tablecomments{
Bolometric corrections are computed using the denominator ($N$) in each quantity above, according to the formula AGN\%$(L(bol))/100 \equiv L(bol)^{agn}/(L(bol)^{agn}+L(bol)^{sb}) = 1 / \{ 1 + [100 / AGN\%(N) - 1] \times (N/L(bol))_{agn} / (N/L(bol))_{sb} \}$.  Col.(1): Range of quantity over which AGN contribution is computed.  Col.(2): Number of galaxies in each bin.  Col.(3): Average AGN contribution to the bolometric luminosity computed from the [\ion{O}{4}]/[\ion{Ne}{2}] line ratio, with standard deviation listed in parentheses.  Individual upper limits are included in the average calculation, and those categories where upper limits dominate the average are labeled as upper limits.  $\mathrm{AGN}\%/100 \equiv$ [\ion{Ne}{2}]$_{agn}$/([\ion{Ne}{2}]$_{starburst}+[$\ion{Ne}{2}]$_{agn}) =$ [\ion{O}{4}]/[\ion{Ne}{2}]$_{observed}$/([\ion{O}{4}]/[\ion{Ne}{2}]$_{agn} - [$\ion{O}{4}]/[\ion{Ne}{2}]$_{starburst})$.  In the latter expression, we assume constant line ratios for a pure AGN or starburst.  Col.(4): Same as column 3, but for [\ion{Ne}{5}]/[\ion{Ne}{2}].  Col.(5): Average AGN contribution computed from the equivalent width of the PAH~7.7\micron\ feature, $W_{eq}$(PAH~7.7\micron), with standard deviation listed in parentheses.  $\mathrm{AGN}\% \equiv f^{agn}_{8\micron~\mathrm{continuum}}/[f^{sb}_8+f^{agn}_8] = 1 - \sqrt{W_{eq}^{obs}(\mathrm{PAH}~7.7\micron)/W_{eq}^{sb}(\mathrm{PAH}~7.7\micron)}$.  In this calculation, we assume PAH destruction due to AGN radiation, such that $f_{obs}(\mathrm{PAH}~7.7\micron) = f_{sb}(\mathrm{PAH}~7.7\micron)\times(1-\mathrm{AGN}\%)$.  Col.(6): Average AGN contribution computed from the Laurent et al. 2000 diagram, as modified by Armus et al. 2007, with standard deviation listed in parentheses.  For 3-component, 2-ratio mixing between an AGN, \ion{H}{2} region, and PDR using quantities $A = f(\mathrm{PAH}~6.2\micron)/f(5.3-5.8\micron)$ and $B = f(14-16\micron)/f(5.3-5.8\micron)$, $\mathrm{AGN}\%/100 \equiv f_{agn}(5.3-5.8\micron)/[f_{agn}(5.3-5.8\micron)+f_{h2}(5.3-5.8\micron)+f_{pdr}(5.3-5.8\micron)] = (A_{obs}B_{h2}+A_{pdr}B_{obs}+A_{h2}B_{pdr}-A_{obs}B_{pdr}-A_{h2}B_{obs}-A_{pdr}B_{h2}) / (A_{agn}B_{h2}+A_{pdr}B_{agn}+A_{h2}B_{pdr}-A_{agn}B_{pdr}-A_{h2}B_{agn}-A_{pdr}B_{h2})$.  Col.(7): Average AGN contribution computed from the $L$(MIR)/$L$(FIR) luminosity ratio, with standard deviation listed in parentheses.  The PAH and silicate emission have been removed from the measured MIR luminosity; only the blackbody dust emission remains.  $\mathrm{AGN}\%/100 \equiv L(FIR)_{agn}/[L(FIR)_{sb}+L(FIR)_{agn}] = [L(MIR)/L(FIR)_{obs} - L(MIR)/L(FIR)_{sb}] / [L(MIR)/L(FIR)_{agn} - L(MIR)/L(FIR)_{sb}]$.  Col.(8): Average AGN contribution computed from the $f_{30\micron}/f_{15\micron}$ flux density ratio, with standard deviation listed in parentheses.  The formula used is the same as for the $L$(MIR)/$L$(FIR) diagnostic.  Col.(8) Average-of-averages, with standard deviation listed in parentheses.  We first average over the 6 methods for each galaxy, and then average over all galaxies.  [See Appendix A for more information on the individual methods.]
}
\label{tab:agnfrac_avg}
\end{deluxetable}

\begin{deluxetable}{crrrrrrr}
\tablecolumns{8}
\tabletypesize{\scriptsize}
\tablecaption{AGN Contributions}
\tablewidth{0pt}
\tablehead{
\colhead{} & \multicolumn{6}{c}{Method} & \colhead{} \\
\colhead{Galaxy} & \colhead{1} & \colhead{2} & \colhead{3} & \colhead{4} & \colhead{5} & \colhead{6} & \colhead{Avg} \\
\colhead{(1)} & \colhead{(2)} & \colhead{(3)} & \colhead{(4)} & \colhead{(5)} & \colhead{(6)} & \colhead{(7)} & \colhead{(8)}
}
\startdata
\cutinhead{ULIRGs}
                 F00091$-$0738 &      $<$   50.6 &       $<$   63.9 & \phm{$<$}   54.6 & \phm{$<$}   16.6 & \phm{$<$}   47.8 & \phm{$<$}   26.1 &    \phm{$<$}   43.3 \\
                 F00188$-$0856 &      $<$   21.0 &       $<$   22.1 & \phm{$<$}   57.4 & \phm{$<$}   73.1 & \phm{$<$}   52.5 & \phm{$<$}   22.4 &    \phm{$<$}   41.4 \\
                 F00397$-$1312 &      $<$   26.6 &       $<$   33.6 & \phm{$<$}   27.6 & \phm{$<$}   44.1 & \phm{$<$}   67.1 & \phm{$<$}   59.0 &    \phm{$<$}   43.0 \\
              F00456$-$2904:SW &      $<$   18.5 &       $<$   12.7 & \phm{$<$}    8.1 & \phm{$<$}    6.6 & \phm{$<$}   15.6 & \phm{$<$}   11.2 &    \phm{$<$}   12.1 \\
                 F00482$-$2721 &\phm{$<$}   36.6 &       $<$   29.2 & \phm{$<$}   26.1 & \phm{$<$}   29.4 & \phm{$<$}    0.0 & \phm{$<$}    5.0 &    \phm{$<$}   21.0 \\
\enddata
\tablecomments{
Col.(1): Galaxy name.  Col.($2-8$): Percent of the bolometric luminosity produced by the AGN, as determined using six different methods, as well as the average percent over all six methods.  Methods used: 1 = [\ion{O}{4}]/[\ion{Ne}{2}]; 2 = [\ion{Ne}{5}]/[\ion{Ne}{2}]; 3 = $W_\mathrm{eq}$(PAH 7.7\micron); 4 = Laurent; 5 = $L$(MIR)/$L$(FIR); and 6 = $f_{30}/f_{15}$.  [See Appendix A for more information on the individual methods.]
}
\label{tab:agnfrac_indiv}
\end{deluxetable}

\begin{deluxetable}{crrrrrr}
\tablecaption{Statistics on Regions of $W_{eq}$(PAH 7.7\micron) vs. $\tau_{9.7}$ Space}
\tablecolumns{7}
\tablewidth{0pt}
\tablehead{
\colhead{} & \multicolumn{3}{c}{\# of Galaxies} & \multicolumn{3}{c}{\% of Total} \\
\colhead{Quantity} & \colhead{R1} & \colhead{R2} & \colhead{R3} & \colhead{R1} & \colhead{R2} & \colhead{R3} \\
\colhead{(1)} & \colhead{(2)} & \colhead{(3)} & \colhead{(4)} & \colhead{(5)} & \colhead{(6)} & \colhead{(7)}
}
\startdata
\cutinhead{Regions Subdivided by AGN \%}
$<$20\%   & 15 &  2 &  0 &    65\% &     8\% &     0\% \\
$20-40$\% &  8 & 12 &  1 &    35\% &    48\% &     4\% \\
$40-60$\% &  0 & 11 & 11 &     0\% &    44\% &    44\% \\
$>$60\%   &  0 &  0 & 13 &     0\% &     0\% &    52\% \\
All       & 23 & 25 & 25 &   100\% &   100\% &   100\% \\
\cutinhead{Regions Subdivided by Nuclear Separation}
$>$6 kpc  &  8 &  4 &  5 &    36\% &    24\% &    26\% \\
$1-6$ kpc &  6 &  5 &  2 &    27\% &    29\% &    11\% \\
$<$1 kpc  &  8 &  8 & 12 &    36\% &    47\% &    63\% \\
All       & 22 & 17 & 19 &   100\% &   100\% &   100\% \\
\cutinhead{Regions Subdivided by Interaction Class}
IIIab     & 13 & 10 &  7 &    68\% &    43\% &    29\% \\
IVab      &  5 &  9 & 13 &    26\% &    39\% &    54\% \\
V         &  1 &  4 &  4 &     5\% &    17\% &    17\% \\
All       & 19 & 23 & 24 &   100\% &   100\% &   100\% \\
\enddata
\tablecomments{
Col.(1): Quantity by which regions R1 -- R3 in $W_{eq}$(PAH 7.7\micron) vs. $\tau_{9.7}$ phase space are subdivided.  Col.($2-4$): Number of galaxies in each region.  Col.($5-7$): Percentage of galaxies in each region.
}
\label{tab:weqpah7_v_tau}
\end{deluxetable}

\begin{deluxetable}{ccrrrrrrr}
\tablecolumns{9}
\tablecaption{Binned Eddington Ratio, Based on Photometric Black Hole Mass Estimates}
\tablewidth{0pt}
\tabletypesize{\footnotesize}
\tablehead{
\colhead{Bin} & \colhead{No.} & \colhead{[\ion{O}{4}]/[\ion{Ne}{2}]} & \colhead{[\ion{Ne}{5}]/[\ion{Ne}{2}]} & \colhead{$W_\mathrm{eq}$(PAH 7.7\micron)} & \colhead{Laurent} & \colhead{$L$(MIR)/$L$(FIR)} & \colhead{$f_{30}/f_{15}$} & \colhead{All}\\
\colhead{(1)} & \colhead{(2)} & \colhead{(3)} & \colhead{(4)} & \colhead{(5)} & \colhead{(6)} & \colhead{(7)} & \colhead{(9)}
}
\startdata
          All ULIRGs  &$58-61$  &-1.07($\pm$0.38) &-1.03($\pm$0.41) &-1.04($\pm$0.47) &-1.11($\pm$0.48) &-1.21($\pm$0.48) &-0.97($\pm$0.39) &-1.07($\pm$0.40)\\
\cutinhead{Spectral Type}
          \ion{H}{2}  &$11-12$  &-1.09($\pm$0.42) &-1.03($\pm$0.50) &-1.01($\pm$0.62) &-1.15($\pm$0.58) &-1.11($\pm$0.60) &-0.83($\pm$0.43) &-1.03($\pm$0.50)\\
               LINER  &$24-26$  &-1.06($\pm$0.39) &-1.07($\pm$0.42) &-1.00($\pm$0.41) &-1.09($\pm$0.45) &-1.32($\pm$0.46) &-1.02($\pm$0.37) &-1.10($\pm$0.37)\\
                Sey2  &     11  &-0.95($\pm$0.30) &-0.86($\pm$0.29) &-1.05($\pm$0.32) &-1.08($\pm$0.31) &-1.16($\pm$0.30) &-1.01($\pm$0.36) &-1.02($\pm$0.27)\\
                Sey1  &$ 8- 9$  &-1.12($\pm$0.41) &-1.04($\pm$0.40) &-0.97($\pm$0.39) &-0.91($\pm$0.43) &-1.00($\pm$0.47) &-0.95($\pm$0.46) &-1.03($\pm$0.44)\\
\cutinhead{log($f_{25}/f_{60}$)}
       $X \leq -1.2$  &$ 7- 8$  &-1.04($\pm$0.39) &-0.87($\pm$0.34) &-1.05($\pm$0.65) &-1.05($\pm$0.47) &-1.42($\pm$0.42) &-1.07($\pm$0.39) &-1.06($\pm$0.38)\\
$-1.2 < X \leq -1.0$  &$21-25$  &-1.11($\pm$0.36) &-1.13($\pm$0.42) &-1.04($\pm$0.50) &-1.14($\pm$0.50) &-1.35($\pm$0.49) &-1.00($\pm$0.37) &-1.14($\pm$0.42)\\
$-1.0 < X \leq -0.8$  &     11  &-1.03($\pm$0.30) &-1.00($\pm$0.29) &-1.06($\pm$0.39) &-1.12($\pm$0.48) &-1.22($\pm$0.41) &-1.05($\pm$0.40) &-1.08($\pm$0.31)\\
          $-0.8 < X$  &$16-18$  &-1.05($\pm$0.46) &-0.98($\pm$0.48) &-1.01($\pm$0.42) &-1.11($\pm$0.50) &-0.92($\pm$0.42) &-0.86($\pm$0.41) &-0.99($\pm$0.44)\\
\cutinhead{log($L(IR)/L_\sun$)}
       $X \leq 12.2$  &$15-17$  &-1.10($\pm$0.42) &-1.10($\pm$0.48) &-1.16($\pm$0.40) &-1.34($\pm$0.55) &-1.28($\pm$0.42) &-1.08($\pm$0.37) &-1.16($\pm$0.42)\\
$12.2 < X \leq 12.4$  &$20-23$  &-1.12($\pm$0.35) &-1.05($\pm$0.40) &-1.10($\pm$0.58) &-1.14($\pm$0.49) &-1.22($\pm$0.54) &-0.92($\pm$0.42) &-1.09($\pm$0.42)\\
$12.4 < X \leq 12.6$  &     13  &-0.98($\pm$0.30) &-0.96($\pm$0.28) &-0.94($\pm$0.35) &-0.98($\pm$0.32) &-1.17($\pm$0.41) &-0.92($\pm$0.29) &-0.99($\pm$0.28)\\
          $12.6 < X$  &$ 8- 9$  &-0.98($\pm$0.51) &-0.94($\pm$0.51) &-0.79($\pm$0.36) &-0.83($\pm$0.36) &-1.10($\pm$0.57) &-0.98($\pm$0.50) &-0.98($\pm$0.47)\\
\cutinhead{Number of Nuclei}
      Binary Systems  &$21-25$  &-1.09($\pm$0.34) &-1.11($\pm$0.39) &-1.14($\pm$0.45) &-1.24($\pm$0.44) &-1.40($\pm$0.49) &-1.11($\pm$0.39) &-1.18($\pm$0.39)\\
      Single Systems  &$34-35$  &-1.04($\pm$0.41) &-0.96($\pm$0.43) &-0.96($\pm$0.49) &-1.03($\pm$0.51) &-1.07($\pm$0.44) &-0.89($\pm$0.38) &-0.99($\pm$0.40)\\
\cutinhead{Nuclear Separation (kpc)}
          $X \leq 1$  &$21-23$  &-0.99($\pm$0.41) &-0.92($\pm$0.45) &-0.91($\pm$0.44) &-0.99($\pm$0.55) &-0.98($\pm$0.45) &-0.85($\pm$0.40) &-0.95($\pm$0.42)\\
      $1 < X \leq 6$  &$10-12$  &-0.95($\pm$0.29) &-1.05($\pm$0.37) &-1.02($\pm$0.38) &-1.20($\pm$0.49) &-1.35($\pm$0.57) &-1.09($\pm$0.51) &-1.11($\pm$0.41)\\
             $X > 6$  &$11-13$  &-1.21($\pm$0.34) &-1.17($\pm$0.42) &-1.22($\pm$0.50) &-1.27($\pm$0.40) &-1.44($\pm$0.42) &-1.14($\pm$0.28) &-1.25($\pm$0.37)\\
\cutinhead{Interaction Class}
                IIIa  &      8  &-1.18($\pm$0.32) &-1.08($\pm$0.44) &-1.14($\pm$0.42) &-1.17($\pm$0.36) &-1.34($\pm$0.30) &-1.13($\pm$0.28) &-1.17($\pm$0.32)\\
                IIIb  &$13-17$  &-1.04($\pm$0.35) &-1.13($\pm$0.38) &-1.14($\pm$0.48) &-1.27($\pm$0.48) &-1.43($\pm$0.56) &-1.10($\pm$0.46) &-1.19($\pm$0.42)\\
                 IVa  &$ 7- 8$  &-1.13($\pm$0.46) &-1.07($\pm$0.46) &-1.14($\pm$0.67) &-1.25($\pm$0.59) &-1.44($\pm$0.46) &-1.11($\pm$0.36) &-1.19($\pm$0.42)\\
                 IVb  &$17-18$  &-1.11($\pm$0.39) &-1.03($\pm$0.44) &-1.01($\pm$0.42) &-1.04($\pm$0.45) &-1.06($\pm$0.37) &-0.93($\pm$0.35) &-1.02($\pm$0.38)\\
                   V  &      9  &-0.81($\pm$0.36) &-0.76($\pm$0.38) &-0.72($\pm$0.37) &-0.80($\pm$0.50) &-0.81($\pm$0.41) &-0.64($\pm$0.35) &-0.76($\pm$0.37)\\
\enddata
\tablecomments{
Col.(1): Bin in which Eddington ratio is averaged.  Col.(2): Number of galaxies in each bin.  Col.(3-9): Average Eddington ratios, computed from the measured AGN luminosity and the photometrically-determined black hole mass.  The AGN luminosity is measured as a fraction of the bolometric luminosity; this AGN contribution is estimated from each of 6 different mid-infrared diagnostics.  Col.(10): Eddington ratio averaged over the 6 different ways of computing the AGN contribution.
}
\label{tab:eddrat_phot_avg}
\end{deluxetable}

\begin{deluxetable}{ccrrrrrrr}
\tablecolumns{9}
\tablecaption{Binned Eddington Ratio, Based on Dynamical Black Hole Mass Estimates}
\tablewidth{0pt}
\tabletypesize{\footnotesize}
\tablehead{
\colhead{Bin} & \colhead{No.} & \colhead{[\ion{O}{4}]/[\ion{Ne}{2}]} & \colhead{[\ion{Ne}{5}]/[\ion{Ne}{2}]} & \colhead{$W_\mathrm{eq}$(PAH 7.7\micron)} & \colhead{Laurent} & \colhead{$L$(MIR)/$L$(FIR)} & \colhead{$f_{30}/f_{15}$} & \colhead{All}\\
\colhead{(1)} & \colhead{(2)} & \colhead{(3)} & \colhead{(4)} & \colhead{(5)} & \colhead{(6)} & \colhead{(7)} & \colhead{(9)}
}
\startdata
          All ULIRGs  &$22-25$  &-0.34($\pm$0.61) &-0.33($\pm$0.67) &-0.31($\pm$0.66) &-0.42($\pm$0.69) &-0.33($\pm$0.70) &-0.22($\pm$0.67) &-0.35($\pm$0.63)\\
\cutinhead{Spectral Type}
          \ion{H}{2}  &$ 7- 9$  &-0.34($\pm$0.63) &-0.36($\pm$0.70) &-0.25($\pm$0.73) &-0.43($\pm$0.76) &-0.18($\pm$0.72) & 0.01($\pm$0.67) &-0.34($\pm$0.67)\\
               LINER  &$ 5- 6$  &-0.54($\pm$0.66) &-0.60($\pm$0.76) &-0.57($\pm$0.66) &-0.70($\pm$0.59) &-0.63($\pm$0.77) &-0.61($\pm$0.64) &-0.61($\pm$0.63)\\
                Sey2  &$ 4- 5$  &-0.34($\pm$0.63) &-0.23($\pm$0.65) &-0.33($\pm$0.82) &-0.32($\pm$0.81) &-0.58($\pm$0.79) &-0.40($\pm$0.78) &-0.38($\pm$0.70)\\
                Sey1  &      4  & 0.02($\pm$0.62) & 0.07($\pm$0.58) & 0.07($\pm$0.51) & 0.09($\pm$0.50) & 0.13($\pm$0.47) & 0.17($\pm$0.46) & 0.09($\pm$0.51)\\
\cutinhead{log($f_{25}/f_{60}$)}
       $X \leq -1.2$  &$ 1- 3$  &-0.59($\pm$0.88) &-0.46($\pm$0.96) &-0.25($\pm$1.16) &-0.30($\pm$1.16) &-0.00($\pm$0.00) & 0.45($\pm$0.00) &-0.51($\pm$0.83)\\
$-1.2 < X \leq -1.0$  &$ 5- 6$  &-0.29($\pm$0.68) &-0.37($\pm$0.77) &-0.34($\pm$0.74) &-0.45($\pm$0.81) &-0.34($\pm$0.96) &-0.32($\pm$0.89) &-0.36($\pm$0.78)\\
$-1.0 < X \leq -0.8$  &$ 4- 5$  &-0.48($\pm$0.50) &-0.53($\pm$0.53) &-0.48($\pm$0.67) &-0.66($\pm$0.50) &-0.71($\pm$0.70) &-0.50($\pm$0.68) &-0.55($\pm$0.54)\\
          $-0.8 < X$  &$10-11$  &-0.23($\pm$0.60) &-0.19($\pm$0.66) &-0.23($\pm$0.63) &-0.33($\pm$0.69) &-0.17($\pm$0.60) &-0.11($\pm$0.56) &-0.22($\pm$0.60)\\
\cutinhead{log($L(IR)/L_\sun$)}
       $X \leq 12.2$  &$11-13$  &-0.55($\pm$0.48) &-0.60($\pm$0.54) &-0.58($\pm$0.56) &-0.69($\pm$0.60) &-0.65($\pm$0.64) &-0.54($\pm$0.58) &-0.60($\pm$0.52)\\
$12.2 < X \leq 12.4$  &$ 5- 6$  &-0.37($\pm$0.53) &-0.24($\pm$0.51) &-0.29($\pm$0.54) &-0.47($\pm$0.41) &-0.20($\pm$0.50) &-0.06($\pm$0.48) &-0.33($\pm$0.46)\\
$12.4 < X \leq 12.6$  &$ 1- 2$  &-0.25($\pm$1.32) &-0.37($\pm$1.46) &-0.18($\pm$1.26) &-0.21($\pm$1.29) & 0.65($\pm$0.00) & 0.73($\pm$0.00) &-0.25($\pm$1.33)\\
          $12.6 < X$  &      4  & 0.36($\pm$0.37) & 0.42($\pm$0.41) & 0.36($\pm$0.46) & 0.27($\pm$0.64) & 0.26($\pm$0.62) & 0.37($\pm$0.64) & 0.34($\pm$0.51)\\
\cutinhead{Number of Nuclei}
      Binary Systems  &$ 4- 6$  &-0.36($\pm$0.30) &-0.48($\pm$0.43) &-0.43($\pm$0.53) &-0.70($\pm$0.35) &-0.53($\pm$0.29) &-0.53($\pm$0.50) &-0.50($\pm$0.34)\\
      Single Systems  &$14-15$  &-0.16($\pm$0.64) &-0.07($\pm$0.61) &-0.10($\pm$0.63) &-0.19($\pm$0.69) &-0.13($\pm$0.77) &-0.00($\pm$0.69) &-0.14($\pm$0.65)\\
\cutinhead{Nuclear Separation (kpc)}
          $X \leq 1$  &$16-19$  &-0.32($\pm$0.68) &-0.29($\pm$0.74) &-0.28($\pm$0.70) &-0.36($\pm$0.74) &-0.24($\pm$0.80) &-0.18($\pm$0.76) &-0.32($\pm$0.71)\\
      $1 < X \leq 6$  &      1  & 0.15($\pm$0.00) & 0.25($\pm$0.00) & 0.18($\pm$0.00) &-0.33($\pm$0.00) &-0.14($\pm$0.00) & 0.13($\pm$0.00) & 0.04($\pm$0.00)\\
             $X > 6$  &$ 3- 5$  &-0.48($\pm$0.18) &-0.61($\pm$0.24) &-0.65($\pm$0.41) &-0.83($\pm$0.31) &-0.62($\pm$0.20) &-0.43($\pm$0.30) &-0.57($\pm$0.21)\\
\cutinhead{Interaction Class}
                IIIa  &$ 1- 2$  &-0.56($\pm$0.23) &-0.80($\pm$0.13) &-1.08($\pm$0.00) &\nodata & -0.69($\pm$0.36) &-0.47($\pm$0.47) &-0.68($\pm$0.37)\\
                IIIb  &$ 3- 4$  &-0.26($\pm$0.30) &-0.33($\pm$0.45) &-0.21($\pm$0.37) &-0.55($\pm$0.20) &-0.43($\pm$0.26) &-0.56($\pm$0.58) &-0.41($\pm$0.34)\\
                 IVa  &      3  &-0.43($\pm$0.60) &-0.45($\pm$0.58) &-0.40($\pm$0.57) &-0.76($\pm$0.51) &-0.57($\pm$0.63) &-0.38($\pm$0.39) &-0.50($\pm$0.51)\\
                 IVb  &$ 7- 8$  &-0.33($\pm$0.66) &-0.19($\pm$0.60) &-0.31($\pm$0.62) &-0.31($\pm$0.62) &-0.28($\pm$0.82) &-0.19($\pm$0.74) &-0.31($\pm$0.64)\\
                   V  &      4  & 0.38($\pm$0.35) & 0.47($\pm$0.35) & 0.47($\pm$0.31) & 0.44($\pm$0.49) & 0.45($\pm$0.52) & 0.61($\pm$0.43) & 0.47($\pm$0.39)\\
\enddata
\tablecomments{
Col.(1): Bin in which Eddington ratio is averaged.  Col.(2): Number of galaxies in each bin.  Col.(3-9): Average Eddington ratios, computed from the measured AGN luminosity and the dynamically determined black hole mass.  The AGN luminosity is measured as a contribution of the bolometric luminosity; this AGN contribution is estimated from each of 6 different mid-infrared diagnostics.  Col.(10): Eddington ratio averaged over the 6 different ways of computing the AGN contribution.
}
\label{tab:eddrat_dyn_avg}
\end{deluxetable}

\begin{deluxetable}{crrrrrrrrrrrrrr}
\tablecolumns{15}
\tabletypesize{\scriptsize}
\tablecaption{Eddington Ratios from Photometry and Dynamics}
\tablewidth{0pt}
\tablehead{
\colhead{} & \multicolumn{7}{c}{log($L_{AGN}/L_{Edd})$, photometry} & \multicolumn{7}{c}{log($L_{AGN}/L_{Edd})$, dynamics} \\
\colhead{Galaxy} & \colhead{1} & \colhead{2} & \colhead{3} & \colhead{4} & \colhead{5} & \colhead{6} & \colhead{Avg} & \colhead{1} & \colhead{2} & \colhead{3} & \colhead{4} & \colhead{5} & \colhead{6} & \colhead{Avg} \\
\colhead{(1)} & \colhead{(2)} & \colhead{(3)} & \colhead{(4)} & \colhead{(5)} & \colhead{(6)} & \colhead{(7)} & \colhead{(8)} & \colhead{(9)} & \colhead{(10)} & \colhead{(11)} & \colhead{(12)} & \colhead{(13)} & \colhead{(14)} & \colhead{(15)}
}
\startdata
\cutinhead{ULIRGs}
                 F00091$-$0738 &      $<$  -0.80 &       $<$  -0.70 & \phm{$<$}  -0.77 & \phm{$<$}  -1.29 & \phm{$<$}  -0.83 & \phm{$<$}  -1.09 &    \phm{$<$}  -0.91 &      $<$   0.15 &       $<$   0.25 & \phm{$<$}   0.18 & \phm{$<$}  -0.33 & \phm{$<$}   0.13 & \phm{$<$}  -0.14 &    \phm{$<$}   0.04 \\
                 F00188$-$0856 &      $<$  -1.19 &       $<$  -1.17 & \phm{$<$}  -0.75 & \phm{$<$}  -0.65 & \phm{$<$}  -0.79 & \phm{$<$}  -1.16 &    \phm{$<$}  -0.95 &         \nodata &          \nodata &          \nodata &          \nodata &          \nodata &          \nodata &             \nodata \\
                 F00397$-$1312 &      $<$  -0.63 &       $<$  -0.53 & \phm{$<$}  -0.61 & \phm{$<$}  -0.41 & \phm{$<$}  -0.23 & \phm{$<$}  -0.29 &    \phm{$<$}  -0.45 &      $<$   0.85 &       $<$   0.95 & \phm{$<$}   0.86 & \phm{$<$}   1.07 & \phm{$<$}   1.25 & \phm{$<$}   1.19 &    \phm{$<$}   1.03 \\
              F00456$-$2904:SW &      $<$  -1.41 &       $<$  -1.57 & \phm{$<$}  -1.76 & \phm{$<$}  -1.85 & \phm{$<$}  -1.48 & \phm{$<$}  -1.62 &    \phm{$<$}  -1.62 &      $<$  -0.73 &       $<$  -0.89 & \phm{$<$}  -1.08 & \phm{$<$}  -1.17 & \phm{$<$}  -0.80 & \phm{$<$}  -0.94 &    \phm{$<$}  -0.94 \\
                 F00482$-$2721 &\phm{$<$}  -0.89 &       $<$  -0.99 & \phm{$<$}  -1.04 & \phm{$<$}  -0.98 &          \nodata & \phm{$<$}  -1.75 &    \phm{$<$}  -1.13 &         \nodata &          \nodata &          \nodata &          \nodata &          \nodata &          \nodata &             \nodata \\
\enddata
\tablecomments{
Col.(1): Galaxy name.  Col.($2-8$): log Eddington ratio, using AGN luminosity computed from six different methods and black hole mass from galaxy photometry, as well as the average ratio over all six methods of computing the AGN luminosity.  Methods used: 1 = [\ion{O}{4}]/[\ion{Ne}{2}]; 2 = [\ion{Ne}{5}]/[\ion{Ne}{2}]; 3 = $W_\mathrm{eq}$(PAH 7.7\micron); 4 = Laurent; 5 = $L$(MIR)/$L$(FIR); and 6 = $f_{30}/f_{15}$.  Col.($9-15$): log Eddington ratio, using AGN luminosity computed from the same six methods and black hole mass from galaxy dynamics, as well as the average ratio over all six methods of computing the AGN luminosity.
}
\label{tab:eddrat_indiv}
\end{deluxetable}

\clearpage

\clearpage

\clearpage

\clearpage

\begin{figure*}[ht]
\epsscale{1.0}
\plotone{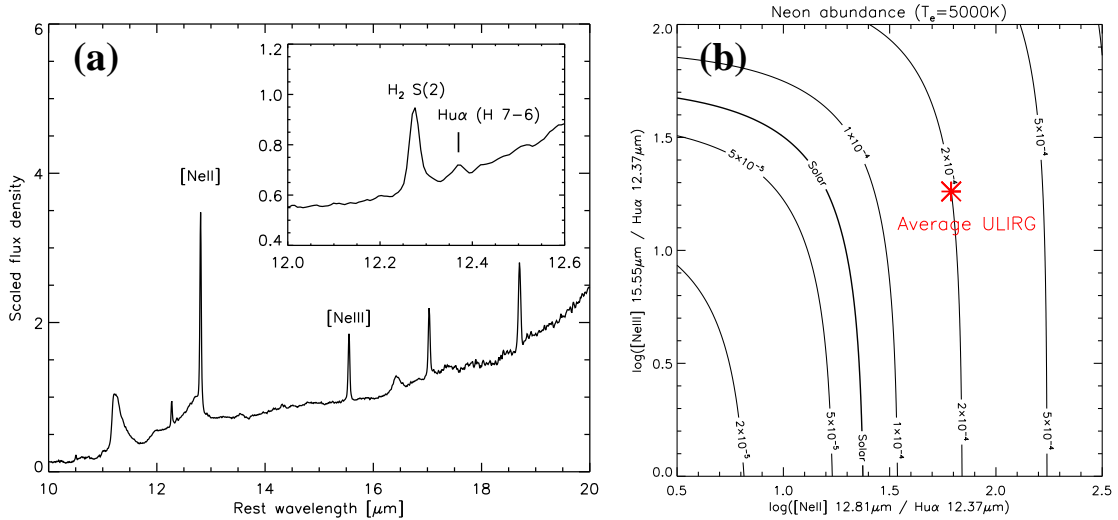}
\caption{ ($a$) Average IRS spectrum of PAH-dominated ULIRGs from
  Figure \ref{fig:avgspec-pah}, minus one object with high obscuration
  (F00397$-$1312).  The average spectrum is from individual spectra
  normalized at 15 \micron.  The zoomed-in insert shows the detection
  of Hu$\alpha$ 12.4 $\mu$m. ($b$) [Ne~III] 15.5 $\mu$m/Hu$\alpha$ and
  [Ne~II] 12.8 $\mu$m/Hu$\alpha$ ratios derived from this average
  spectrum. The neon abundance relative to hydrogen increases from the
  lower-left portion of this diagram to the upper-right, as indicated
  by the solid iso-metallicity curves. The line ratios of the average
  spectrum suggest a neon abundance $\sim$2.9 $\times$ solar, based on
  the Asplund et al.\ (2004) normalization (see text for a discussion
  of the uncertainties on this value). }
\label{fig:neabund}
\end{figure*}

\clearpage

\begin{figure*}[ht]
\epsscale{0.7}
\plotone{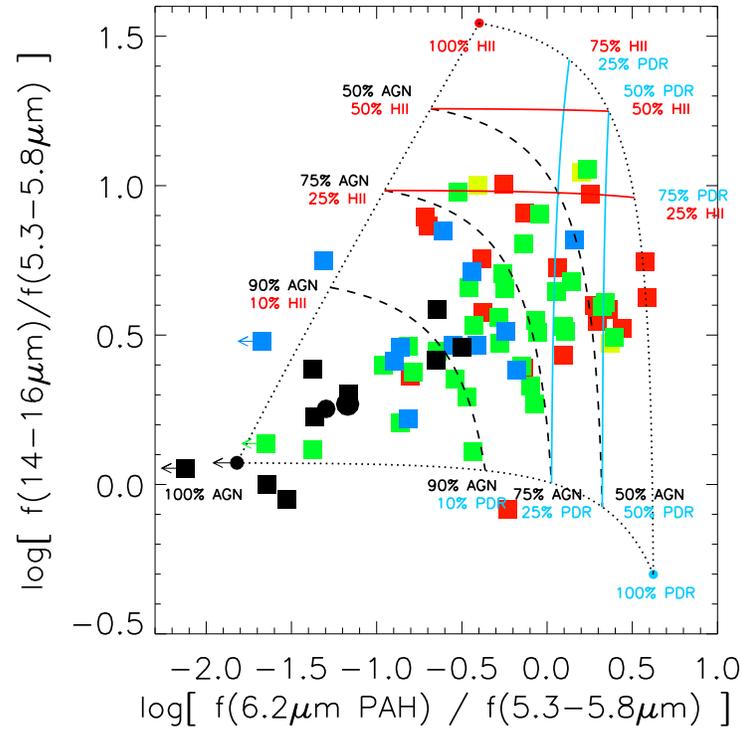}
\caption{AGN/HII/PDR mixing diagram based on the Laurent et al.\
  (2000) method, as modified by Armus et al. (2007): PAH (6.2 $\mu$m)
  to continuum (5.3 -- 5.8 $\mu$m) flux ratios {\em vs. \/} the
  continuum (14 -- 16 $\mu$m) / (5.3 -- 5.8 $\mu$m) flux ratios. The
  meaning of the symbols is the same as in Figure \ref{fig:fluxcomp}.
  The zero points for the pure HII region (upper-right) and PDR
  (lower-right) are from Armus et al.\ (2007) and the zero point for
  the pure AGN (lower-left) corresponds to the average value for the
  FIR-undetected PG~QSOs to reduce possible starburst contributions to
  the continuum emission (Paper II).  {\it Note that the percentages
    included here are percentages of the $5.3-5.8$ \micron\
    continuum}; actual AGN fractional contributions to the bolometric
  luminosities will be lower. }
\label{fig:laurent}
\end{figure*}

\begin{figure*}[ht]
\epsscale{1.0}
\plotone{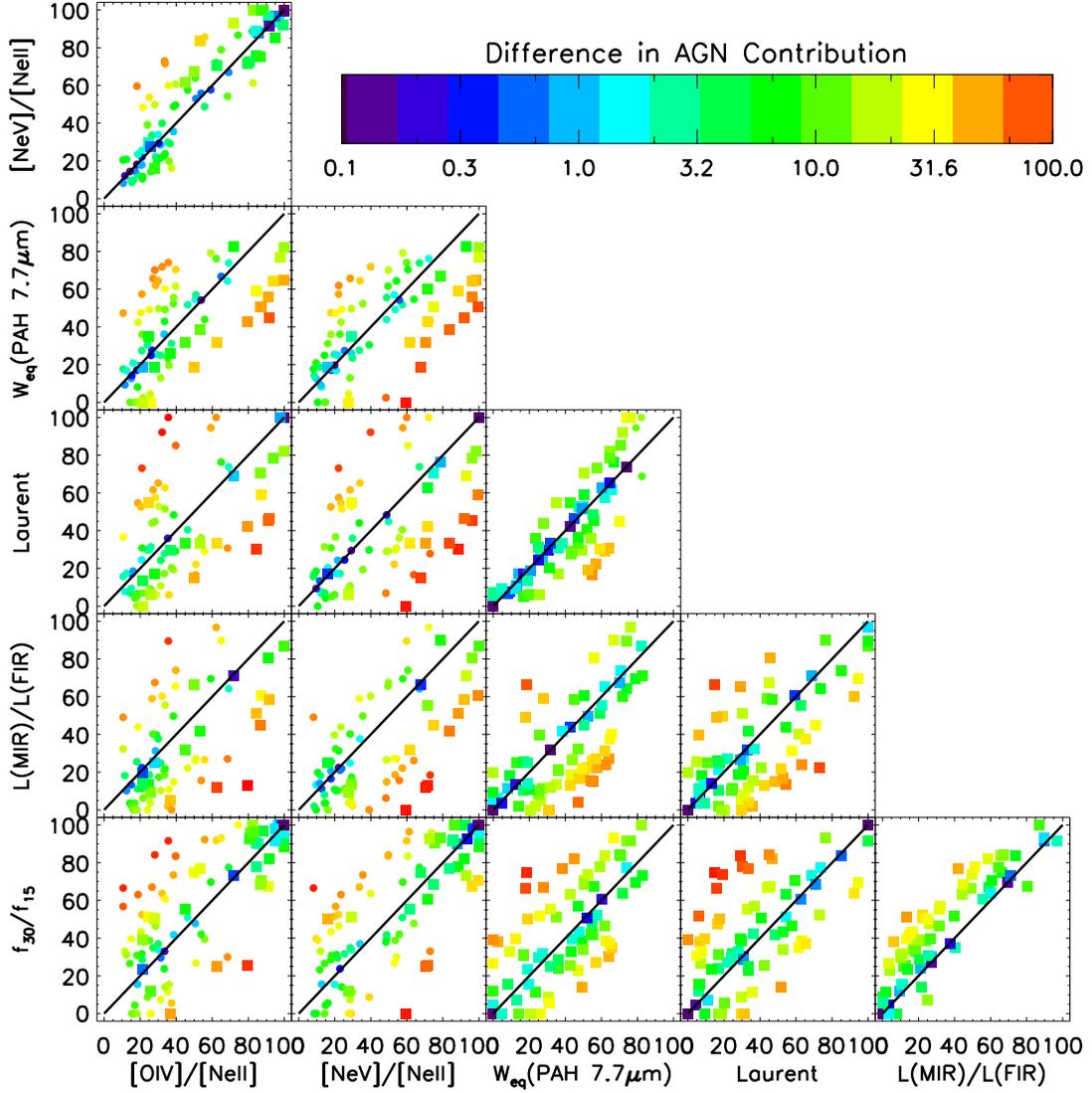}
\caption{ Comparison of AGN fractional contributions to the bolometric
  luminosities of ULIRGs and PG~QSOs derived from 6 different methods:
  the ([O~IV] 25.9 $\mu$m/[Ne~II] 12.8 $\mu$m) ratio, the ([Ne~V] 14.3
  $\mu$m/[Ne~II] 12.8 $\mu$m) ratio, the equivalent width of PAH 7.7
  $\mu$m, the PAH (5.9 - 6.8 $\mu$m) to continuum (5.1 -- 6.8 $\mu$m)
  flux ratio combined with the continuum (14 -- 15 $\mu$m) / (5.1 --
  5.8 $\mu$m) flux ratio, the MIR blackbody to FIR flux ratio, and the
  $f_{30}/f_{15}$ continuum flux ratio. Squares are actual AGN
  fractional contributions, while small circles are upper limits on
  the AGN fractional contributions derived from the fine structure
  line ratios. The colors represent deviations ($x - y$) from the line
  of equality, on a logarithmic scale. Good agreement is seen on
  average between the various methods. See Appendix A, Section 7.1.1,
  and Tables $\ref{tab:agnfrac_zp}-\ref{tab:agnfrac_indiv}$ for a
  description of each method and the results of these comparisons.}
\label{fig:agnfrac_comp}
\end{figure*}

\begin{figure*}[ht]
\epsscale{0.6}
\plotone{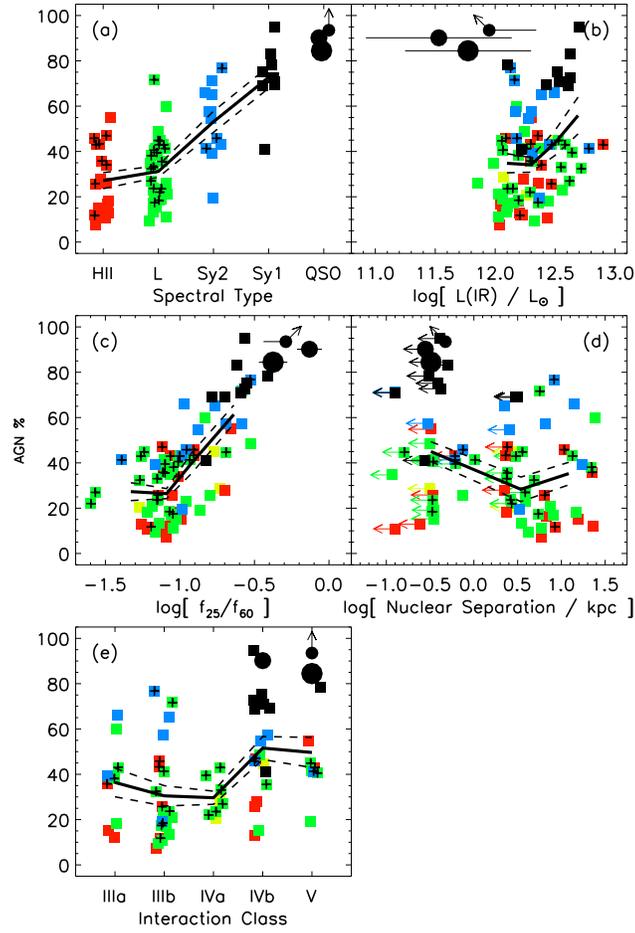}
\caption{AGN fractional contributions to the bolometric luminosities
  of ULIRGs and PG~QSOs, averaged over all 6 methods, {\it vs.} (a)
  optical spectral type; (b) infrared luminosity; (c) $f_{25}/f_{60}$;
  (d) nuclear separation; and (e) interaction class. The meaning of
  the small square (large circle) symbols is the same as in
  Figure \ref{fig:fluxcomp} (\ref{fig:mirfir_v}).  Overlaid crosses
  indicate ULIRGs with higher than average MIR extinction.
  The average ($\pm$1 standard error) AGN contributions in each
  horizontal bin are connected by the solid (dashed) lines.  These
  binned points include ULIRGs only.  The scatter in spectral type and
  interaction class [panels ($a$) and ($e$)] is added artificially for
  clarity. Strong positive correlations are detected between AGN
  contribution and optical spectral type / continuum slope: more
  Seyfert-like and warmer galaxies are increasingly AGN-dominated.
  Weaker, though significant, trends are seen for the other 3
  independent variables: AGN contribution is highest at the highest
  infrared luminosities, smallest nuclear separations, and latest
  interaction classes.  In all cases but infrared luminosity, the
  addition of PG QSOs either extends or enhances these trends.}
\label{fig:agnfrac_v}
\end{figure*}

\clearpage

\begin{figure*}[ht]
\epsscale{1.0}
\plotone{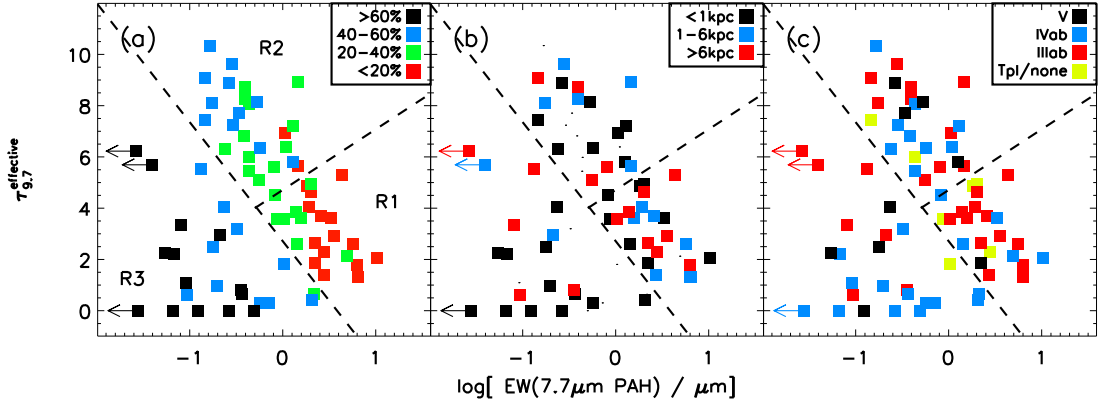}
\caption{ Effective optical depth of the 9.7 $\mu$m silicate feature
  {\em vs.} equivalent width of the 7.7 $\mu$m PAH feature as a
  function of (a) AGN fractional contribution to bolometric
  luminosity, (b) nuclear separation, and (c) interaction class. The
  meaning of the symbols is the same as in Figure \ref{fig:fluxcomp}.
  For clarity, we divide this diagram into three regions of roughly
  equal numbers.  AGN contribution increases as one moves from Region
  1 through Region 2 and into Region 3.  There is also a trend toward
  smaller nuclear separations and later merger stage along the
  sequence R1 $-$ R2 $-$ R3, but it is a weak trend with significant
  scatter (see Table \ref{tab:weqpah7_v_tau}).}
\label{fig:weqpah7_v_siltau_v}
\end{figure*}

\begin{figure*}[ht]
\epsscale{0.6}
\plotone{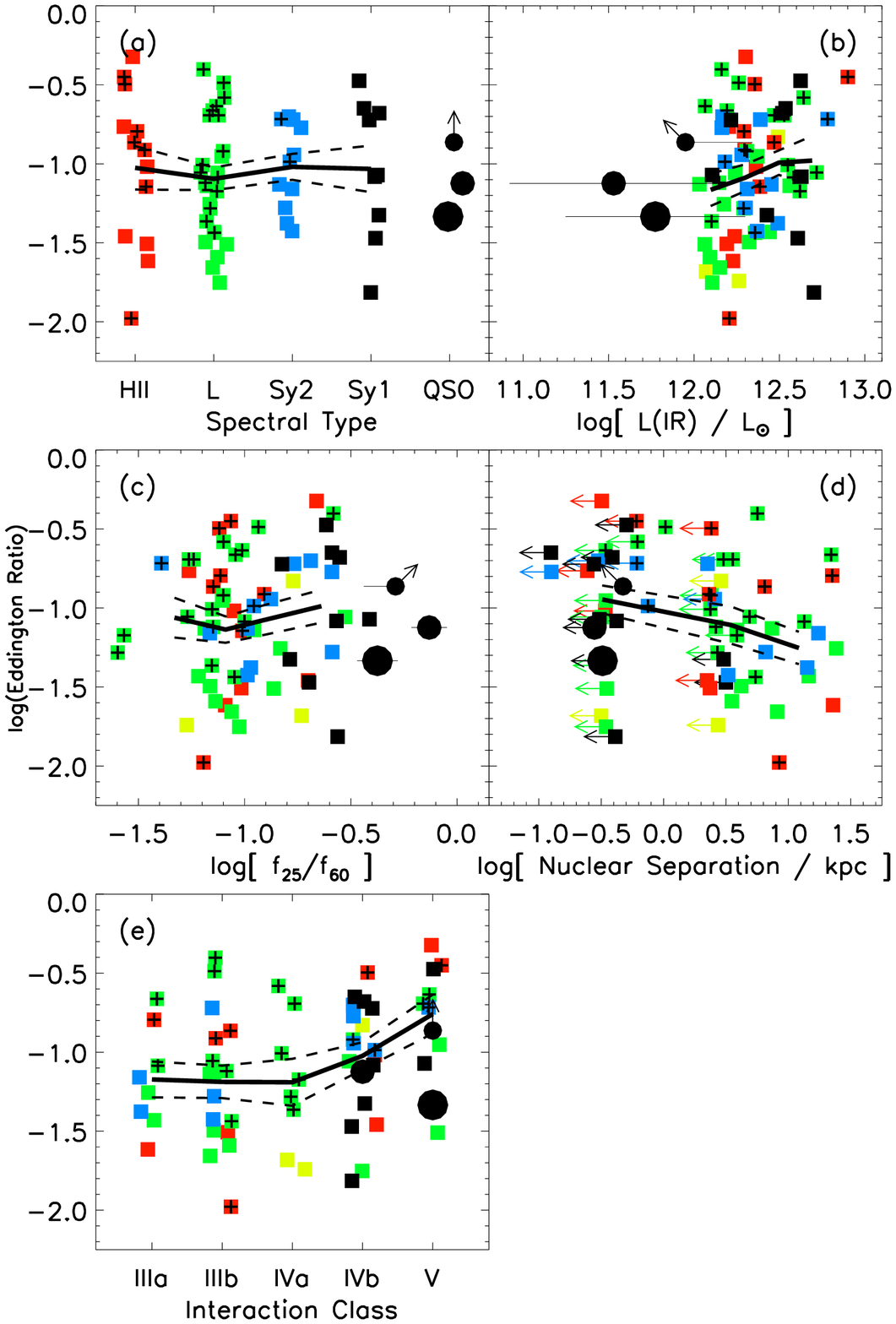}
\caption{Eddington ratio from photometry {\it vs.} (a) optical
  spectral type; (b) infrared luminosity; (c) $f_{25}/f_{60}$; (d)
  nuclear separation; and (e) interaction class. The meaning of
  the small square (large circle) symbols is the same as in
  Figure \ref{fig:fluxcomp} (\ref{fig:mirfir_v}).  
  Overlaid crosses indicate ULIRGs with higher than average
  MIR extinction.  The average ($\pm$1 standard error)
  Eddington ratios in each horizontal bin are connected by the solid
  (dashed) lines.  These binned points include ULIRGs only.  The
  scatter in spectral type and interaction class [panels ($a$) and
  ($e$)] is added artificially for clarity. There are no obvious
  correlations between Eddington ratios and any of these
  paramters. The Eddington ratios appear to be highest at the smallest
  nuclear separations and latest interaction classes, and the addition
  of PG~QSOs seems to either extend or enhance these trends, but these
  results cannot be confirmed in a more rigorous statistical
  analysis.}
\label{fig:eddrat_v}
\end{figure*}

\clearpage

\begin{figure*}[ht]
\epsscale{0.6}
\plotone{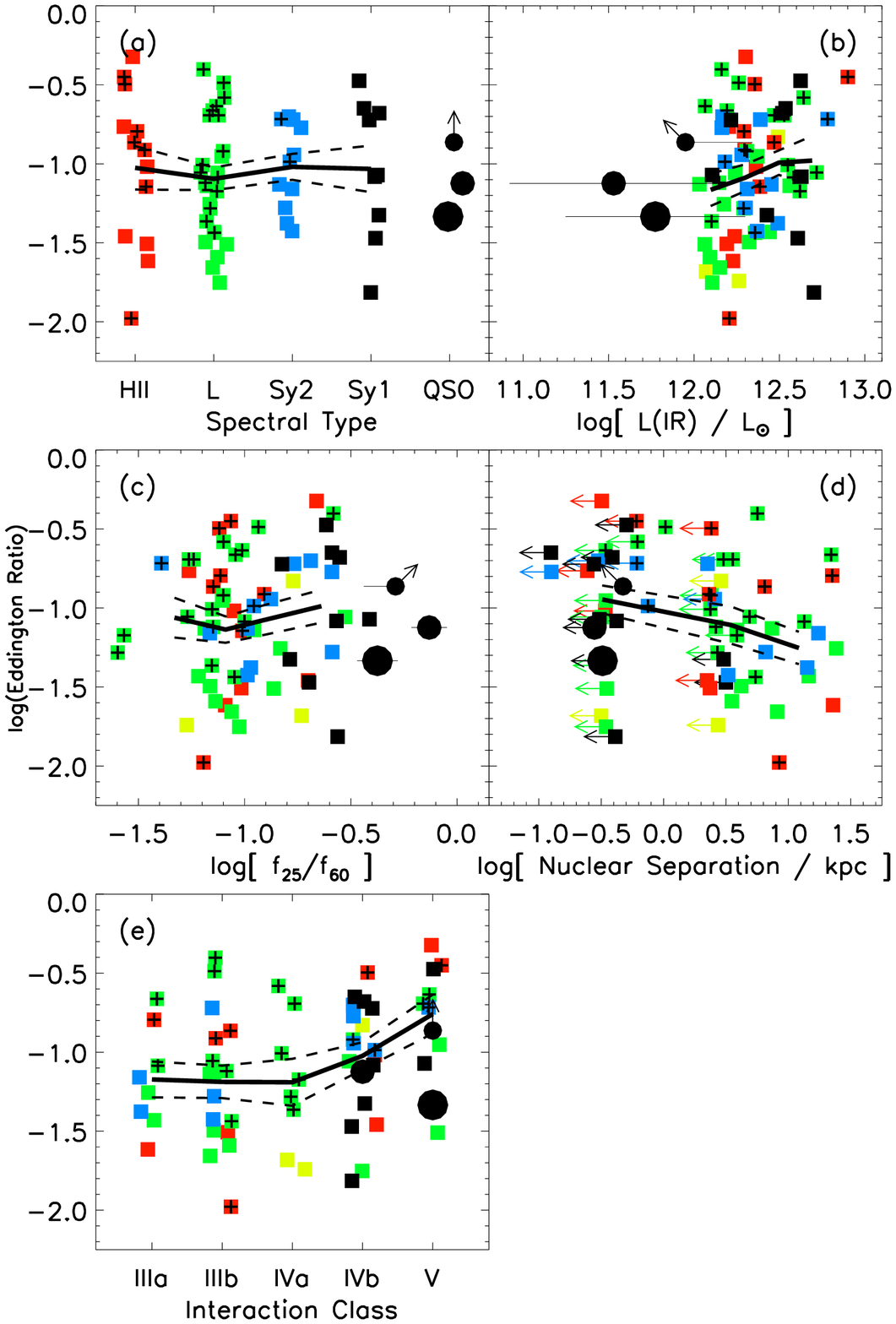}
\caption{Eddington ratio from photometry {\it vs.} (a) optical
  spectral type; (b) infrared luminosity; (c) $f_{25}/f_{60}$; (d)
  nuclear separation; and (e) interaction class. The meaning of
  the small square (large circle) symbols is the same as in
  Figure \ref{fig:fluxcomp} (\ref{fig:mirfir_v}).  
  Overlaid crosses indicate ULIRGs with higher than average
  MIR extinction.  The average ($\pm$1 standard error)
  Eddington ratios in each horizontal bin are connected by the solid
  (dashed) lines.  These binned points include ULIRGs only.  The
  scatter in spectral type and interaction class [panels ($a$) and
  ($e$)] is added artificially for clarity. There are no obvious
  correlations between Eddington ratios and any of these
  paramters. The Eddington ratios appear to be highest at the smallest
  nuclear separations and latest interaction classes, and the addition
  of PG~QSOs seems to either extend or enhance these trends, but these
  results cannot be confirmed in a more rigorous statistical
  analysis.}
\label{fig:eddrat_v}
\end{figure*}

\end{document}